\documentclass[a4paper,11pt]{article}
\pdfoutput=1 

\usepackage{jheppub} 

\usepackage[T1]{fontenc} 

\usepackage{booktabs,multirow}
\usepackage{amsfonts}
\usepackage{amsmath}
\usepackage{amssymb}
\usepackage{multirow}
\usepackage{dsfont}

\usepackage{caption}
\usepackage{subcaption}

\DeclareMathOperator*{\IM}{Im}
\DeclareMathOperator*{\res}{res}

\usepackage{graphicx}

\title{Bounding Violations of the Weak Gravity Conjecture}

\author[\gamma]{J. Henriksson,}
\author[\gamma]{B. McPeak,}
\author[\gamma]{F. Russo,}
\author[\gamma]{A. Vichi,}

\affiliation[\gamma]{Department of Physics, University of Pisa and INFN, \\Largo Pontecorvo 3, I-56127 Pisa, Italy}

\emailAdd{johan.henriksson@df.unipi.it}
\emailAdd{brian.mcpeak@df.unipi.it}
\emailAdd{francesco.russo@phd.unipi.it}
\emailAdd{alessandro.vichi@unipi.it}

\newcommand{\vc}[1]{\boldsymbol{#1}}
\newcommand{\mt}[1]{\mathbf{#1}}

\newcommand{\1}{{\mathds 1}}

\abstract{The black hole weak gravity conjecture (WGC) is a set of linear inequalities on the four-derivative corrections to Einstein--Maxwell theory. Remarkably, in four dimensions, these combinations appear in the $2 \to 2$ photon amplitudes, leading to the hope that the conjecture might be supported using dispersion relations. However, the presence of a pole arising in the forward limit due to graviton exchange greatly complicates the use of such arguments. In this paper, we apply recently developed numerical techniques to handle the graviton pole, and we find that standard dispersive arguments are not strong enough to imply the black hole WGC. Specifically, under a fairly typical set of assumptions, including weak coupling of the EFT and Regge boundedness, a small violation of the black hole WGC is consistent with unitarity and causality. We quantify the size of this violation, which vanishes in the limit where gravity decouples and also depends logarithmically on an infrared cutoff. We discuss the meaning of these bounds in various scenarios. We also implement a method for bounding amplitudes without manifestly positive spectral densities, which could be applied to any system of non-identical states, and we use it to improve bounds on the EFT of pure photons in absence of gravity.}

\begin{document} 
\maketitle
\flushbottom

\section{Introduction}

The effective field theory (EFT) describing the known universe at the lowest energies includes only photons and gravitons. The broad array of massive particles in the Standard Model and beyond leave their imprints on the low-energy world in the form of higher-derivative operators. The resulting EFT includes the Einstein--Hilbert term of gravity and the Maxwell term of electromagnetism, plus an infinite number of higher-dimensional operators,
\begin{align}
    \label{eq:EMaction}
    \mathcal{L} =  \sqrt{-g} \Big( \frac{M_{\mathrm{P}}^2}{2} R - \frac{1}{4} F_{\mu \nu} F^{\mu \nu} + \alpha_1 ( F_{\mu \nu} F^{\mu \nu})^2 + \alpha_2 ( F_{\mu \nu} \tilde{F}^{\mu \nu})^2  + \beta W_{\mu \nu \rho \sigma} F^{\mu \nu} F^{\rho \sigma} + \ldots \Big) \, .
\end{align}
In general, an $n$-derivative operator will introduce corrections to the observables which are suppressed by a factor of $(E / M)^{n-2}$ compared to the leading two-derivative contribution. Here $M$ refers to the scale of new physics -- it is the energy at which new poles or cuts appear in the amplitude. The Planck mass $M_{\mathrm{P}}$ determines the strength of the gravitational interaction;\footnote{In our conventions, the metric expands as $g_{\mu\nu}=\eta_{\mu\nu}+ \frac{2}{M_{\mathrm{P}}} h_{\mu\nu}$, and Newton's constant is given by $8 \pi G_{\text{N}} = \frac{1}{M_{\mathrm{P}}^2}$.} gravity decouples in the limit where $M_{\mathrm{P}} / M \to \infty$. The coefficients $\alpha_1$, $\alpha_2$, $\beta$, and so on are dimensionful. 
In examples where their contribution from UV physics is known, such as the Euler--Heisenberg EFT where they arise from integrating out a massive electron, they are order one numbers times powers of the coupling constant, in units of the scale of new physics $M$. It is a general expectation that this dimensional analysis should hold universally. 

Recent developments have made it possible to put this general expectation on a more rigorous footing. It has been clear for some time that not every EFT is consistent with some of the most basic principles of physics. Unitarity and causality imply positivity bounds \mbox{\cite{Pham:1985cr, Pennington:1994kc,  Ananthanarayan:1994hf, Comellas:1995hq, Dita:1998mh, Adams:2006sv}} -- constraints on the signs of EFT coefficients; such bounds are most efficiently derived with the aid of dispersion relations. 
An enormous amount of effort has gone into exploring the extent of these constraints, applying them broadly to EFTs across particle physics, quantum gravity, and cosmology \cite{Manohar:2008tc, Mateu:2008gv, Nicolis:2009qm, Baumann:2015nta, Bellazzini:2015cra, Bellazzini:2016xrt, Cheung:2016yqr, Bonifacio:2016wcb,  Cheung:2016wjt, deRham:2017avq, Bellazzini:2017fep, deRham:2017zjm, deRham:2017imi, Hinterbichler:2017qyt, Bonifacio:2017nnt, Bellazzini:2017bkb, Bonifacio:2018vzv, deRham:2018qqo, Zhang:2018shp, Bellazzini:2018paj, Bellazzini:2019xts, Melville:2019wyy, deRham:2019ctd, Alberte:2019xfh, Alberte:2019zhd, Bi:2019phv, Remmen:2019cyz, Ye:2019oxx, Herrero-Valea:2019hde, Zhang:2020jyn}. 
Recently, the methods for extracting constraints on EFTs from these basic requirements have been given a more systematic foundation \mbox{\cite{Arkani-Hamed:2020blm, Bellazzini:2020cot, Tolley:2020gtv, Caron-Huot:2020cmc, Sinha:2020win, Trott:2020ebl}}. This has led to a number of important outcomes, including a demonstration that $S$-matrix consistency implies two-sided bounds on ratios of EFT coefficients, essentially ``proving'' the intuition of dimensional analysis above, as well as a precise numerical recipe for obtaining optimal bounds. These methods have since been used to bound the Standard Model EFT \cite{Zhang:2021eeo}, systems of scalars \cite{Wang:2020jxr, Li:2021lpe, Du:2021byy} and spinning particles \cite{Davighi:2021osh, Chowdhury:2021ynh} including photons \cite{Henriksson:2021ymi} and gravitons \cite{Bern:2021ppb}. See also \cite{deRham:2022hpx} for a recent review.

Gravity presents a particular challenge for these methods, due to the so-called graviton pole, a $1 / u$ divergence in the forward limit that arises from graviton exchange. This obstacle may be surmounted by considering a dispersion relation with more subtractions, which simply removes the pole entirely \cite{Arkani-Hamed:2020blm, Bern:2021ppb}. 
This is not, however, entirely satisfactory because including more subtractions in the sum rule will typically remove the four-derivative interactions from the sum rule as well. These are the leading corrections, and they often have considerable theoretical interest. For an example of relevance to this paper, the four-derivative corrections to Einstein--Maxwell theory are required to obey a certain inequality if the weak gravity conjecture \cite{Arkani-Hamed:2006emk} is to be satisfied by the spectrum of black holes alone \cite{Kats:2006xp}. We shall review this in more detail below, but essentially this requires that 
\begin{align}
    4 \alpha_1  \pm \frac{\beta}{M_{\mathrm{P}}^2} > 0 \, , \qquad \alpha_2 > 0
    \label{eq:wgcbounds}
    \,,
\end{align}
in the parametrization of~\eqref{eq:EMaction}. Remarkably, it was shown \cite{Cheung:2014ega, Hamada:2018dde} that this so-called ``black-hole weak gravity conjecture'' immediately follows if the graviton pole may be safely ignored. 
A number of consequences of this observation were subsequently explored \cite{Bellazzini:2019xts, Alberte:2020jsk, Alberte:2020bdz}, with the conclusion that such bounds are probably not applicable, as they would imply impossibly strong constraints or a unrealistically low EFT cutoff. 
An alternative possibility was conjectured in \cite{Alberte:2020bdz}: \eqref{eq:wgcbounds} may be violated by a small amount without spoiling the consistency of the $S$-matrix. This insight was supported by recent results \cite{Hollowood:2015elj, deRham:2019ctd, deRham:2020zyh} where a weakening of the causality criteria was observed in EFT coupled to gravity. 
In fact, problems of superluminality in EFTs with gravity had been understood since 1980, when Drummond and Hathrell \cite{Drummond:1979pp} showed that the EFT that arises from integrating out an electron with dynamical gravity can allow light to travel superluminally on some backgrounds (see \cite{Goon:2016une} for nice recent analysis). Roughly, the resolution seems to be that gravitational interactions universally cause a time delay, so EFT operators that cause a time advance are allowed in principle as long as the advance is smaller than the gravitational time delay. 

In fact, many of these ideas are implicit in the work of \cite{Camanho:2014apa}, where three-point couplings such as $\beta$ are shown to cause a time advance which overwhelms the gravitational time delay unless there is an infinite tower of higher-spin particles. For the theory described by~\eqref{eq:EMaction}, this time delay argument requires that these new particles must enter with masses satisfying $M^2_{\text{HS}} \lesssim \frac{1}{\beta}$. This may be thought of as a bound on $\beta$. We must have $M \lesssim M_{\text{HS}}$, which suggests the naive scalings
\begin{align}
    \alpha_1 \sim \frac{1}{M^4} \qquad \alpha_2 \sim \frac{1}{M^4}, \qquad \beta \lesssim \frac{1}{M^2}
    \,,
\end{align}
where $M$ is the scale of new physics. Provided that $M$ is lower than the Planck mass, the inequalities~\eqref{eq:wgcbounds} will hold provided $\alpha_1$ and $\alpha_2$ are positive.\footnote{This is known to be the case without gravity: see \cite{Cheung:2014ega, Bellazzini2016talk, Falkowski, Henriksson:2021ymi}. 
Furthermore, as we show in this paper, $\beta$ gives a unique contribution to the scattering amplitudes which must be absent without gravity.} However, as anticipated above, we shall see in this paper that that is not the case. 

Our goal is to use $2 \to 2$ photon scattering amplitudes to derive bounds on Einstein--Maxwell theory, including on the four-derivative coefficients appearing in~\eqref{eq:EMaction}. A general method for finding such bounds in the presence of a graviton pole was given in \cite{Caron-Huot:2021rmr} and \cite{Caron-Huot:2022ugt}, where it was shown how to extract bounds on the leading four-derivative coefficients by acting on the dispersion relations with a more general class of functional. 
The result is, as expected, that a small amount of negativity is tolerated, but this negativity is essentially proportional to $M^2/M_{\mathrm{P}}^2$, and thus vanishes in the limit $M_{\mathrm{P}} \to \infty$, where gravity decouples. Applying this method to 4d requires care because infrared divergences preclude the existence of the positive functional needed for the argument. 
However, it was shown in \cite{Caron-Huot:2021rmr} that this issue can be circumvented in some cases by regulating the divergences with an infrared cutoff (IR), leading to a number of interesting bounds on modifications to Einstein gravity in four dimensions. We shall use the same approach to handling the graviton pole in this paper, though we shall see that there are a few issues plaguing us which did not appear in \cite{Caron-Huot:2022ugt} (essentially because corrections to Einstein gravity in 4d do not include any four-derivative operators). 

Another technical improvement we make in this paper, especially relative to \cite{Henriksson:2021ymi}, is to show how to bound amplitudes which do not have manifestly positive partial wave expansions. This may be accomplished using a more general approach to unitarity constraints, sometimes called the ``generalized optical theorem''. A similar method has been used recently in the case of gravity \cite{Bern:2021ppb,Caron-Huot:2022ugt} and more explicitly in \cite{Du:2021byy} for a system of multiple scalars. 
In the present case, this will allow us to obtain bounds on helicity amplitudes without positive partial wave expansions, such as $\mathcal{M}^{+++-}$, in terms of other amplitudes with manifest positivity. For the case when gravity decouples, we shall see that these bounds are stronger than the bounds we previously obtained in \cite{Henriksson:2021ymi}. For the case with gravity, we shall see that some negativity is allowed in the coefficients $\alpha_1$ and $\alpha_2$, and we find that $\beta$ is bounded by $\alpha_1$ and $\alpha_2$. 

\subsection{The black hole weak gravity conjecture}

Let us review the black hole weak gravity conjecture, and why our work is relevant to it. For a recent review of the literature, see \cite{Harlow:2022gzl}. 

The weak gravity conjecture (WGC) \cite{Arkani-Hamed:2006emk} was formulated as a criterion for determining which EFTs can be consistently coupled to quantum gravity. Such EFTs are said to live in the ``Landscape,'' in contrast with the EFTs which are inconsistent with quantum gravity and therefore live in the ``Swampland.'' The original version of the WGC states that there must be a particle whose charge is greater than its mass in Planck units, meaning 
\begin{align}
    q \geqslant \frac{m}{\sqrt2M_{\mathrm{P}}} \, . 
    \label{eq:wgc_ext}
\end{align} 
In this case, the electric repulsion of two equally charged particles would be stronger (or equal, if the equality is saturated) than the gravitational attraction; hence it is a state for which ``gravity is the weakest force.'' The requirement that such a state exists was motivated by the requirement that any non-supersymmetric black hole should be able to decay. The simplest possible case where such electrically charge black holes exist is Einstein--Maxwell theory, described by the leading terms in \eqref{eq:EMaction},
\begin{align}
    \label{eq:pureEM}
    \mathcal{L} =  \sqrt{-g} \left( \frac{M_{\mathrm{P}}^2}{2} R - \frac{1}{4} F_{\mu \nu} F^{\mu \nu} \right) \, .
\end{align}
Ignoring rotation and magnetic charges, this theory has a two-parameter family of black hole solutions, parametrized by mass $m$ and electric charge $q$:
\begin{align}
\begin{split}
    & ds^2 -f(r) dt^2 + g(r)^{-1} dr^2 + r^2 d\Omega^2 \, ,  \\
    & f(r) = g(r) = 1 - \frac{m}{M_{\mathrm{P}}^2 r} + \frac{q^2}{ 2 M_{\mathrm{P}}^2 r^2}\, , \qquad F^{01}(r) = \frac{ q}{r^2}
    \,.
\end{split}
\end{align}
The curvature of these spacetimes blows up as $r$ approaches zero; only those solutions where this point is hidden behind an event horizon are physically sensible. This means that the functions $f(r)$ and $g(r)$ must have a zero, which only happens when 
\begin{align}
\label{eq:chargetomassratio}
    m \geqslant \sqrt{2} M_{\mathrm{P}} \, q \, .
\end{align}
This is the black hole extremality bound: states satisfying it are called subextremal, those saturating it are extremal, and those violating it are called superextremal. Consider now the decay of a black hole with mass $m\geqslant m_1+m_2$ and charge $q=q_1+q_2$ into two daughter states: 
\begin{align}
    (m, q) \qquad \to \qquad (m_1, q_1) \quad \mathrm{and} \quad (m_2, q_2) \, .
\end{align}
If the initial black hole is extremal, \emph{i.e.} $\sqrt 2 q = m/M_{\mathrm{P}}$, then one of two options holds. Either both of the daughter states are exactly extremal (and $m=m_1+m_2$), or at least one of them is superextremal. This leads to the conclusion~\eqref{eq:wgc_ext}. More precisely, the WGC, in the original form of~\eqref{eq:wgc_ext}, states that theories of quantum gravity must have superextremal states so that their nearly extremal black holes can decay.

It is important to stress that there is no proof of this (or any) version of the WGC. For one, it is not at all clear why all black holes must be able to decay. Original arguments have included issues with large numbers of species \cite{Banks:2006mm} or remnants \cite{Susskind:1995da}. Another hint is the conceptual consistency with the no-global-symmetry conjecture \cite{Banks:2010zn, Harlow:2018tng}. The charge $q$ depends implicitly on the gauge coupling $g$, so the WGC will be violated if one takes $g \to 0$. In a sense, the WGC may be thought of as forbidding ``nearly global symmetries.'' None of these arguments amount to a proof of the conjecture. Nonetheless, the WGC has been observed in every
UV complete model known. In our own universe, it is resoundingly satisfied by the electron, where $\sqrt{2} q M_{\mathrm{P}} / m \simeq 2 \times 10^{21}$.

The WGC has been given a number of interesting extensions and generalizations (see \cite{Harlow:2022gzl}). One possibility, considered almost as early as the WGC itself, is that the superextremal states satisfying~\eqref{eq:wgc_ext} are black holes themselves \cite{Kats:2006xp}. The key idea of this work is that higher-derivative operators will shift the solution to the equations of motion, which may introduce corrections to the extremality bound. For the case of the Lagrangian given in~\eqref{eq:EMaction}, the equations of motion get corrected to
\begin{align}
\begin{split}
    & g(r) = 1 - \frac{ m}{ M_{\mathrm{P}}^2 r} + \frac{q^2}{2 M_{\mathrm{P}}^2 r^2}  + \Delta g(r) \, , \\
    & \Delta g(r) = - \frac{q^2}{2 M_{\mathrm{P}}^4 r^6} \left( \frac{8}{5} \alpha_1 q^2 M_{\mathrm{P}}^2 + \beta \left( \frac{8}{5} q^2 + \frac{8}{3} r^2 M_{\mathrm{P}}^2 - \frac{10}{3} m r \right) \right) + \mathcal{O}(\alpha_1^2, \ldots )  \   
    \,,
\end{split}
\end{align}
with analogous corrections to $f(r)$ and $F^{01}$ which are not important. Working to first order in the coefficients $\alpha_1$ and $\beta$, one finds that the shifted solution leads to a shifted extremality condition:
\begin{align}
\label{eq:massshift}
    \frac{m}{\sqrt{2} q M_{\mathrm{P}}} = 1 - \frac{4 M_{\mathrm{P}}^4}{5 q^2} \left( 4 \alpha_1 - \frac{\beta}{M_{\mathrm{P}}^2} \right) \,  +  \mathcal{O}(\alpha_1^2, \ldots)\,.
\end{align}
Let us imagine comparing two black holes, in the shifted and unshifted theory, which have the same charge and the minimal possible mass. Then the black hole in the theory with higher-derivative corrections can have a \emph{superextremal} charge-to-mass ratio, compared to \eqref{eq:chargetomassratio}, and still have an event horizon, if the mass shift in \eqref{eq:massshift} is \emph{negative}. We see that this occurs when 
\begin{align}
    4 \alpha_1 - \frac{\beta}{M_{\mathrm{P}}^2} > 0 \, .
\end{align}
We derived this inequality by considering only electric black holes: the other two inequalities of~\eqref{eq:wgcbounds} come from considering purely magnetic and dyonic black holes \cite{Jones:2019nev}. One of the main points of this paper is that causality constraints alone allow these inequalities to be violated by corrections proportional to $M^2 / M^2_{\mathrm{P}}$. 

\subsection{Overview of results}

\begin{figure}
     \centering
         \includegraphics[width=0.5\textwidth]{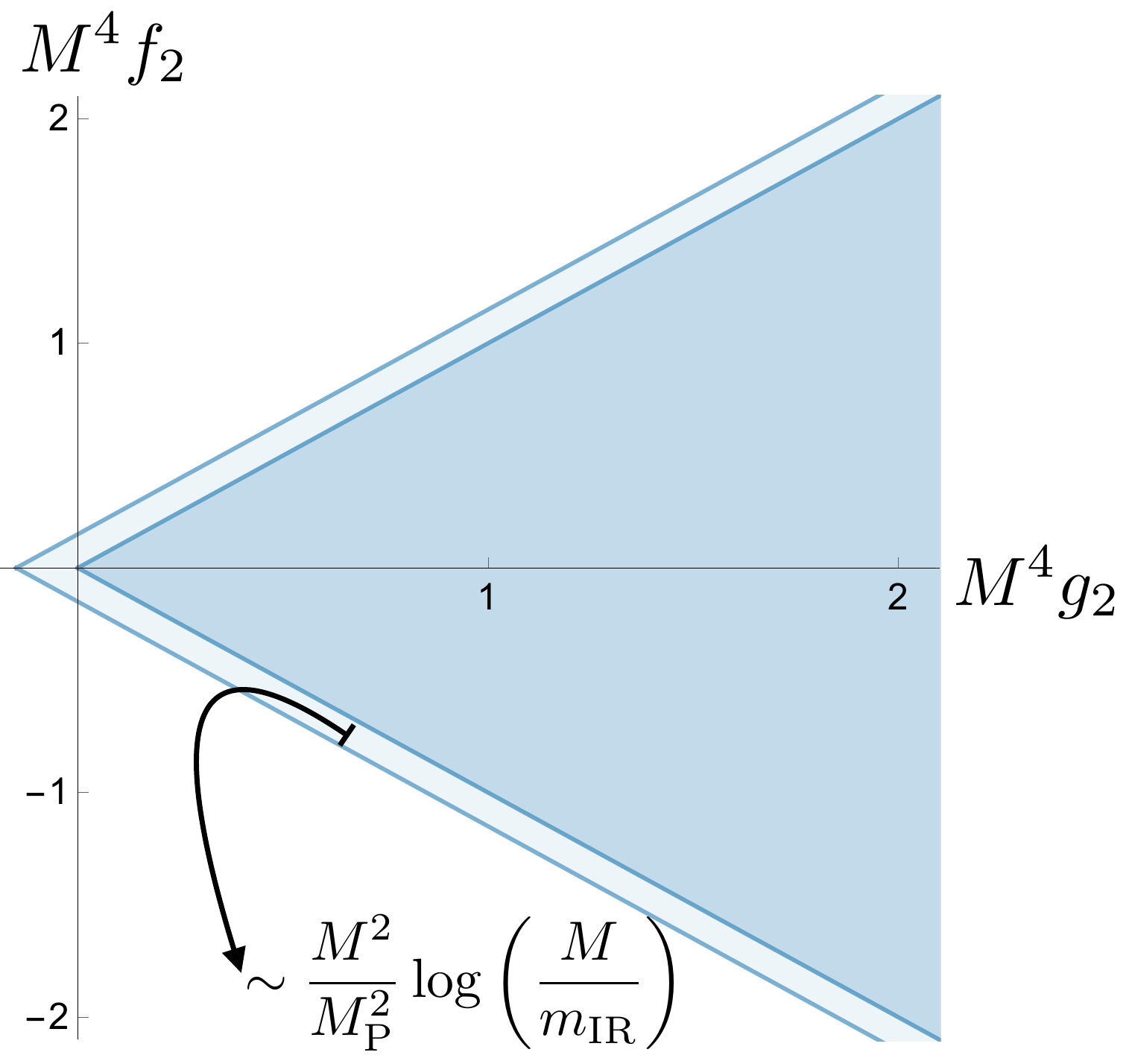}
        \caption{Schematic representation of our bounds: the weak gravity conjectures is satisfied at leading order  (dark shaded region), but violations are still admissible at sub-leading order in $M/M_\mathrm{P}$ (light shaded region).}
         \label{fig:schematicBounds}
\end{figure}

The purpose of this paper is to explore the use of dispersion relations and positivity bounds in the Einstein--Maxwell EFT. Our main conclusions are

\begin{itemize}
\item Using the generalized optical theorem, we bound quantities without manifestly positive spectral densities. This allows us to derive positivity bounds involving all three independent amplitudes $f\sim\mathcal{M}^{++++}$, $g\sim \mathcal{M}^{++--}$ and $h\sim\mathcal{M}^{+++-}$.
\item In the limit where gravity decouples, it is easy to prove the WGC inequalities~\eqref{eq:wgcbounds} by expanding in the forward limit. This is consistent with previous work \cite{Hamada:2018dde, Bellazzini:2019xts}, where it was shown that the WGC immediately follows if the graviton pole is discarded.
\item We show how to derive corrections to the $M_{\mathrm{P}} \to \infty$ limit. The strongest possible bounds with our approach allow for a violation of the WGC: introducing the notation $16 \alpha_{1,2}=g_2\pm f_2$, the WGC would require $g_2\geqslant |f_2|\geqslant 0$. Instead we find
\begin{equation}
g_2>-\frac{c_1}{M^2M_{\mathrm P}^2}\log(M/m_{\mathrm{IR}})\,,
\end{equation}
where we give upper bounds on the $O(1)$ constant $c_1< 24.2571$. 
A schematic representation of our bounds in the $(g_2, f_2)$ plane is shown in figure~\ref{fig:schematicBounds}. The WGC appears to be satisfied at leading order in $M/M_\textrm{P}$, but zooming in on the boundaries of the allowed region unveils a region where it is violated. The size of the violation is suppressed by $(M/M_\textrm{P})^2$ but enhanced by the logarithm of an infrared cut-off.
\end{itemize}

This paper is organized as follows: In section \ref{sec:method}, we review the fundamentals of $2 \to 2$ photon scattering and the assumptions we use. Among these are (1) weak coupling: the requirement that loops are suppressed in the EFT, and (2) Regge boundedness: the requirement that, at fixed $u$, the amplitude grows slower than $s^2$ at large $|s|$. We also describe the approach to scattering non-identical states known as the ``generalized optical theorem,'' and show how it improves the bounds obtained in the limit without gravity.

In section \ref{sec:resultswithgravity}, we consider the problem of bounding the EFT coefficients in the presence of gravity. We derive a number of improved sum rules and use them to bound the four-derivative coefficients. Some explicit examples of functionals which yield these bounds are given. We end the section with a discussion on the relevance of our bounds to the WGC.

\section{Bounding Photon Scattering}
\label{sec:method}

Let us first review the technical ingredients we will need in order to derive bounds. The goal will be to apply dispersion relations to $2 \to 2$ scattering amplitudes of photons. The result will be a set of sum rules which depend on Mandelstam invariants $s$ and $u$. Semi-definite programming may then be used to derive optimal constraints on EFT coefficients from these sum rules. This numerical approach to deriving EFT constraints was pioneered in \cite{Caron-Huot:2020cmc}, and generalized to handle the graviton pole in \cite{Caron-Huot:2021rmr}. 

At the heart of this method is the $S$-matrix, which maps ingoing states to outgoing states. For the four-particle amplitudes considered here, this amounts to
\begin{align}
    {}_{\text{out}}\left\langle \psi_3 \psi_4 \middle| \psi_1 \psi_2\right\rangle_{\text{in}} = {}_{\text{free}}\left\langle \psi_3 \psi_4 \middle| S \middle| \psi_1 \psi_2\right\rangle_{\text{free}}
    \,,
\end{align}
for particles $\psi_1$, $\psi_2$, $\psi_3$, and $\psi_4$. The $S$-matrix can be split into the identity operator and the interacting part as 
\begin{align}
    S = \1 + i T \,.
\end{align}

For the four-particle amplitude, we consider the external states to be two-particle center-of-mass plane waves. We will be concerned with the scattering of photons, hence the amplitudes will be depend on the helicities of the external particles, $\mathcal{M}^{\lambda_1 \lambda_2 \lambda_3 \lambda_4}$, where $\lambda_i=\pm$ denote states of circular polarization. We define the amplitude by
\begin{align}
   \left\langle  p_f ,\theta, \phi, \lambda_3, \lambda_4 \middle| T \middle| p_i, 0, 0, \lambda_1, \lambda_2 \right\rangle = (2 \pi)^4 \, \delta^{(4)} \left(\sum p_i^\mu \right) \,  \mathcal{M}^{\lambda_1 \lambda_2 \lambda_3 \lambda_4}(s, t, u)
    \,.
\end{align}
This describes two photons with helicities $\lambda_1$ and $\lambda_2$ coming in along the $z$-axis, and scattering to two photons with helicities $\lambda_3$ and $\lambda_4$, going in the direction $(\theta, \phi)$ and $(\pi-\theta, \pi + \phi)$. We shall use all-ingoing conventions in this paper. The dynamics are symmetric with respect to rotating $\phi$, so we will set it to $0$.

It will be convenient to package the individual helicity amplitudes into a matrix 
\begin{equation}
\mt M=\begin{pmatrix}
\mathcal M^{++--} &\mathcal  M^{++-+} & \mathcal M^{+++-} & \mathcal M^{++++}
\\
\mathcal M^{+---} & \mathcal M^{+--+} & \mathcal M^{+-+-} & \mathcal M^{+-++}
\\
\mathcal M^{-+--} & \mathcal M^{-+-+} & \mathcal M^{-++-} & \mathcal M^{-+++}
\\
\mathcal M^{----} & \mathcal M^{---+} & \mathcal M^{--+-} & \mathcal M^{--++}
\end{pmatrix}
    \,.
\end{equation}
Here we see 16 amplitudes, but in the scattering of identical indistinguishable particles, there are discrete symmetries which reduce the number of independent functions which these depend on. For our case, where all particles have spin 1, the amplitudes are related by
\begin{align}
    \mathcal{P}:& \qquad \mathcal{A}^{\lambda_1 \lambda_2 \lambda_3 \lambda_4} = \mathcal{A}^{- \lambda_1 - \lambda_2 -\lambda_3 -\lambda_4}\,, \\
    \mathcal{T}:& \qquad \mathcal{A}^{\lambda_1 \lambda_2 \lambda_3 \lambda_4} = \mathcal{A}^{-\lambda_3 -\lambda_4 -\lambda_1 -\lambda_2 }\,, \\
    \mathcal{B}:& \qquad \mathcal{A}^{\lambda_1 \lambda_2 \lambda_3 \lambda_4} = \mathcal{A}^{\lambda_2 \lambda_1 \lambda_4 \lambda_3}\,,
\end{align}
following from parity, time-reversal and boson exchange respectively.

In addition to the helicities, these amplitudes are functions of the momenta of the external particles, parametrized by the usual Mandelstam invariants, $s = -(p_1 + p_2)^2$, $t = -(p_1 - p_4)^2$, and $u = -(p_1 - p_3)^2$. The amplitudes are also related by crossing symmetry, which acts on their helicities and permutes the Mandelstam invariants (\emph{e.g.} $\mathcal{M}^{++--}(s,t,u) = \mathcal{M}^{+-+-}(t,s,u)$, and by complex conjugation of the helicities which relates $\mathcal{M}^{-\lambda_1-\lambda_2-\lambda_3-\lambda_4}(s,t,u)=\mathcal{M}^{\lambda_1\lambda_2\lambda_3\lambda_4*}(s,t,u)\equiv (\mathcal{M}^{\lambda_1\lambda_2\lambda_3\lambda_4}(s^*,t^*,u^*))^*$.

Now let us count the number of independent amplitudes. In the most general situation, we only allow two symmetries $\mathcal B$ and $\mathcal{P}\mathcal{T}$. This reduces the number of independent functions from 16 to 7 (real) functions, which reduces to 5 after crossing symmetry:
\begin{equation}
\mt M=\begin{pmatrix}
\label{eq:matrixM}
g(s| t, u) & h(s,t,u) & h(s,t,u) & f(s, t, u)
\\
h^*(s,t,u) & g(t| s, u) & g(u| s, t) & h(s,t,u)
\\
h^*(s,t,u) & g(u| s, t) & g(t| s, u) & h(s,t,u)
\\
f^*(s,t,u) & h^*(s,t,u) & h^*(s,t,u) & g(s| t, u)
\end{pmatrix}
    \,.
\end{equation}
In this case, $g(s|t, u)$ is a real function with $t$-$u$ symmetry. $f(s,t,u)$ and $h(s,t,u)$ are complex functions, fully symmetric under any permutation of $s$-$t$-$u$. Their real and imaginary parts reflect the parity-even and parity-odd parts respectively. In the rest of this paper, we will restrict ourselves to parity-even interactions. In this case, $f^*(s,t,u) = f(s,t,u)$ and $h^*(s,t,u) = h(s,t,u)$. As a result, we are left with only three independent real functions, $f$, $g$, and $h$. 

\subsection{Dispersion relations}

Dispersion relations are a standard technique for deriving positivity bounds. The typical strategy is the following: Consider an amplitude $\mathcal{M}(s,u)$ which obeys the Froissart bound
\begin{align}
    \lim_{s \to \infty} \frac{\mathcal{M}(s,u)}{s^2} = 0 \, ,
    \label{eq:Froissart}
\end{align}
at fixed $u$ in the physical region where $u<0$. This behavior has been demonstrated for gapped systems \cite{Froissart:1961ux, Martin:1965jj}, however for scattering of massless particles its status is less clear -- see \cite{Chowdhury:2019kaq} for an interesting recent discussion, and \cite{Haring:2022cyf} for a proof of the required property for scalar amplitudes in $d>4$. In this paper, we will take \eqref{eq:Froissart} as an assumption.

\begin{figure}
     \centering
         \includegraphics{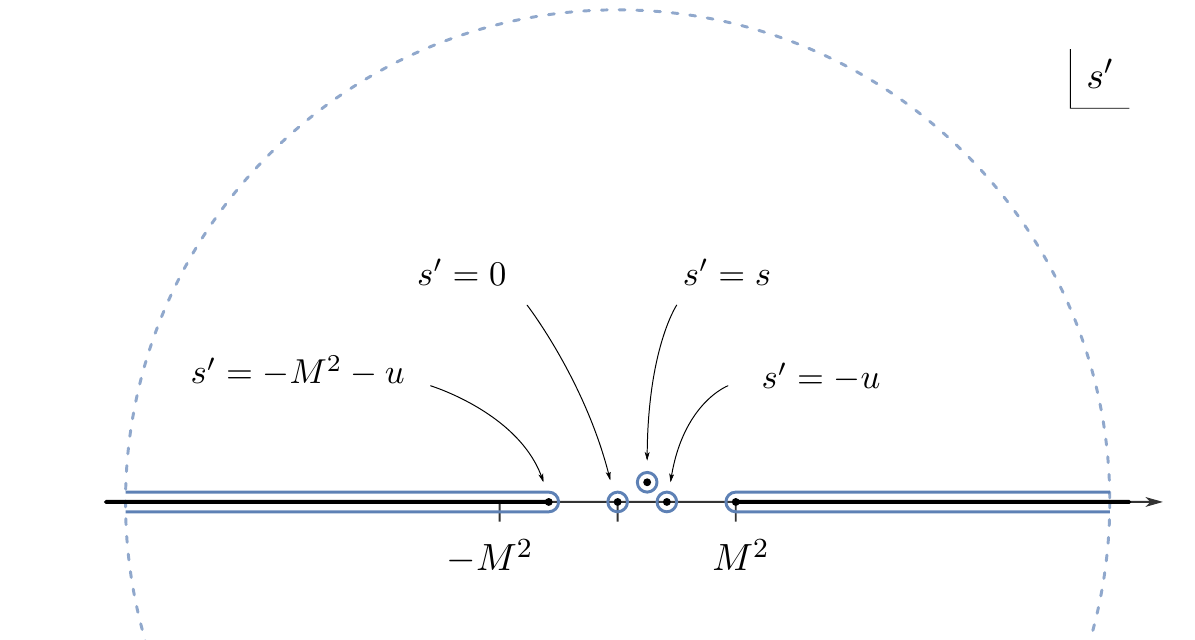}
        \caption{The starting point for our dispersive argument is the dashed contour at infinity in the $s'$ plane. It is deformed inwards to pick up contributions from three low-energy poles and two infinite cuts.}
         \label{fig:contour}
\end{figure}

Equation \eqref{eq:Froissart} implies that the following doubly-subtracted contour integral vanishes,
\begin{align}
\label{eq:lowresiduesum}
    \int_\infty \frac{ds'}{2 \pi i} \frac{\mathcal{M}(s',u)}{(s' - s)s'(s'+u)}  = 0\, .
\end{align}
If the amplitude is analytic in the upper half $s$-plane, which follows from causality, then the contour can be deformed towards the real-$s$ axis, defining the amplitude on the lower-half plane via $\mathcal{M}(s,u)\equiv \mathcal{M}(s^*,u^*)^*$. Then there are two contributions to the integral, which must therefore cancel: one contribution from three simple poles, and one contribution from the discontinuity across two cuts along the real axis, see figure~\ref{fig:contour}. In a parity-respecting theory, the discontinuity picks up the imaginary part of the amplitude, and we get
\begin{align}
\begin{split}
    0 \ &= \ \left(\res_{s'=s}+\res_{s'=0}+\res_{s'=-u}\right) \frac{\mathcal{M}(s',u)}{(s - s')s'(s'+u)} \\
    & \ \quad - \int_{M^2}^{\infty} \frac{ds'}{\pi} \IM \left[\frac{\mathcal{M}(s', u)}{(s-s')s'(s'+u)} \right] - \int_{-\infty}^{-M^2 - u} \frac{ds'}{\pi}  \IM \left[\frac{\mathcal{M}(s', u)}{(s-s')s'(s'+u)} \right] \, .
    \label{eq:dispersion}
\end{split}
\end{align}

The strategy we will follow below is to parametrize the amplitude in the top line of~\eqref{eq:dispersion} using the EFT, where it is given as a sum of undetermined coefficients. The amplitude in the bottom line will be parametrized using the partial wave expansion. Unitarity implies that the partial wave densities are positive (or, more generally, form positive definite matrices). This will allow us to convert \eqref{eq:dispersion} into an equation of the form $\mt L = \mt H$, where $\mt L$ and $\mt H$ are the low- and high-energy results of the dispersion integral and will be defined below. 

\subsection{Low-energy: EFT expansion}

For parity-respecting theories, the scattering of photons is described by three independent real functions, $f$, $g$, and $h$. At low energies, these functions are approximated by an effective field theory, \emph{i.e.}~\eqref{eq:EMaction}, and they may be expanded in powers of the Mandelstam invariants. 
The functions $f(s,t,u)$ and $h(s,t,u)$ are individually $s$-$t$-$u$ symmetric, while $g(s|t,u)$ is $t$-$u$ symmetric. These functions have expansions in small values of the Mandelstam invariants\footnote{In graviton scattering, the terms in $g(s|t, u)$ are multiplied by an universal helicity factor of $s^4$, so once this is stripped, the scaling of $g \sim s^2$ at large $s$ means that unsubtracted or even antisubtracted sum rules are possible. In our case, the universal helicity factor is $s^2$, so the stripped amplitude still requires an inverse power of $s$ to kill the pole at infinity that appears in the dispersion integral. The result is that, unlike \cite{Caron-Huot:2022ugt}, we can not immediately read off improved sum rules by using unsubtracted dispersion relations. Instead we will need to derive improved sum rules by systematically subtracting off higher-derivative coefficients, as is done in \cite{Caron-Huot:2020cmc, Caron-Huot:2021rmr}.
}
\begin{align}
\begin{split}
    f(s, t, u)  &=  10 \frac{\beta^2}{M_{\mathrm{P}}^2} \,  stu + f_2\, (s^2 + t^2 + u^2) + f_3\,  stu + f_4 \, (s^2 + t^2 + u^2)^2 + \ldots 
    ,
    \\
    g(s|t, u) &=  \frac{1}{M_{\mathrm{P}}^2} \frac{s^3}{tu} +  \frac{\beta^2}{M_{\mathrm{P}}^2}\! \left(\!4 stu -\frac{2}{3} s^3 \right) + g_2   s^2 + g_3   s^3 + g_{4,1}   s^4 + g_{4,2}  s^2 (s^2 + t^2 + u^2) + \ldots
    ,
    \\
    h(s, t, u)  &= \frac{\beta}{M_{\mathrm{P}}^2} (s^2 + t^2 + u^2)+ h_3 \, stu + h_4 \,  (s^2 + t^2 + u^2)^2 + \ldots.
    \label{eq:lowamplitudes}
\end{split}
\end{align}
By a direct computation, we note that the first terms in these expressions agree with the action~\eqref{eq:EMaction} upon identifying $ 16 \alpha_1 = g_2 + f_2$ and $16  \alpha_2 = g_2 - f_2$. The remaining parametrization allows for all possible terms consistent with the mentioned symmetries, and the assumption that contact interactions give a contribution to $g(s|t,u)$ proportional to $s^2$. 

Gravity decouples in the $M_{\mathrm{P}} \to \infty$ limit, and the terms that involve graviton propagators go to zero in this limit. Let us comment on the terms with $\beta$, which are a little special: $\beta$ arises from the Feynman diagrams with a single or double insertion of the operator $F_{\mu\nu}F_{\rho\sigma}W^{\mu\nu\rho\sigma}$, together with a graviton propagator in the diagram. From the way we have written it, it is clear that these terms vanish in the $M_{\mathrm{P}}\to\infty$ limit. But one might ask why not also include an independent $h_2$ term, \emph{i.e.}\ a term $h_2\,(s^2+t^2+u^2)$ in $h(s,t,u)$. In fact, we shall see that forward-limit sum rules, which are applicable in the limit where gravity has decoupled, preclude the existence of any such term (and in fact also show that $h_4=0$). The $h_2$-type interaction must shut off in that limit.

In terms of the matrix of amplitudes, we will introduce $\mt L$, the low-energy matrix, by
\begin{equation}
\mathbf L\equiv\left(\res_{s'=s}+\res_{s'=0}+\res_{s'=-u}\right) \frac{\mt M(s',u)}{(s' - s)s'(s'+u)}  \,.
\label{eq:defL}
\end{equation}
Using ``prime'' to denote this sum over residues, this can be written as
\begin{equation}
\mt L=\begin{pmatrix}
g'(s|t,u) & h'(s,t,u)& h'(s,t,u)& f'(s,t,u)
\\
h'(s,t,u) & g'(t|s,u) & g'(u|s,t) & h'(s,t,u)
\\
h'(s,t,u) & g'(u|s,t) & g'(t|s,u) & h'(s,t,u)
\\
f'(s,t,u) & h'(s,t,u) & h'(s,t,u) & g'(s|t,u) 
\end{pmatrix}
\label{eq:primemat}\, ,
\end{equation}
where
\begin{align}
\begin{split}
    f'(s,t,u) \ & = \ 2 f_2 - \left( f_3 + 10 \frac{\beta^2}{M_{\mathrm{P}}^2} \right) \,  u + 4 f_4  (s^2 + s u + 2u^2) + \ldots
    ,
     \\ 
    g'(s|t,u) \ & = \ -\frac{1}{M_{\mathrm{P}}^2 u} + g_2 + g_3(s - u) - \frac{2}{3} \frac{\beta^2}{M_{\mathrm{P}}^2} \, (s + 5 u)
     \\
    & \qquad \qquad + g_{4,1}(s^2 - s u + u^2) +2 g_{4,2}(s^2 + u^2) + \ldots 
    ,
     \\
    g'(t|s,u) \ & = \ -\frac{1}{M_{\mathrm{P}}^2 u} + g_2 - g_3(s + 2 u) + \frac{2}{3} \frac{\beta^2}{M_{\mathrm{P}}^2} \,  (s - 4 u) 
    \\
    & \qquad \qquad + g_{4,1} \, (s^2 + 3 s u + 3 u^2) + g_{4,2} \, (2s^2 + 4 su + 4 u^2)+ \ldots 
    ,
    \\
    g'(u|s,t) \ & = \ -4 \frac{\beta^2}{M_{\mathrm{P}}^2} \,  u + 2 g_{4,2} \, u^2 + \ldots 
    ,
    \\
    h'(s,t,u) \ & = \  2 \frac{\beta}{M_{\mathrm{P}}^2} - h_3 u + 4 h_4 (s^2 + s u + 2 u^2) + \ldots.
    \label{eq:primes}
\end{split}
\end{align}
The low-energy part is entirely determined by these functions $f'$, $g'$, and $h'$.

\subsection{High-energy: partial waves and unitarity}

Now we turn to the high-energy part of the dispersion relation, defined by
\begin{align}
\begin{split}
    \mt H & = \int_{-\infty}^{-u-M^2} \, \frac{ds'}{\pi} \, \IM{\left[\frac{\mt M(s',u)}{(s' - s)s'(s'+u)}\right]} + \int_{M^2}^{\infty} \, \frac{ds'}{\pi} \, \IM{\left[\frac{\mt M(s',u)}{(s' - s)s'(s'+u)}  \right]} \, \\
    & \qquad = \int_{M^2}^{\infty} \, \frac{ds'}{\pi} \, \IM{\left[\frac{\mt M(s',u)}{(s' - s)s'(s'+u)}  +  \frac{\mt M(-s'-u,u)}{(s + s' + u )s'(s'+u)} \right]}\,.
    \end{split}
    \label{eq:defH}
\end{align}
The amplitudes $\mt M(s',u)$ and $\mt M(-s'-u,u)$ can be related by crossing, and we will use this fact to derive the exact form of the sum rules.\footnote{In our previous work \cite{Henriksson:2021ymi}, only $s$-$t$ symmetric combinations of the amplitudes were considered. Here we will derive a larger set of sum rules compared to that paper, and consequently stronger constraints from dispersion relations.}

At high-energies, \emph{i.e.} above the scale $M$, the EFT no longer applies, and we are forced to be more agnostic about the form of the amplitude. However the symmetries alone strongly constrain the possible form the amplitude can take. This motivates the use of the partial wave expansion. For spinning particles in four dimensions, this takes the form \cite{Jacob:1959at}
\begin{align}
     M^{IJ} = 16 \pi \sum_\ell (2\ell + 1) A_\ell^{IJ}(s) d^\ell_{\lambda_I \lambda_J}(\theta)\, .
\end{align}
Here $I$ and $J$ label the rows and columns of the matrix, or equivalently pairs of helicities.\footnote{Specifically, $I$ ranges over $\{++,+-,-+,--\}$, and $J$ over $\{--,-+,+-,++\}$. We will use the labels $IJ$ and $\lambda_1\lambda_2\lambda_3\lambda_4$ interchangeably.} The partial wave densities $A^{IJ}_{\ell}(s)$ are given by
\begin{align}
    A^{\lambda_1\lambda_2\lambda_3\lambda_4}_\ell(s) = \left\langle s, \ell, \lambda_3, \lambda_4 \middle|T\middle| s, \ell, \lambda_1, \lambda_2 \right\rangle \, ,
    \label{eq:partialdef}
\end{align}
where $| s \ell \lambda_1 \lambda_2 \rangle$ refers to a two-particle state with definite angular momentum $\ell$ and energy $\sqrt s$.
The Wigner $d$ functions $d_{\lambda_I \lambda_J}$ are functions of $\lambda_I = \lambda_1(I) - \lambda_2(I)$, $\lambda_J = \lambda_4(J) - \lambda_3(J)$, and the scattering angle $\theta = \arccos\left(1 + \frac{2u}{m^2}\right)$. 
The set of allowed values of the spin $\ell$ of exchanged states depends on the external helicities. As we will see below, for $f(s, t, u)$ and $g(s|t, u)$ we have $\ell =0,2,4,\ldots$, for $g(t| s, u)$ and $g(u|s, t)$, $\ell=2,3,4, \ldots$, and for $h(s,t,u)$, $\ell =2,4,6,\ldots $.

Now we define the spectral densities $\rho^{\lambda_1\lambda_2\lambda_3\lambda_4}_\ell(s) = \IM A^{\lambda_1\lambda_2\lambda_3\lambda_4}_\ell(s)$. Using the partial wave expansion in the integral, and combining the right-hand and left-hand cuts using crossing symmetry, we find an expression for the components of $\mt H$. From here on, we will use $s' = m^2$. Then we have
\begin{align}
\begin{split}
   \mt{H}^{\lambda_1 \lambda_2 \lambda_3 \lambda_4}
&=\int_{M^2}^\infty \frac{dm^2}{m^2}\sum_{\ell}16(2\ell+1)\bigg(
\rho_{\ell}^{\lambda_1\lambda_2\lambda_3\lambda_4}(m^2)\frac{d_{\lambda_1-\lambda_2,\lambda_4-\lambda_3}^\ell(\theta)}{(m^2-s)(m^2+u)}
\\&\quad\qquad\qquad\qquad\qquad\qquad
+\rho_{\ell}^{\lambda_1\lambda_4\lambda_3\lambda_2}(m^2) \frac{d^\ell_{\lambda_1-\lambda_4,\lambda_2-\lambda_3}(\theta)}{(m^2+u)(m^2+s+u)}
\bigg)
\label{eq:Mhighpart1}
    \,.
\end{split}
\end{align}
For convenience, let us denote the integrand by $h^{IJ}$, so that

\begin{align}
\label{eq:Hhigh}
 H^{IJ} = \int_{M^2}^\infty \frac{dm^2}{m^2} \sum_\ell h^{IJ} \, .
\end{align}

\subsubsection{Positivity from unitarity}
\label{sec:posfromunit}

Having defined $h^{IJ}$, we shall now write it in a way that separates the dynamical and kinematical content of the high-energy amplitude. The aim is to do it in such a way that it makes the positivity conditions manifest.
The key to finding positivity condition is to invoke unitarity, which implies that
\begin{align}
    S^{\dagger}S = 1 \quad \implies \quad 2 \IM T = T^{\dagger} T  \, .
\end{align}
Contracting the second equation with external states of definite helicity and angular momentum $\ell$ gives
\begin{align}
    \IM A^{IJ}_\ell(s) =  \frac{1}{2} \langle s, \ell, \lambda_3, \lambda_4 | T^\dagger T | s, \ell, \lambda_1, \lambda_2 \rangle  = \frac{1}{2} \sum_X \langle s, \ell, \lambda_3, \lambda_4 | T^\dagger | X \, \ell\rangle \langle X \, \ell  | T | s, \ell, \lambda_1, \lambda_2 \rangle
    \,,
\end{align}
where we have inserted a complete set of intermediate states of spin $\ell$, labeled by $X$, which accounts for all other labeling of the state.
If we define
\begin{align}
    c^{\lambda_i \lambda_j}_{\ell,X}(s) =
    \sqrt{8(2\ell+1) }
     \langle X \ell
     | T | s,\ell, \lambda_i, \lambda_j \rangle
    \,,
\end{align}
and use $\rho_\ell = \IM A_\ell$, the result is %
\begin{equation}
16(2\ell+1)\rho_\ell^{\lambda_1\lambda_2\lambda_3\lambda_4}(s)=\sum_X c^{\lambda_1\lambda_2}_{\ell,X}(s)\left(c^{-\lambda_3-\lambda_4}_{\ell,X}(s)\right)^*
    \,.
\end{equation}

From this result, sometimes called the generalized optical theorem (\emph{e.g.} \cite{Du:2021byy}), we can see that $\rho^{IJ}_\ell(s)$ is a positive definite Hermitian matrix. In what follows, we shall use the positivity of $\rho^{IJ}_\ell(s)$ to show that specific linear combinations of the high-energy part of the dispersion relation are positive. Specifically, they will be constructed by acting on $\mt H$ by certain linear functionals which will involve contracting $\mt{H}^{IJ}$ with vectors $v^I$, and taking various integrals over $u$. To make this concrete, let us start by rewriting $h^{IJ}$ as
\begin{equation}
\label{eq:hIJintocs}
h^{IJ}(\ell,m^2,s,u)=\sum_Q\sum_X \vc c_{\ell,X}^\dagger(m^2) V^{IJ}_Q(\ell,m^2,s,u) \vc c_{\ell,X}(m^2)
    \,,
\end{equation}
where for generality we consider a vector $\vc c_{\ell,X}$ and a matrix $V_Q^{IJ}$. The $V^{IJ}_Q$ may be determined from~\eqref{eq:Mhighpart1} and are given in explicit form in appendix~\ref{app:moredetailssumrules}. The purpose of defining $h^{IJ}$ this way is that it makes the positivity constraints easier to deal with. 
This follows from the fact that for a real-valued symmetric matrix $V$, the condition $V \succcurlyeq 0$ implies that $\vc c^\dagger V \vc c\geq0$ for all complex vectors $\vc c$. Since all the dynamical information of the high-energy amplitude is contained in the vectors $\vc c$, we can analyze its positivity without making any further assumptions than those following from analyticity, unitarity and symmetry. Below we shall see how to construct sum rules by contracting \eqref{eq:hIJintocs} with vectors different vectors $\vc v$, and derive positivity constraints by applying linear functionals on such sum rules.

In \eqref{eq:hIJintocs}, the sum over $Q$ represents a sum over ``selection sectors'' -- defined by the parity $P_X$ and spin $\ell_X$ of the exchanged state. Invariance under parity and boson exchange imply\footnote{See for instance \cite{Hebbar:2020ukp}. Using a basis of two particle states which transforms as irreducible representations of the Poincaré group $| s \ell \lambda_1 \lambda_2 \rangle$, in the center of mass frame we have the following relations
\begin{equation*}
\mathcal{P}|s \ell \lambda_1 \lambda_2\rangle=(-1)^\ell |s \ell  (-\lambda_1) (-\lambda_2) \rangle,
\hspace{1cm} 
\mathcal{B}|s \ell \lambda_1 \lambda_2\rangle=(-1)^{\ell +\lambda_1-\lambda_2}|s\ell  \lambda_2 \lambda_1\rangle\,.
\end{equation*} }
\begin{align}
\label{eq:parityonalphamain}
c_{\ell,X}^{\lambda_1\lambda_2}&=P_Xc_{\ell,X}^{-\lambda_2-\lambda_1}
    \,,
\\
\label{eq:exchangeonalphamain}
c_{\ell,X}^{\lambda_1\lambda_2}&=(-1)^{\ell}c_{\ell,X}^{\lambda_2\lambda_1}
    \,.
\end{align}
From this, we can see that $c^{++} = c^{--} = 0$ for odd spins, and $c^{+-} = c^{-+} = 0$ for odd parity. Thus, we have that $Q$ ranges over the following sectors:
\begin{itemize}
\item Spin zero and parity-even, denoted $Q=0$. We have $c^0_{0,X}=(c_{0,X}^{++})^*$, and $V_0^{IJ}$ is a number for any fixed $I,J$.
\item Even spin $\ell=2,4,\ldots$ and parity-even, denoted  $Q=+$. We have $\vc c^+_{\ell,X}=(c_{\ell,X}^{++},c^{+-}_{\ell,X})^\dagger$ and $V_+^{IJ}$ is a $2\times 2$ matrix for any fixed $I,J$.
\item Even spin $\ell=0,2,4,\ldots$ and parity-odd, denoted $Q=-$. We have $c^-_{\ell,X}=(c_{\ell,X}^{++})^*$, and $V_-^{IJ}$ is a number for any fixed $I,J$.
\item Odd spin $\ell=3,5,\ldots$ and parity-even, denoted $Q=o$. We have $c^o_{\ell,X}=(c_{\ell,X}^{+-})^*$, and $V_o^{IJ}$ is a number for any fixed $I,J$.
\end{itemize}
There are no parity-odd exchanges for odd spin.
With these considerations at hand, we are able to write the matrix entries of the high-energy integrand $\mt h$ as
\begin{align}
h^{IJ} (m^2,s,u)= &\sum_X   |c_{0,X}^{0}|^2   V_0^{IJ}(m^2,s,u)  +\sum_{\ell=2,4,\ldots}\sum_X \sum_{a,b=1}^2  (c_{\ell,X}^+)^a ({c_{\ell,X}^+}^*)^b   (V^{IJ}_+)^{ab}(\ell,m^2,s,u) \nonumber  \\
&+  \sum_{\ell=0,2,\ldots}\sum_X  |c^-_{\ell,X} |^2   V^{IJ}_{-}(\ell,m^2,s,u) +   \sum_{\ell=3,5,\ldots}\sum_X |c^o_{\ell,X} |^2    V_o^{IJ}(\ell,m^2,s,u)\,.
\label{eq:hIJwrittenout}
\end{align}

\subsection{Sum rules}

We may now express the result of our dispersive arguments in the form
\begin{align}
    \mt L = \mt H \, ,
\end{align}
where $\mt L$ and $\mt H$, as defined above, are matrices of functions of $s$ and $u$. To proceed we will first contract the matrices with real vectors $v^I$, which will lead to 
\begin{align}
     v^I  L^{IJ}  v^J =  v^I  H^{IJ}  v^J \, .
\end{align}
The left and right sides of this equation are both functions of $s$ and $u$. From here, we make use of two basic ways to derive sum rules:

\begin{enumerate}
\item Choose powers $s^{p}u^{q}$ and look at 
\begin{equation}
\label{eq:low-highpq}
\left.\vc v^T \mt L \vc v\right|_{s^{p}u^{q}} =  \left. \vc v^T \mt H\vc v\right|_{s^{p}u^{q}}  
    \,.
\end{equation}
\item Choose the power $s^0$ and a \emph{function} $\phi(p)=\phi(\sqrt{-u})$
\begin{equation}
\label{eq:low-highint}
\int_0^{M} dp\,\phi(p)\left.\vc v^T \mt L \vc v\right|_{s^0,u=-p^2} = \int_0^{M} dp\,\phi(p)  \left. \vc v^T \mt H\vc v\right|_{s^0,u=-p^2}
    \,.
\end{equation}
\end{enumerate}
We refer to the first type of sum rules as ``forward-limit sum rules'' because they essentially amount to a series expansion around $u = 0$. This means that they are not valid in the presence of a $1 / u$ graviton pole. The second type we shall call ``integral sum rules.'' These are more general: they include the forward-limit sum rules when $\phi(u)$ has a $\delta(u)$ factor.

In practice, we shall consider linear combinations of such sum rules. The primary reason to do this is to derive ``improved sum rules,'' where the low-energy part only depends on a finite number of EFT coefficients. We shall show in detail how these are constructed below. A general linear combination of rules can be constructed from a linear combination of sum rules formed from different vectors $\vc v_i$. The practical algorithm will therefore be
\begin{align}
&\text{If}  &&
\forall Q,\ \forall \ell\in Q,\ \forall m^2\geqslant M^2
\nonumber
\\
&&& \sum_i \left.v_i^IV^{IJ}_Q(\ell,m^2,s,u)v_i^J\right|_{s^{p_i}u^{q_i}}+\sum_i \int_0^Mdp\, \phi_i(p)\left.v_i^IV^{IJ}_Q(\ell,m^2,s,u)v_i^J\right|_{s^{0},u=-p^2} \succcurlyeq 0,
\label{eq:RHShighIf}
\\
&\text{Then} && \sum_i\left.\vc v_i^I  L^{IJ} \vc v^J_i\right|_{s^{p_i}u^{q_i}}+\sum \int_0^Mdp\,\phi_i(p)\left.\vc v_i^I  L^{IJ} \vc v^J_i\right|_{s^{0},u=-p^2}  \geqslant0
    \,.
\label{eq:RHSlowThen}
\end{align}
The argument is as follows. It follows from \eqref{eq:hIJwrittenout} that if the condition \eqref{eq:RHShighIf} is true, then
\begin{equation}
\sum_i \left.v_i^I h^{IJ} v_i^J\right|_{s^{p_i}u^{q_i}}+\sum_i\int_0^Mdp\,\left.\phi_i(p)v_i^I h^{IJ}v_i^J\right|_{s^{0},u=-p^2} \geqslant0\,.
\end{equation}
Then by \eqref{eq:Hhigh}, the high-energy part of the dispersion relation is positive. This implies that the low-energy part must also be positive, implying the positivity in \eqref{eq:RHSlowThen}.

To connect with the numerical bootstrap philosophy, we will think of the (weighted) sum over different choices of $i$ in the argument above as acting on a set of sum rules with a linear functional $\Lambda$. Specifically, on matrix-valued functions $f^{IJ}(s,u,m^2)$
\begin{equation}
\Lambda[f^{IJ}] = \sum_i \left.v_i^I f^{IJ}(m^2,s,u) v_i^J\right|_{s^{p_i}u^{q_i}}+\sum_i\int_0^Mdp\left.\phi_i(p)v_i^I f^{IJ}(m^2,s,-p^2)v_i^J\right|_{s^{0}}
    \,.
\end{equation}
Then the algorithm can be reformulated in terms of searching for optimal functionals, a problem that can be implemented as a semi-definite program. Specifically, 
\begin{align}
\text{If}\qquad  \Lambda[  V_0^{IJ} ]& \geqslant 0 \quad \text{for all }m^2\geqslant M^2\,\nonumber
\\
\Lambda[  V_+^{IJ}(\ell)  ] &\succcurlyeq 0 \quad \text{for all }m^2\geqslant M^2,\ \ell=2,4,6,\ldots\, ,\nonumber
\\
 \Lambda[  V_-^{IJ}(\ell)  ] &\geqslant 0 \quad \text{for all }m^2\geqslant M^2,\ \ell=0,2,4,\ldots\, ,
\\
 \Lambda[  V_o^{IJ}(\ell) ] &\geqslant 0 \quad \text{for all }m^2\geqslant M^2,\ \ell=3,5,7,\ldots\, ,\nonumber
\\
\text{Then} \qquad  \Lambda[L^{IJ}]&\geqslant0
    \,.
\end{align}

\subsubsection{Null constraints}

Another important part of the numerical method of this paper is the addition of null constraints, first introduced in \cite{Tolley:2020gtv, Caron-Huot:2020cmc}. The are equations that arise when a single EFT coefficient can be written in terms of the high-energy expansion in two different ways. As such, they take the form of constraints only on the high-energy data. Including them in the numerics significantly improves the possible bounds. The use of null constraints for photon scattering in the forward limit was explained in \cite{Henriksson:2021ymi}: in practice one expands the dispersion relations as in \eqref{eq:low-highpq}, with $p$ sufficiently large to kill the graviton pole, and equates high-energy expansions leading to the same low energy expression. In this paper, however, we find more null constraints than in our previous work.
Consider the sum rules derived from the ``$g$-type'' amplitudes only: $g(s|t,u)$, $g(t|s,u)$ and $g(u|s,t)$. Each of these amplitudes enters a dispersion relation, not manifestly positive. In \cite{Henriksson:2021ymi}, only $s$-$t$ symmetric combinations were considered, effectively reducing the number of amplitudes to consider to two: $g(s|t,u)+g(t|s,u)$ and $g(u|s,t)$.

\begin{table}[ht]
	\centering
	\caption{Number of sum rules and null constraints constructed from the $g$-type amplitudes only. The coefficients $\tilde g_{p,q}$ are not important for our discussion.} \label{tab:gtypesumrules}
	{
		\renewcommand{\arraystretch}{1.5}
		\begin{tabular}{|c|lcc|}
			\hline
		Order	&  Sum rules & Null constraints: \cite{Henriksson:2021ymi}   & Null constraints: this work
			\\
			\hline
			$\frac1{m^4}$ & 1: $g_2$ 
			  &   0  & 0
			\\
			$\frac1{m^6}$ & 1: $g_3$  &   0  & 1
\\
			$\frac1{m^8}$ & 2: $g_{4,1}$, $g_{4,2}$  &   1  &   2
			\\
			$\frac1{m^{10}}$ & 2: $\tilde g_{3,0}$, $\tilde g_{3,1}$  &   1  &   3
			\\
			$\frac1{m^{12}}$ & 3: $\tilde g_{4,0}$, $\tilde g_{4,1}$, $\tilde g_{4,2}$  &   1  &4   
			\\\hline
		\end{tabular}
	}
\end{table}

The counting of null constraints is given in table~\ref{tab:gtypesumrules}. When considering only $g$-type sum rules, this leads to stronger constraints than the previous work.
Moreover, we now have sum rules and null constraints involving the ``$h$-type'' amplitudes. 

Another addition to the previous work is the null constraints of the form of integral sum rules. We will return to them in section~\ref{sec:resultswithgravity}.

\subsection{Bounds without gravity}
\label{sec:resultswithoutgravity}

The rest of this section will be devoted to applying the methodology laid out above to the case where $M_{\mathrm{P}} \to \infty$. In this limit, gravity decouples and the graviton pole vanishes. This means that we may apply forward-limit sum rules with as few as two subtractions, \emph{i.e.} $p \geqslant 0$ in \eqref{eq:low-highpq}. After reviewing these sum rules in more depth, we present a number of bounds derived from them. Among other things, we show how this immediately implies the WGC inequalities~\eqref{eq:wgcbounds}. This is consistent with the known fact that the WGC is directly provable from forward-limit bounds if the graviton pole is ignored \cite{Bellazzini:2019xts}.

The problem of bounding $2 \to 2$ photon amplitudes in the absence of gravity was addressed using a less general method in \cite{Henriksson:2021ymi}. As such, we also include a discussion of the difference between the results obtained here and the results of that paper. We see that they are significantly stronger and we are also able to bound coefficients, such as $h_3$, which remain unconstrained in \cite{Henriksson:2021ymi}.

\subsubsection{Forward-limit sum rules}
\label{subsubsection: Forward sum rules}

Recall that the sum rules take the form $ \mt L  =    \mt H $, with components
\begin{align}
       L^{IJ}\ =  \ 
          \int_{M^2}^\infty \frac{dm^2}{ m^2} \sum_Q \sum_\ell \sum_X  \vc c_{\ell,X}^\dagger V_Q^{IJ}(\ell, m^2,s,u) \vc c_{\ell,X}\, .
\end{align}
Contracting with a vector $\vc v$ and writing out the sum over $Q \in \{0, +, -, o\}$ explicitly gives
%
\begin{align}
\vc v^T\mt L \vc v|_{s^pu^q} = & \int_{M^2}^\infty \frac{dm^2}{m^2}\bigg(  \sum_X  |c_{X,0}^{0}|^2   v^IV^{IJ}_0v^J +\sum_{\ell=2,4,\ldots}\sum_X \sum_{a,b=1}^2 (c_{\ell,X}^+)^a ({c_{\ell,X}^+}^*)^b  v^I (V_+^{IJ})^{ab}v^J \nonumber  \\
&+  \sum_{\ell=0,2,\ldots}\sum_X  |c^-_{\ell,X} |^2  v^I V_-^{IJ}v^J+   \sum_{\ell=3,5,\ldots}\sum_X |c^o_{\ell,X} |^2   v^I V_o^{IJ}v^J  \bigg)
\label{eq:Dispersion}
    \,.
\end{align}

\paragraph{Positivity of $\boldsymbol{g_2}$} Let us illustrate this with a simple example. If we choose $\vc v=(1,0,0,0)^T$ then we find the low-energy part
\begin{align}
    \vc v^T\mt L \vc v = g'(s|t, u)  = \ g_2 + g_3(s - u) + g_{4,1} (s^2 - su + u^2) + g_{4,2} (2 s^2 + 2 u^2) + \ldots
    .
\end{align}
Consider the lowest-order sum rule by specifying $p = 0$, $q = 0$. This picks out $ \left. \vc v^T\mt L \vc v \right|_{s^0 u^0} = g_2 $.
For the high-enery part, we use the explicit formulas in appendix~\ref{app:moredetailssumrules}. For our choice of $\vc v$ we get
\begin{align}
\begin{split}
    v^IV^{IJ}_0v^J \ &= \ \frac{1}{(m^2 - s)(m^2 + u)}
    \,, \\
    v^IV^{IJ}_+v^J \ &= \ \begin{pmatrix}
     \frac{d^\ell_{0,0}(\theta)}{(m^2 - s)(m^2 + u)} & 0 \\
     0 & \frac{d^\ell_{2,2}(\theta)}{(m^2 + u)(m^2 + s + u)}
    \end{pmatrix}
    \,, \\
    v^IV^{IJ}_-v^J  \ &= \ \frac{d^\ell_{0,0}(\theta)}{(m^2 - s)(m^2 + u)} 
    \,,\\
    v^IV^{IJ}_ov^J \ &= \ \frac{d^\ell_{2,2}(\theta)}{(m^2 + u)(m^2 + s + u)}
    \,,
\end{split}
\end{align}
where $\theta=1+\frac{2u}s$ is the scattering angle.
Now the forward limit $u\to0$ imples $\theta\to0$, in which case $d^\ell_{0,0}(\theta) \to 1$ and $d^\ell_{2,2}(\theta) \to 1$. So we find that $\left. v^IV^{IJ}_Qv^J \right|_{s^0 u^0} =  m^{-4}$ for all of the sectors (except $\left. v^I V^{IJ}_+v^J\right|_{s^0 u^0} = m^{-4} \mathbb{I}_2$). The result is a sum rules for $g_2$,
\begin{align}
g_2 = & \int_{M^2}^\infty \frac{dm^2}{m^6}   \bigg (\sum_X|c_{0,X}^{0}|^2    +\!\!\sum_{\ell=2,4,\ldots} \!\sum_X\sum_{a=1}^2 |(c_{\ell,X}^+)^a|^2 + \! \!\sum_{\ell=0,2,\ldots}\! \sum_X|c^-_{\ell,X} |^2   +\! \!  \sum_{\ell=3,5,\ldots}\!\sum_X |c^o_{\ell,X} |^2  \bigg).
\label{eq:g2sumrule}
\end{align}
It is clear that this is a sum over positive terms. Hence $g_2$ must be positive! This example, therefore, turns out to be a translation of known results \cite{Falkowski, Bellazzini2016talk, Henriksson:2021ymi} into the language of this paper. Likewise, if one picks the power $s^{2k}u^0$, one finds a sum rule that implies positivity of the coefficient of $s^{2k+2}u^0$ in $g(s|t,u)$, in agreement with \cite{Arkani-Hamed:2020blm}, see \eqref{eq:tildegp0rule} in appendix~\ref{app:moredetailssumrules}.

\paragraph{WGC bounds without gravity}

Let us consider a slightly more complicated example, which will give us a very interesting result. Let us choose $\vc v^T = \frac14( 1, -1, -1, 1 )$, and again look at the leading (four-derivative) coefficients by specifying $p = q = 0$. From the low-energy expansion, we can see that 
\begin{equation}
\mt L|_{s^2u^0}=\begin{pmatrix}
g_2 &- 2h_2 & -2h_2 & 2f_2 \\
-2h_2 & g_2 & 0 & -2h_2\\
-2h_2 & 0 & g_2 &- 2h_2\\
2f_2 &- 2h_2 & -2h_2 & g_2
\end{pmatrix}
    \,,
\end{equation}
so we find 
\begin{align}
     \left. \vc v^T\mt L \vc v \right|_{s^0 u^0} \ = \ g_2 +  f_2 -4  h_2  \ = \ 4\left(4 \alpha_1  - \frac{\beta}{M_{\mathrm{P}}^2}\right)
    \,.
\end{align}
This is proportional to the exact combination that appears in the (electric) WGC bound in~\eqref{eq:wgcbounds}. Here we have defined $h_2 = \beta / M_{\mathrm{P}}^2$. The reason is that $h_2 = \beta / M_{\mathrm{P}} \to 0$ in the decoupling limit. However for the moment, we would like to be agnostic about the source of $h_2$ in the amplitude, and instead think of it as the nothing more than the coefficient of $s^2 + t^2 + u^2$ in the $h(s,t,u)$ amplitude. 

Now let us look at the high-energy parts 
\begin{align}
    \left.  v^I V^{IJ}_0v^J  \right|_{s^0 u^0} \ &= \ \frac{2}{ m^4} \, , \quad 
    &
    \left.v^I V^{IJ}_+v^J  \right|_{s^0 u^0} \ &= \  \frac1{m^4}
    \begin{pmatrix}
     2 & 0 \\
     0 & 1 \\
    \end{pmatrix}  \, ,  \\
    \left.v^I V^{IJ}_-v^J \right|_{s^0 u^0} \ &= \ 0 \, , \quad 
    &
    \left.v^I V^{IJ}_ov^J\right|_{s^0 u^0} \ &= \ \frac1 {m^4} 
    \,.
\end{align}
The consequence of this is a sum rule for the WGC combination: 
\begin{align}
\begin{split}
    &  g_2 +  f_2 -4 h_2  =  \\
& \qquad \sum_X \int_{M^2}^\infty \frac{dm^2}{m^6}   \bigg ( 2 |c_{0,X}^{+}|^2    +\sum_{\ell=2,4,\ldots} \left( 2 |(c_{\ell,X}^{+})^1|^2 +  |(c_{\ell,X}^{+})^2|^2 \right)   +    \sum_{\ell=3,5,\ldots} |c^o_{\ell,X} |^2  \bigg)\,.
\end{split}
\label{eq:wgcsumrule}
\end{align}
The result is a sum of squares, and must therefore be positive. This result is also known in the literature. It was pointed out in \cite{Hamada:2018dde, Bellazzini:2019xts} that these forward limit bounds directly imply the positivity of the WGC combination. Note also that we can easily obtain the magnetic and dyonic WGC inequalities by choosing instead choosing $\vc v = (1, 1, 1, 1)^T$ and $\vc v= (1, 1, 1, -1)^T$, respectively.

Let us be clear that this does not prove the WGC: we are required to take the decoupling limit $M_{\mathrm{P}} \to \infty$ before we are allowed to use the forward-limit sum rules in the first place. As a result, it is sort of a silly example. Of course ``gravity is the weakest force'' in the limit where the strength of gravity goes to zero. Still, it serves to illustrate an important point: the bounds we can derive in the absence of gravity by expanding in the forward limit are \emph{stronger} than the bounds available when the graviton pole is present. This shall be a major theme of section \ref{sec:resultswithgravity}, where we will explore the bounds in the presence of gravity. 

\paragraph{Vanishing of $\boldsymbol{h_2}$ without gravity}

Let us define $h_2$ to be the term proportional to $s^2+t^2+u^2$ in the amplitude $h(s,t,u)$. The parametrization $h_2=\beta/M_{\mathrm P}^2$ used in \eqref{eq:lowamplitudes} indicates that $h_2=0$ in the absence of gravity, and in fact it is easy to derive a sum rule that shows this fact. Consider for instance the entry $L^{12}|_{s^0}=2h_2-h_3u+\ldots$ in the low-energy amplitude. The corresponding entry $H^{12}|_{s^0}$ in the high-energy amplitude is in fact proportional to $u$, giving
\begin{equation}
\label{eq:forh20}
2h_2-h_3u+\ldots = H^{12}|_{s^0} =0+ O(u)\,,
\end{equation}
which shows that $h_2=0$. Note that argument is not valid in the presence of gravity, since it requires expanding the twice-subtracted dispersion relation in the forward limit.\footnote{One might believe that it would be safe to expand \eqref{eq:forh20} in the forward limit even with gravity, since the dangerous gravity pole is not present in this particular amplitude. This example makes it clear that this is not allowed, since there are known partial UV-completions of Einstein--Maxwell theory with non-zero values of $\beta=M^2_{\mathrm{P}}h_2$. An example is the theory of a charged spin-$\frac12$ fermion (QED), which will be discussed in section~\ref{sec:interpretation}.}

\subsubsection{Numerical results}

The strategy to get bounds is very similar to the one in~\cite{Henriksson:2021ymi}, but with two differences. The first one is that in this case we have a more general set of sum rules, which include the amplitude $\mathcal{M}^{+++-}$, too. Thus, we can now get bounds on the coefficients in the $h$ amplitude; the corresponding sum rules and null constraints are non-diagonal in $V_+^{IJ}$. For instance, the sum rule for $h_3$ reads
\begin{figure}
     \centering
     \begin{subfigure}[t]{0.45\textwidth}
         \centering
         \includegraphics[width=\textwidth]{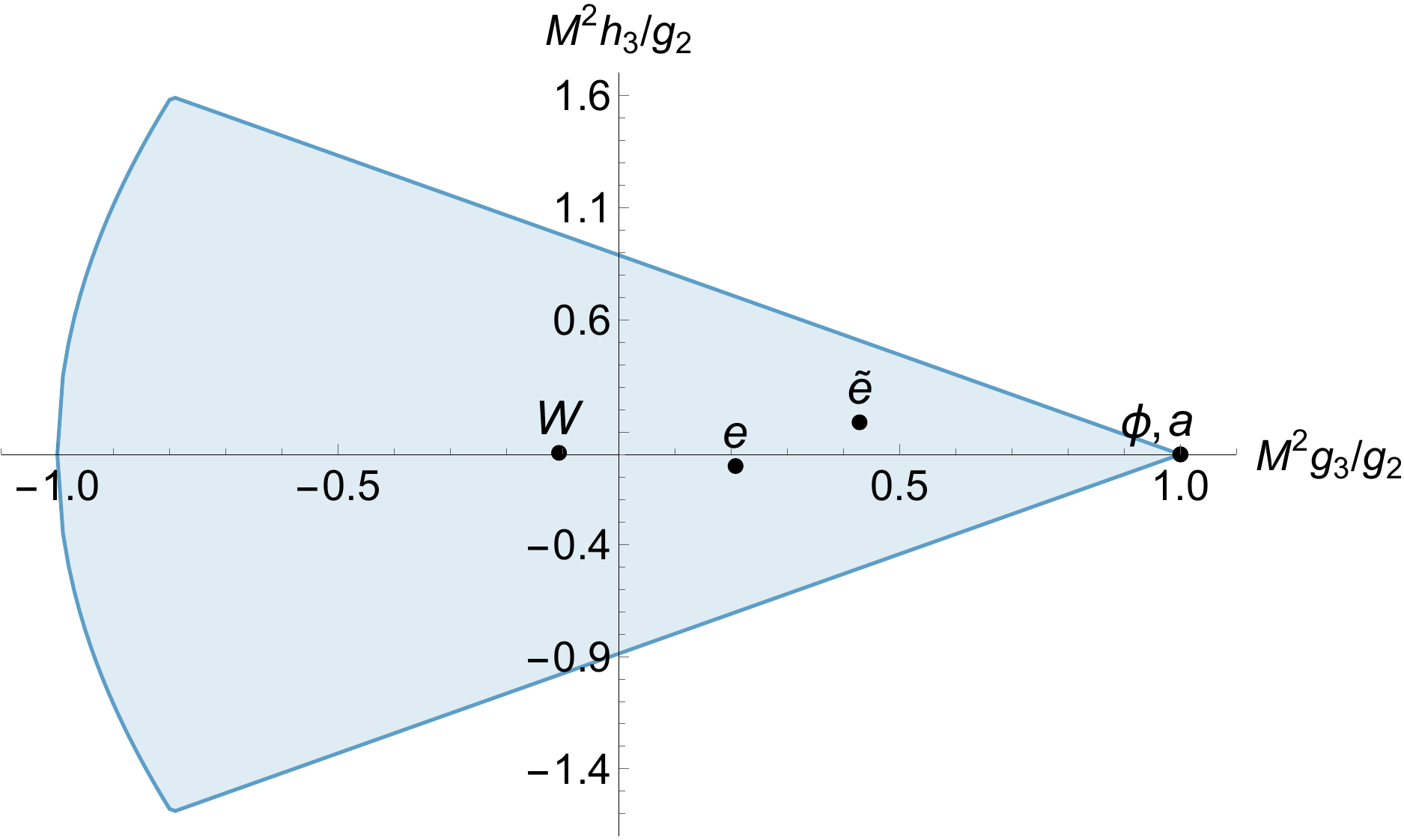}
     \end{subfigure}
     \hfill
     \begin{subfigure}[t]{0.45\textwidth}
         \centering
         \includegraphics[width=\textwidth]{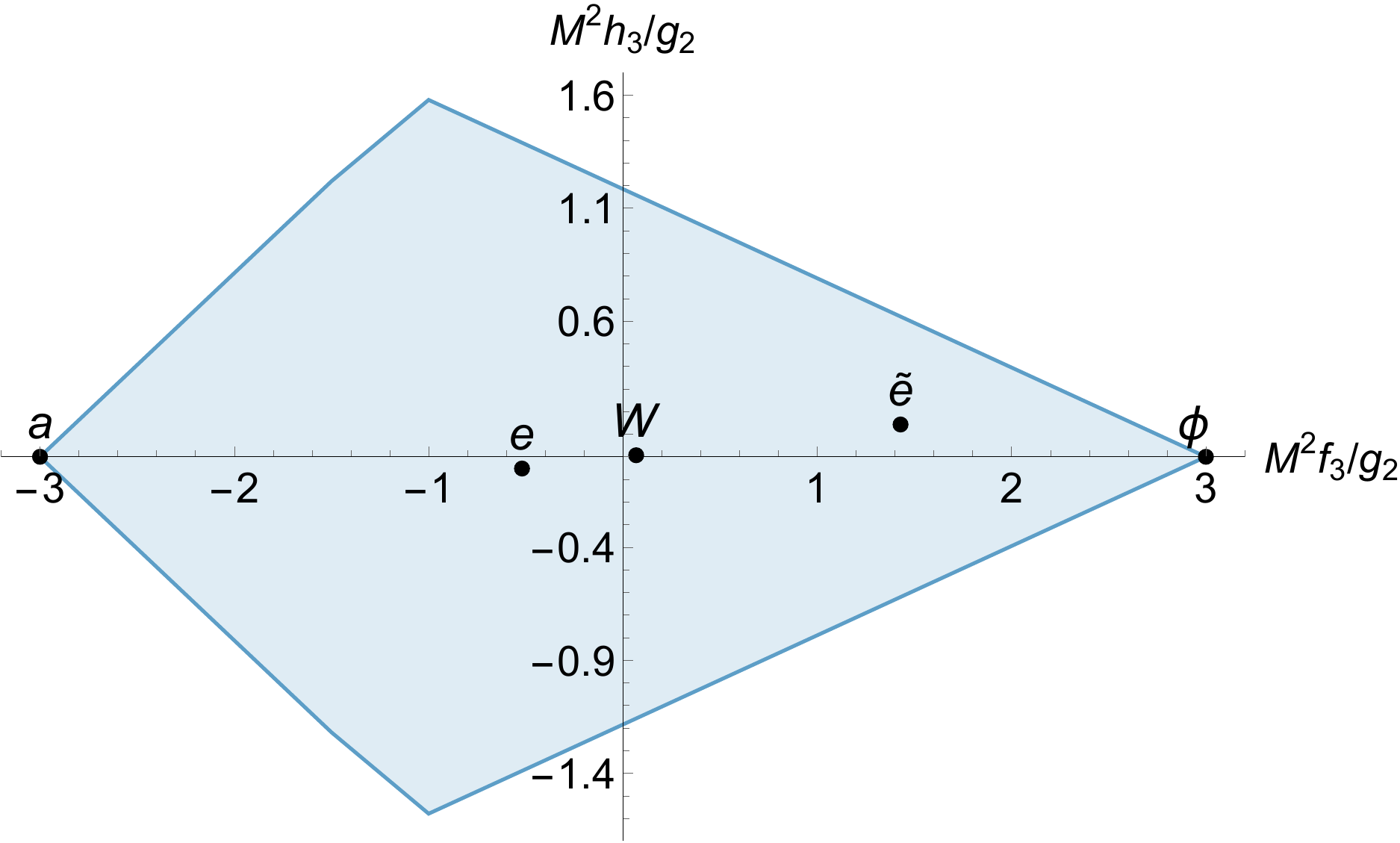}
     \end{subfigure}
     \hfill
        \caption{Some bounds on the six-derivative term $M^2 h_3/g_2$. The dots refer to the partial UV completions as above. }
        \label{fig:2d bound h3}
\end{figure}
\begin{align}
\label{eq:h3sumrule}
	h_3=& \int_{M^2}^\infty \frac{dm^2}{m^8} \bigg(\sum_X  0 |c_{0,X}^{0}|^2  + \sum_{\ell=0,2,\ldots}\sum_X 0 |c^-_{\ell,X} |^2  +   \sum_{\ell=3,5,\ldots} \sum_X0 |c^o_{\ell,X} |^2   \\
	&+\sum_{\ell=2,4,\ldots} \sum_X (\vc c_{\ell,X}^+)^\dagger \frac{1}{2}\left(
	\begin{array}{cc}
		0  & \sqrt{(\ell-1)\ell(\ell+1)(\ell+2)} \\
		\sqrt{(\ell-1)\ell(\ell+1)(\ell+2)} & 0 \\
	\end{array}
	\right)({\vc c_{\ell,X}^+})\bigg) \nonumber
    \,.
\end{align}
In figure~\ref{fig:2d bound h3} we give some examples of bounds derived when including the sum rule \eqref{eq:h3sumrule} in the set of sum rules and null constraints.
We include in the plots the values of some known partial UV completions (table~\ref{tab:EFTcoefValues_trees}), which consist of integrating out massive fields at tree- and loop-level. The notations in the plots for them is the following: massive axion ($a$), scalar ($\phi$), graviton ($h$), QED ($e$), scalar QED ($\tilde e$) and $W^\pm$ sector ($W$).

Another novelty with respect to our previous work is that we are not building crossing-invariant sum rules as before. This leads to a number of new null constraints, so in general the bounds will be stronger than those found in \cite{Henriksson:2021ymi}. For instance, we can make use of a new null constraint of order $m^{-6}$,
\begin{align}
	0&= \int_{M^2}^\infty \frac{dm^2}{m^6} \bigg( \sum_X0 |c_{X,0}^{+}|^2    +\sum_{\ell=2,4,\ldots} \sum_X (\vc c_{\ell,X}^+)^\dagger \left(
		\begin{array}{cc}
		\ell (\ell+1) & 0 \\
		0 & (\ell^2+\ell-7) \\
	\end{array}
    \right)\vc c_{\ell,X}^+ \nonumber  \\
	 &\quad+  \sum_{\ell=0,2,\ldots} \sum_X |c^-_{\ell,X} |^2  \ell (\ell+1) +   \sum_{\ell=3,5,\ldots}\sum_X |c^o_{\ell,X} |^2   (\ell^2+\ell-7) \bigg) 
    \,,
\end{align}
which immediately gives an improved bound for $g_3/g_2$. Previously, while the upper bound $g_3/g_2 \leqslant 1/M^2$ was found without null constraints, the lower bound $g_3/g_2 \geqslant -4.82/M^2$ required the use of null constraints at order $m^{-8}$ and higher, and the precise value of that bound depended on the number of null constraints used. Now, using the non-crossing symmetric sum rules, and therefore the new null constraint at order $m^{-6}$, we are able to obtain both an upper and lower bound which does not improve when adding more null constraints. The new optimal bound is 
\begin{equation}
	-\frac1{M^2}\, \leqslant\, \frac{g_3}{g_2}\, \leqslant \, \frac{1}{M^2} \,.
\end{equation}
In fact, this is just the first instance of an infinite sequence of two-sided bounds involving the coefficients of the powers of $s^{p}$ in the amplitude $g(s|t,u)$, as shown in \eqref{eq:gtilderule} in appendix~\ref{app:moredetailssumrules}.

\begin{figure}
     \centering
     \begin{subfigure}[t]{0.45\textwidth}
         \centering
         \includegraphics[width=\textwidth]{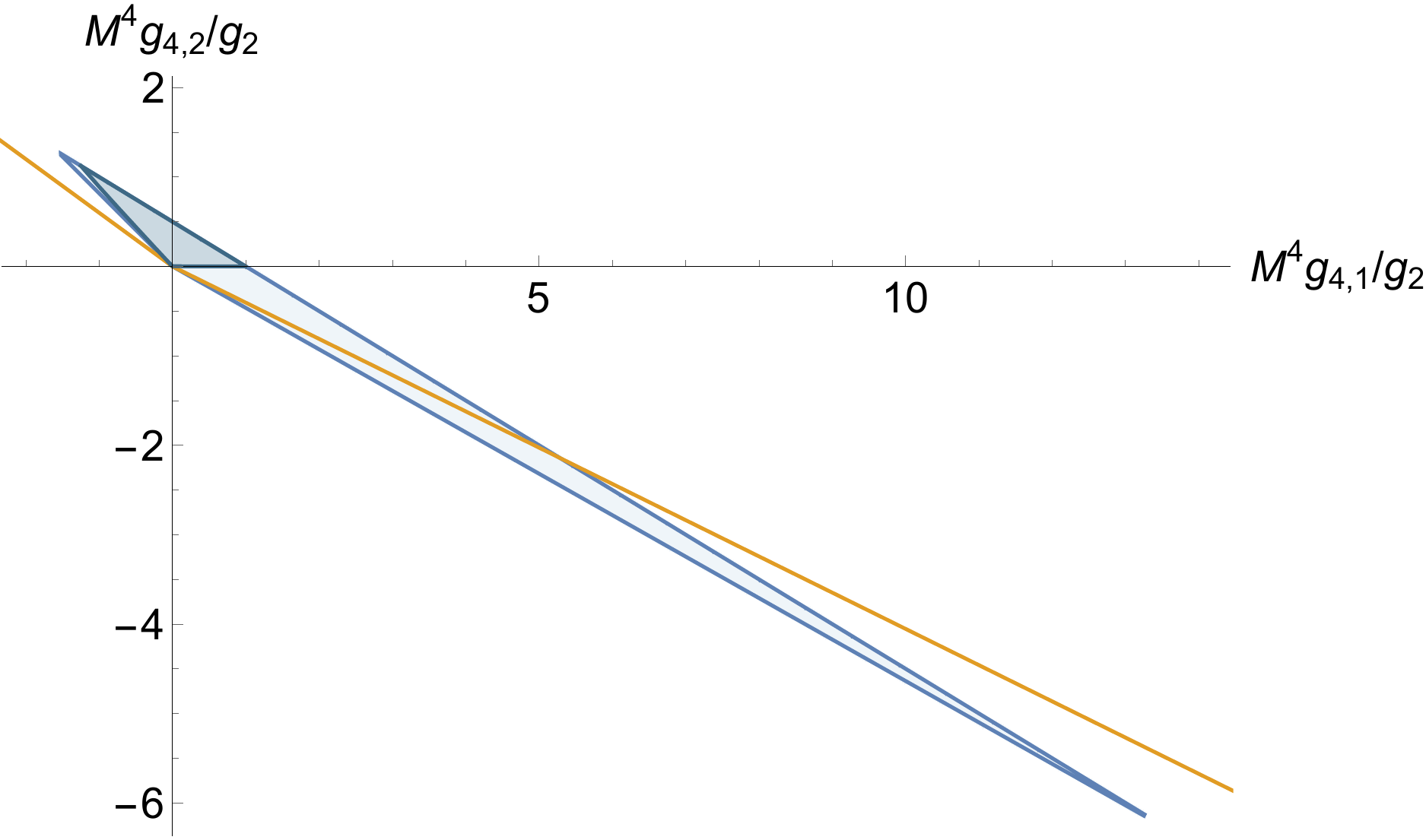}
         \label{fig:y equals x}
     \end{subfigure}
     \hfill
     \begin{subfigure}[t]{0.45\textwidth}
         \centering
         \includegraphics[width=\textwidth]{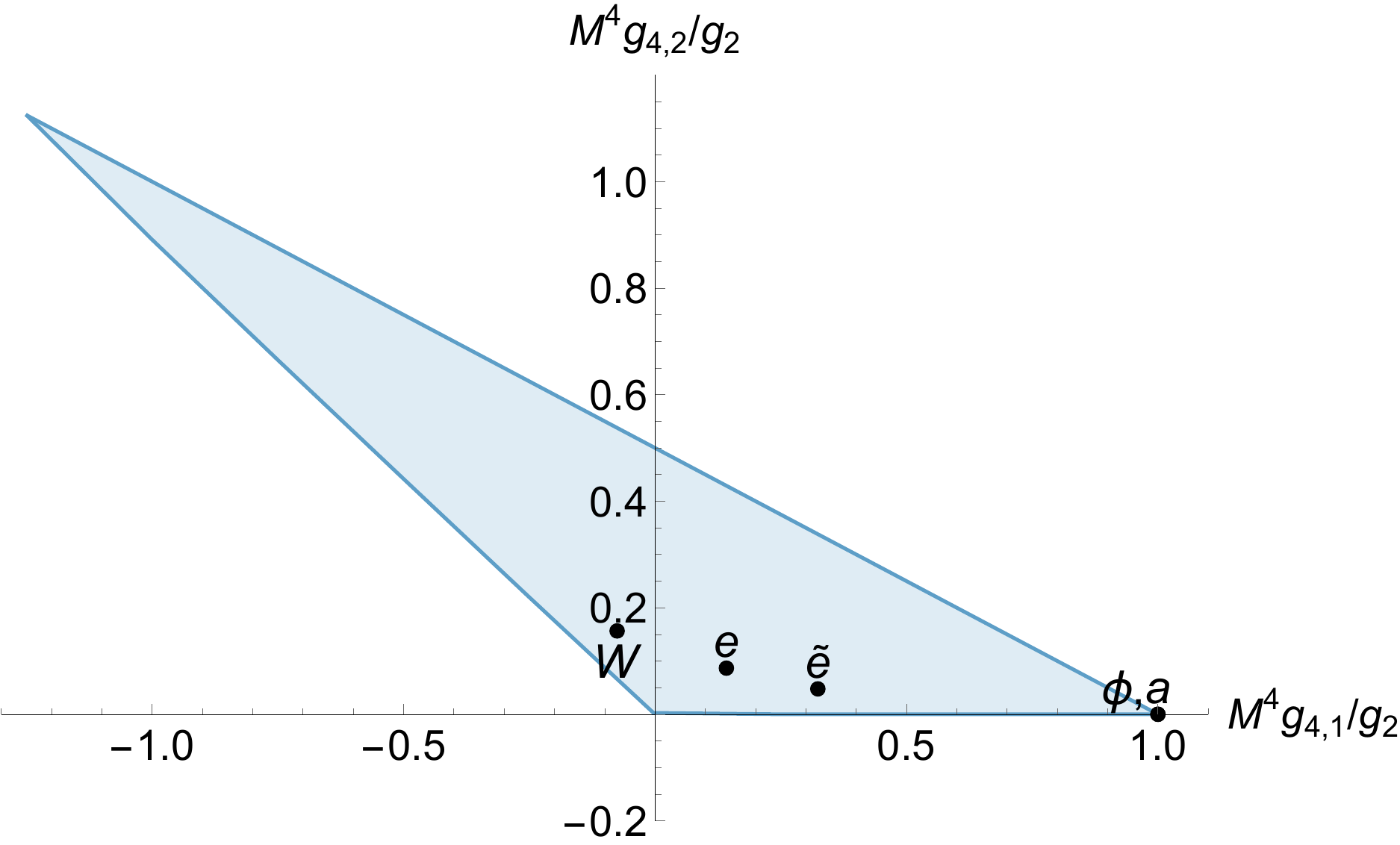}
         \label{fig:three sin x}
     \end{subfigure}
     \hfill
        \caption{Allowed region in the space $(M^4 g_{4,1}/g_2,M^4 g_{4,2}/g_2)$. In the left figure we show in orange the one sided bound (allowed above, disallowed below) from \cite{Arkani-Hamed:2020blm}, in light blue the result from \cite{Henriksson:2021ymi} and in dark blue our new result. In the other figure we zoom in the new allowed region, adding some partial completions \ref{tab:EFTcoefValues_trees}.}
        \label{fig:2d bound g42g41}
\end{figure}

Moving to eight-derivative order, we can see that our new approach significantly reduces the allowed region in the plane given by $g_{4,1}$ and $g_{4,2}$, see figure~\ref{fig:2d bound g42g41}. More precisely, the new allowed region fits into a triangle determined by the inequalities
\begin{equation}
0\, \leqslant\, \frac{2g_{4,2}}{g_{4,1}+2g_{4,2}}\, \leqslant\, \frac{12}5, \qquad 0\, \leqslant \,  \frac{g_{4,1}+2g_{4,2}}{g_2}\, \leqslant\,  \frac1{M^4}
    \,.
\end{equation}
The previous available bounds were $-\frac{370}{29}\leqslant\frac{2g_{4,2}}{g_{4,1}+2g_{4,2}}\leqslant\frac{18}7$, $ 0\leqslant \frac{g_{4,1}+2g_{4,2}}{g_2}\leqslant \frac1{M^4}$ from \cite{Henriksson:2021ymi} (light blue) and $-\frac{30}7\leqslant\frac{2g_{4,2}}{g_{4,1}+2g_{4,2}}\leqslant 6$, $g_{4,1}+2g_{4,2}\geqslant0$ from \cite{Arkani-Hamed:2020blm} (yellow line).

Additional bounds involving more EFT coefficients are given in Appendix~\ref{app: Plots no grav}. The plots are obtained using all the 15 null constraints up to order $m^{-12}$ and setting $\ell_{max}=40$.

\section{Results with Gravity}
\label{sec:resultswithgravity}

In this section we turn to the main novelty of this paper, which is bounds on the EFT coefficients that describe photon amplitudes in the presence of gravity. These bounds are derived via integral sum rules \cite{Caron-Huot:2021rmr,Caron-Huot:2022ugt}, which provide a way to circumvent the problem with the graviton pole. In four dimensions, using such sum rules introduces a logarithmic dependence on an infrared cutoff $m_{\mathrm{IR}}$.
The details on how to derive such bounds will be laid out below; here we will summarize the main results.

\begin{figure}[ht]	
  \centering
\includegraphics[width=0.7\textwidth]{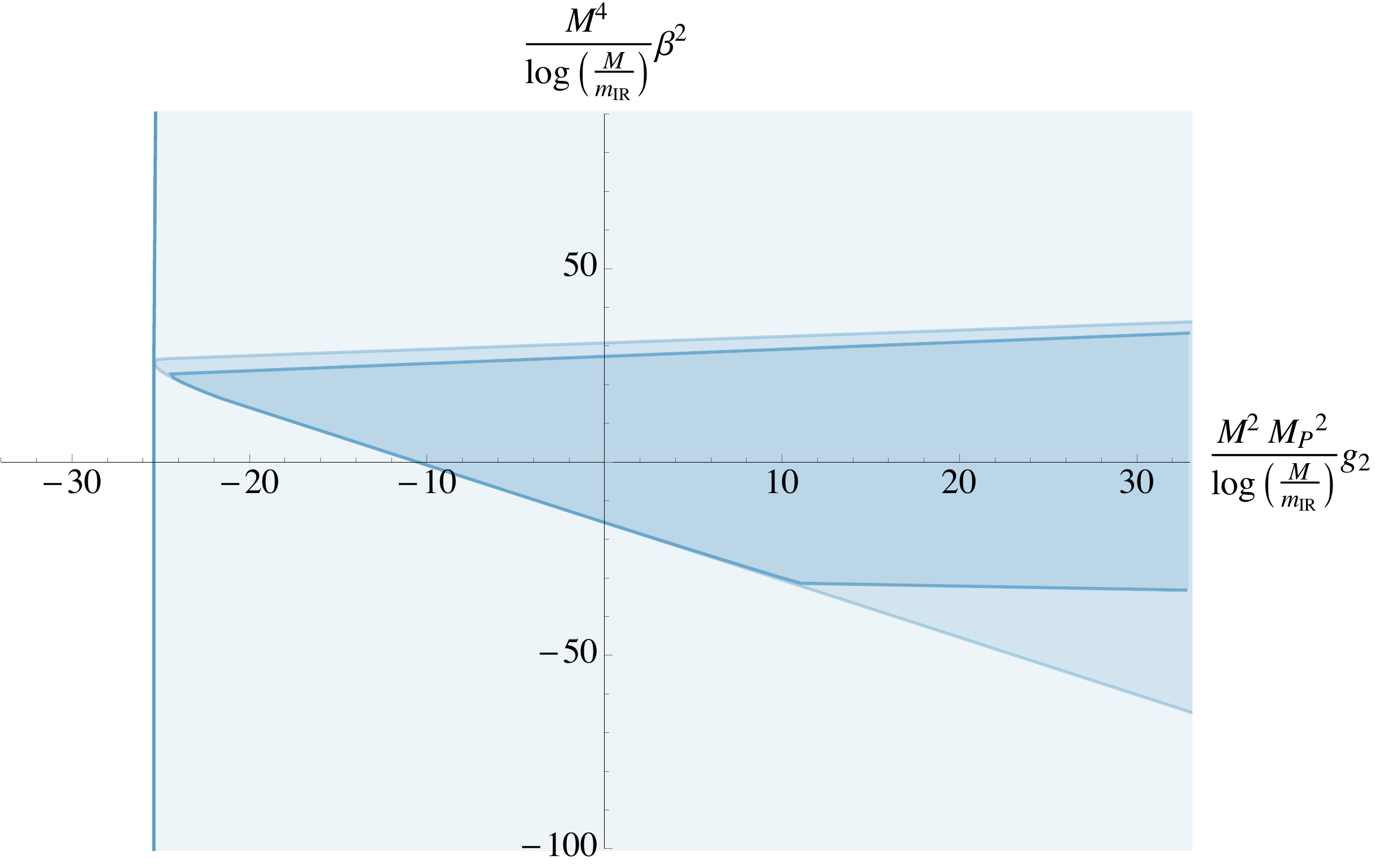}
\caption{Exclusion plot in the plane $(g_2, \beta^2)$ properly normalized and divided by $ \log(M/m_\mathrm{IR})$. The shaded regions represent the allowed values obtained by using various combinations of dispersion relations. In particular the light blue only uses the $\mathcal I_g$ dispersion relation, while the darker blue uses the $\mathcal I_g$,$\mathcal I_0$ and the $\mathcal I_g$, $\mathcal I_0$, $\mathcal I_{\beta^2}$ dispersion relations. The bounds have been obtained in the parametric limit $\log(M/m_\mathrm{IR})\gg 1$.}
\label{fig:g2VSh2sq}
\end{figure}

Our most interesting result is that we find that the positivity approach used in this paper cannot rule out violations to the black hole weak gravity conjecture. 
Specifically, we find that the coefficient $g_2$ must satisfy an inequality of the form
\begin{equation}
g_2\geqslant
- \frac{c_1}{M^2M_{\mathrm P}^2}\log\left(\frac M{m_{\mathrm{IR}}}\right) + \frac{c_0}{M^2M_{\mathrm{P}}^2}\,,
\label{eq:globalg2min}
\end{equation}
where $c_1= 24.257$ and $c_0=33.328$.
Moreover, assuming that $\beta=0$, this inequality is strengthened to
\begin{equation}
g_2\geqslant 
- \frac{\tilde c_1}{M^2M_{\mathrm P}^2}\log\left(\frac M{m_{\mathrm{IR}}}\right) + \frac{\tilde c_0}{M^2M_{\mathrm{P}}^2}\,, \qquad (\beta=0)\,,
\label{eq:g2minAtbeta0}
\end{equation}
where now $\tilde c_1= -10.557$ and $\tilde c_0=11.659$.
We can also construct bounds in the plane $g_2,\,\beta^2$ by considering arbitrary values of the ratio $\beta^2M_{\mathrm P}^2/(g_2M^2)$. In figure~\ref{fig:g2VSh2sq} we present such bounds in the limit $m_{\mathrm{IR}}\to0$, where the IR logarithm dominates.

The rest of this section will be devoted to a detailed description of how to obtain bounds in the presence of gravity. In section~\ref{sec:methodgrav} we outline the method used to generate the bounds in \eqref{eq:globalg2min}--\eqref{eq:g2minAtbeta0} and figure~\ref{fig:g2VSh2sq}. Then in section~\ref{sec:WorkedExample1} we present a completely explicit functional that gives a weaker version of the bound \eqref{eq:globalg2min}. The stronger bound \eqref{eq:globalg2min} is simply found by extending this method to allow for more complicated functionals. 

In section~\ref{sec:interpretation} we make an interpretation of our bounds. It is noteworthy that the violations to the black hole gravity conjecture vanish in the limit $M^2/M_{\mathrm P}^2\to0$. By assuming a scaling that is compatible with integrating out charged matter, which covers the case of QED, we find that in this limit the usual QED positivity bounds such that $g_2\geqslant0$ are recovered. In the limit where the electromagnetic strength becomes comparable to the gravitational, the bound \eqref{eq:g2minAtbeta0} applies, and negative values of $g_2$ cannot be ruled out. 

\subsection{Method}
\label{sec:methodgrav}

By using the general formalism laid out in the last subsection of appendix~\ref{app:moredetailssumrules}, integral sum rules can be derived.
By taking suitable linear combinations of them (and some of the higher sum rules), we write down the following useful integral sum rules, which we compactly collect into a vector equation
\begin{align}
\vec{\mathcal I}  |_{\mathrm{low}} = &\int_{M^2}^\infty \frac{dm^2}{m^2}\bigg( \sum_X |c_{0,X}^{0}|^2  \vec V^0_{0,m^2,u}  +\sum_{\ell=2,4,\ldots}\sum_X \sum_{a,b=1}^2 
(c_{\ell,X}^+)^a ({c_{\ell,X}^+}^*)^b  (\vec V^+_{\ell,m^2,u})^{ab} \nonumber  \\
&+  \sum_{\ell=0,2,\ldots}\sum_X
 |c^-_{\ell,X} |^2   {\vec V^-_{\ell,m^2,u}} +   \sum_{\ell=3,5,\ldots}\sum_X 
  |c^o_{\ell,X} |^2   {\vec V^o_{\ell,m^2,u}}  \, .
\label{eq:vectorDispersion}
\end{align}

In the above equation we have introduced the vector of low-energy coefficients $\vec{\mathcal I}  |_{\mathrm{low}} $, a vector of $2\times2$ matrices $\vec V^+_{\ell,m^2}$ and the two vectors $\vec V^-_{\ell,m^2,u}$ and $\vec V^o_{\ell,m^2,u}$. Notice also that the presence of three positive spectral densities  $ |c_{0,X}^{0}|^2,  |c_{\ell,X}^{-}|^2$ and $ |c_{\ell,X}^{o}|^2$ and the two-dimensional vector $(c_{\ell,X}^+)^a$.  The low-energy part of \eqref{eq:vectorDispersion} is a vector given by
\begin{equation}
\label{eq:I-low}
\vec{\mathcal I}  |_{\mathrm{low}} = \left\{\mathcal I_g |_{\mathrm{low}} ,  \,\mathcal I_f |_{\mathrm{low}},  \,\mathcal I_h |_{\mathrm{low}}, \,\mathcal I_{0} |_{\mathrm{low}} , \, \mathcal I_{\beta^2} |_{\mathrm{low}}  \right\}^T,
\end{equation}
where
\begin{align}
\begin{split}
\mathcal I_g |_{\mathrm{low}} &=- \frac{1}{M_{\mathrm P}^2}\frac 1u+ g_2- 4\frac{\beta^2}{M_{\mathrm P}^2} u\,,
\\
\mathcal I_f |_{\mathrm{low}}&=   f_2-5\frac{\beta^2}{M_{\mathrm P}^2}u
-\frac u2 f_3\,,
\\
\mathcal I_h |_{\mathrm{low}} &=\frac{\beta}{M_{\mathrm P}^2}-\frac u2 h_3\,,
\\
\mathcal I_{0} |_{\mathrm{low}} &=0\,,
\\
\mathcal I_{\beta^2} |_{\mathrm{low}} &=\frac{\beta^2}{M_{\mathrm P}^2}u\,.
\end{split}
\end{align}
The high-energy part of \eqref{eq:vectorDispersion} contains the four vectors
\begin{align}
\label{eq:vectorsV}
\vec V^0_{0,m^2,u}& = \left\{ \frac1{m^4},\frac1{m^4}-\frac{3u}{2m^6}, 0,0,0  \right\}^T
    \,,
\\
\vec V^+_{\ell,m^2,u} &= \left\{ \begin{pmatrix}I_{g,1}&0\\0&I_{g,2}\end{pmatrix} ,\begin{pmatrix}I_{f}&0\\0&0\end{pmatrix}, \begin{pmatrix}0&I_h\\I_h&0\end{pmatrix} , \begin{pmatrix}I_{0,1}&0\\0&I_{0,2}\end{pmatrix},\begin{pmatrix}I_{\beta^2,1}&0\\0&I_{\beta^2,2}+I_{\beta^2,3}\end{pmatrix} \right\}^T
    \,,
\\
\vec V^-_{\ell,m^2,u} &= \left\{ I_{g,1}\, , \,  -I_f \, , \,  0  \, , \,  I_{0,1} \, , \,  I_{\beta^2,1}\right\}^T
    \,,
\\
\vec V^o_{\ell,m^2,u}& = \left\{ I_{g,2} \, , \, 0 \, , \, 0\, , \,  I_{0,2} \, , \,  I_{\beta^2,2}-I_{\beta^2,3}\right\}^T
    \,,
\end{align}
where
\begin{align}
\label{eq:functionsI}
\begin{split}
I_{g,1}&=\frac{u+m^2{_2F_1}\left(-\ell,\ell+1;1;-\frac u{m^2}\right)}{m^4(u+m^2)}
\,,
\\
I_{g,2}&=\frac{u+(u-m^2){_2F_1}\left(2-\ell,\ell+3;1;-\frac u{m^2}\right)}{m^4(u-m^2)}
\,,
\\
I_f&=\frac{\ell(\ell+1)u^3}{m^6(u^2-m^4)}-\frac{u^2(3u+4m^2)}{2m^6(u+m^2)^2}+\frac{m^4(u+2m^2)}{2m^6(u+m^2)^2}{_2F_1}\left(-\ell,\ell+1;1;-\frac u{m^2}\right)
\,,
\\
I_h&=\sqrt{(\ell-1)\ell(\ell+1)(\ell+2)}\frac{2u^3+u(u^2+m^2u-2m^4){_2F_1}\left(2-\ell,3+\ell;3;-\frac u{m^2}\right)}{8m^6(m^4-u^2)}
\,,
\\
I_{0,1}&=\frac{\ell(\ell+1)u^2}{m^6(u+m^2)}+\frac{u-u\,{_2F_1}\left(-\ell,\ell+1;1;-\frac u{m^2}\right)}{m^2(u+m^2)^2}
\,,
\\
I_{0,2}&=\frac{\ell(\ell+1)u^2}{m^6(u-m^2)}+\frac{u\left(m^2-7u+(u-m^2){_2F_1}\left(2-\ell,\ell+3;1;-\frac u{m^2}\right)\right)}{m^6(u-m^2)}
\,,
\\
I_{\beta^2,1}&=\frac{\ell(\ell+1)u^2}{4m^6(m^2-u)}
\,,
\\
I_{\beta^2,2}&=\frac{u^2(7u+8m^2)}{4m^6(u+m^2)^2}-\frac{\ell(\ell+1)u^2}{4m^6(u+m^2)}
\,,
\\
I_{\beta^2,3}&=\frac{(\ell-1)\ell(\ell+1)(\ell+2)u^2(u+2m^2)}{96m^6(u+m^2)^2}{_2F_1}\left(2-\ell,\ell+3;5;-\frac u{m^2}\right)\,.
\end{split}
\end{align}
Here we used the formula \eqref{eq:WignerDsimple} for the Wigner $d$ functions. 

\subsubsection{Algorithm}
\label{sec:algorithm}

In order to obtain bounds on the low-energy parameters \eqref{eq:I-low}, we act on the vector dispersion relation \eqref{eq:vectorDispersion} with a functional $\Lambda$ and demand positivity of each term appearing on the right-hand side of the equation.  If such a functional exists, it will produce a constraint on the low-energy parameters $g_2$, $ f_{2}$, $\beta$, $g_3$, etc., in terms of the parameters $M$, $M_{\mathrm P}$ and $m_{\mathrm{IR}}$.

More concretely, we consider functionals of the form
\begin{align}
\label{eq:functinalDef}
&\Lambda \left[ \vec f \right] = \sum_i \Lambda_i\left[  f_i \right] \,,\qquad  \Lambda_i \left[ f(p^2) \right] = \int_{m_\mathrm{IR}}^M  \phi(p) f(p) dp
    \,,\\
& \phi(p)  = \sum_{q_n \in \mathcal{S}_i} c_{n,i}   \, q_n(p)
    \,,
\end{align}
where $ \mathcal{S}_i$ is a set of functions $q_n(p)$ in the variable $p$. Each $q_n$ contains integer or half-integer powers of $p$. 
We will discuss the choice of these sets in the next section. The index $i$ runs over any non-empty subset of $\{1,2,3,4,5\}$. For instance, in section~\ref{sec:WorkedExample1} and section~\ref{sec:WorkedExample2} we will consider  functionals using respectively only $i=1$ and $i=1,2$.

The lower extreme of integration in \eqref{eq:functinalDef} deserves special attention. We will discuss it in the next sections. Let us first spell out the concrete algorithm:
\begin{enumerate}
\item Choose a subset of dispersion relation to use. This corresponds to selecting which of the $\Lambda_i$ appear in \eqref{eq:functinalDef}.
\item Check if there exists a choice of coefficients $c_{n,i}$ such that:
\begin{align}
\label{eq:sdpb-conditions}
\begin{split}
\Lambda[ \vec V^+_{0,m^2,u} ]    &\geqslant 0 \qquad \text{for all }m^2\geqslant M^2\,,
\\
\Lambda[ \vec V^+_{\ell,m^2,u} ] &\succcurlyeq 0 \qquad \text{for all }m^2\geqslant M^2,\ \ell=2,4,\ldots\, ,
\\
 \Lambda[\vec  V^-_{\ell,m^2,u} ] &\geqslant 0 \qquad \text{for all }m^2\geqslant M^2,\ \ell=0,2,\ldots\, ,
\\
 \Lambda[\vec  V^o_{\ell,m^2,u} ] &\geqslant 0 \qquad \text{for all }m^2\geqslant M^2,\ \ell=3,5,\ldots .
 \end{split}
\end{align}
\item If such a functional exists then we obtain the constraint
\begin{align}
\label{eq:bounds}
 \Lambda[\vec{\mathcal I} |_\mathrm{low}]   = \frac{1}{M^2_\textrm{P}} \Lambda_1[p^{-2}] + g_2 \Lambda_1[1] + \frac{\beta^2}{M^2_\textrm{P}} \Lambda_1[4p^2]+\ldots + \frac{-\beta^2}{M^2_{\textrm{P}}}\Lambda_5[p^2]  \geqslant 0\,,
\end{align}
where the $\ldots$ includes the other low-energy coefficients, if present.
There is a constraint for any functional satisfying \eqref{eq:sdpb-conditions}.
\item In order to obtain the optimal constraint on a given parameter, say $g_2$, one can fix the values of the other parameters ($\beta=\beta_*, f_2=f_{2*}$, etc) 
and choose the functional that optimizes the following conditions:
\begin{align}
\label{eq;sdpb-norm}
&  \Lambda_1[1] = \pm 1 \quad \mathrm{(normalization)}\,, 
 \\
 &\mathrm{maximize}\,\, - \Lambda[\vec{\mathcal I} |_\mathrm{low}]  \bigg|_{g_2=0, \beta=\beta_* ,\ldots}\,.
\end{align}
Then
\begin{align}
\label{eq:exclusion}
g_2 \gtreqless \mp  \Lambda[\vec{\mathcal I} |_\mathrm{low}]  \bigg|_{g_2=0, \beta=\beta_* ,\ldots}\,.
\end{align}
In this way one obtains a bound on $g_2$ for fixed $\beta_*,f_{2*},$ etc. By scanning over them one gets bounds as a function of the other parameters. Similarly one can obtain bounds on any other parameter.\footnote{When only a subset of dispersion relation is considered  (\emph{i.e.} $\Lambda_i\equiv0$ for certain values of $i$), the function $\mathcal F_\Lambda$ does not depend on some parameters and the resulting bounds are independent of them.} \footnote{When  $\beta$ and $\beta^2$ do not appear simultaneously in $\mathcal F_\Lambda$, it is more convenient to fix the ratio between the parameters and get a bound on the overall normalization.}
\end{enumerate}

\subsubsection{Positivity}
\label{sec:positivity}

\begin{figure}[ht]
  \centering
\includegraphics{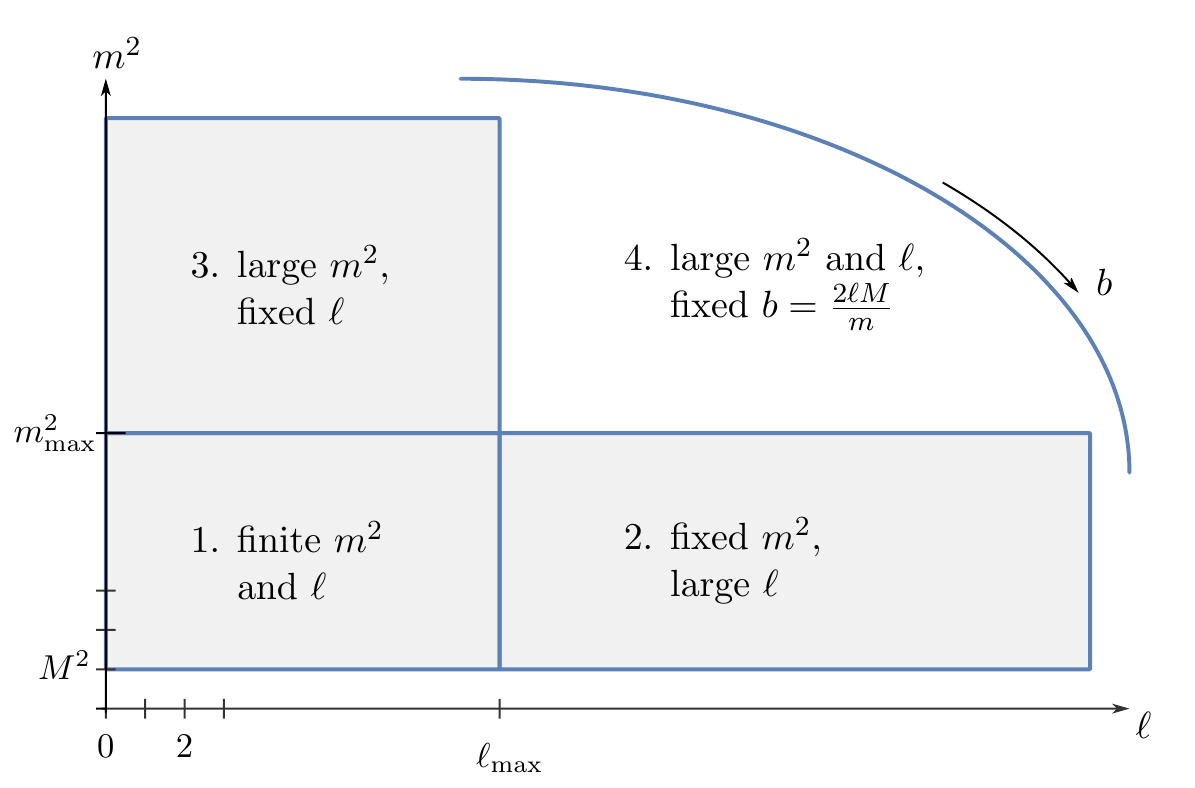}
\caption{We will demand positivity in the regimes indicated.}\label{fig:positivityregimes}
\end{figure}

In order to implement the algorithm, we need to find a method to impose positivity on the whole parameter space in the high-energy regime. Specifically, we need to impose that the functional is positive in all of the following regions, see figure~\ref{fig:positivityregimes},
\begin{enumerate}
\item Finite $\ell$ and $m^2$,
\item Large $\ell$ fixed $m^2$,
\item Fixed $\ell$, large $m^2$,
\item Large $\ell$ and $m^2$, for fixed (dimensionless) impact parameter $b=\frac{2\ell M}m$.
\end{enumerate}
Demanding positivity in the first three regions in the numerical implementation is a standard task and is achieved by a suitable discretization. We give more details about this in appendix~\ref{app:numerical}. The fourth region, namely in the limit of large $\ell$ and $m^2$ for fixed impact parameter $b$, requires a very careful consideration, which will be the topic of the remainder of this subsection.

\subsubsection{Large-$\ell,m^2$ behavior and choice of functionals}

As explained in \cite{Caron-Huot:2021rmr}, the major obstruction in getting a positive functional comes from the tension between the need to cancel the oscillating behavior of the hypergeometric functions at large $\ell$ and the need to have convergent integrals in $p$. Let us review this problem and see how we can choose the functionals $\Lambda_i$.

To study more carefully the limit of large $\ell$, we introduce the (dimensionless) impact parameter $b = 2\ell M/m$ and consider the behavior of the terms in the dispersion relation in the limit of large $\ell$, large $m$ and fixed $b$. In order to extract the leading behavior, we need the known limit of hypergeometric functions. More precisely we need the asymptotics of the expression
\begin{equation}
\label{eq:asymptoticsC}
C_{\nu,n}(b)=\lim_{m^2,\ell \to\infty}\int_0^M dp\,\frac{ p^n}{M^{n+1}}{_2F_1}\left(a_1-\ell,a_2+\ell,\nu,\frac{p^2}{m^2}\right)\,,
\end{equation}
where the limit is taken for fixed $b=\frac{2\ell M}m$, and we have indicated that this limit is independent of the finite shifts $a_1$ and $a_2$. We use the following formula, extracted from \cite{Thorsley2001}\footnote{This formula is known as Hansen's expression for the hypergeometric function, see equation~(1) in section 5$\cdot$7 of \cite{Watson1944}.}
\begin{equation}
J_{\nu-1}(z)=\frac{\left(\frac z2\right)^{\nu-1}}{\Gamma(\nu)}\lim_{|\lambda|,|\mu|\to\infty}{_2F_1}\left(\lambda,\mu;\nu;-\frac{z^2}{4\lambda\mu}\right),
\end{equation}
where $J_n$ is a Bessel function.
Applying it to \eqref{eq:asymptoticsC}, we find
\begin{equation}
\label{eq:intCnu}
C_{\nu,n}(b)=\frac{\Gamma(\nu)}{(b/2)^{\nu-1}}\int_0^Mdp\, \frac{p^{n+1-\nu}}{M^{n+2-\nu}}J_{\nu-1}(bp/M)\,.
\end{equation}
Using the expression for the integral of $p^aJ_k(bp/M)$ (which is a known integral, convergent for $n>-1$ and implemented for instance in Mathematica), we get
\begin{equation}
\label{eq:Cnu}
C_{\nu,n}(b)=\frac{1}{n+1}{_1F_2}\left(\frac{n+1}2;\frac{n+3}2, \nu ;-\frac{b^2}{4}\right)\,.
\end{equation}

Given our ansatz \eqref{eq:functinalDef} for the functional $\Lambda$, we need to consider the large spin limit of the elementary integrals $p^n \vec V^+_{\ell,m^2,-p^2}$, $p^n \vec V^-_{\ell,m^2,-p^2}$ and $ p^n \vec V^o_{\ell,m^2,-p^2}$. 
For the moment let us take the lower extreme of the integral  $m_\mathrm{IR}=0$. Using the above results we obtain
\begin{align}
\label{eq:largespin}
\int_0^1dp\, p^n\! \lim_{j,m\to \infty } \!\vec V^+_{\ell,m^2}  & = {\small \left\{\! \begin{pmatrix}C_{0,n}&0\\0&C_{0,n}\end{pmatrix}\! ,\!\begin{pmatrix}C_{0,n}&0\\0&0\end{pmatrix}\!, \!\begin{pmatrix}0&0\\0&0\end{pmatrix}\! ,\! \begin{pmatrix}C_{2,n}&0\\0&C_{2,n}\end{pmatrix}\!,\!\begin{pmatrix}0&0\\0&C_{4,n}\end{pmatrix}\! \right\}^{\!T} }\! \frac{b^4}{16\ell^4} +\ldots\,,
\\
\int_0^1 dp\, p^n\! \lim_{j,m\to \infty } \! \vec V^-_{\ell,m^2}  & =   \left\{ C_{0,n}, -C_{0,n},0,0 ,0\right\}^T \frac{b^4}{16\ell^4} +\ldots\,,
\\
\int_0^1 dp\, p^n\! \lim_{j,m\to \infty } \!\vec V^o_{\ell,m^2}  & = \left\{ C_{0,n}, 0,0, 0, -C_{4,n}\right\}^T \frac{b^4}{16\ell^4} +\ldots\,,
\end{align}
where for simplicity we put $M=1$ and omitted the $b$ dependence inside $C_{\nu,n}$.

Let us begin by focusing on the $\mathcal I_g$ dispersion relation only, namely the first entry of~\eqref{eq:vectorDispersion}, discarding the others for the moment. In this case the asymptotic behavior is controlled by $C_{0,n}$. The functional $\Lambda\equiv\Lambda_1$ should then satisfy (in addition to other conditions)
\begin{equation}
\label{eq:asympPositivity}
\sum_{n } a_n C_{0,n}(b) \geqslant 0 \,, \quad \text{for any } b\geqslant 0\,,
\end{equation}
where the $a_n$ are linearly related the coefficients $c_{n,1}$ appearing in \eqref{eq:functinalDef} by the actual choice of functions $q_n \in \mathcal S_1$.

Inspecting the large $b$ expansion of  $C_{0,n}(b)$, one observes potentially dangerous oscillating terms:
\begin{align}
\label{eq:leadingAsymp}
C_{0,n}(b)\simeq \frac{1}{b^{n+1}}\frac{2^n \Gamma \left(\frac{n+1}{2}\right)}{\Gamma
   \left(\frac{1-n}{2}\right)}+\frac{\sin (b)-\cos (b)}{\sqrt{\pi } b^{3/2}}+\frac{(8 n-5) (\sin (b)+\cos (b))}{8 \sqrt{\pi }
   b^{5/2}}+O(
   b^{-7/2})\,,
 \end{align}
 where the higher order terms correspond to half-integer powers only. In order to have a chance to fulfill the positivity condition one must suppress the oscillating behavior. One possibility would be to include, in the set $\mathcal S_1$, a power  $n<1/2$ such that the first non-oscillating term in \eqref{eq:leadingAsymp} could dominate over the rest. Unfortunately in four dimensions the integral of such a term would produce a divergence in the low-energy part of the dispersion relation due to the graviton pole:
\begin{equation}
\label{eq:finiteAlow}
\int_0^M \frac{p^n}{p^2} \frac{dp}{M_\mathrm{P}^2} < \infty\ \iff \ n>1\,.
\end{equation}

Alternatively, following \cite{Caron-Huot:2021rmr}, we can engineer a linear combination to cancel the leading oscillating terms. For numerical reasons it is convenient (although not necessary) not to introduce integer powers but to preserve the expansion in half-integer powers only. Hence the smallest power at our disposal is $n=3/2$. In order to have this term dominate at large $b$ we must cancel the first two oscillating terms. We also note that when $n$ is an odd positive integer, the first term in  \eqref{eq:leadingAsymp} vanishes, thus we can use two such powers and create the combinations\footnote{Any two positive odd integers will do.}
\begin{equation}
\label{eq:noOscillations}
C_{0,n}(b)+ \frac12 (n - 5)  C_{0,3}(b) - \frac12 (n - 3) C_{0,5}(b) \simeq  \frac{1}{16 b^{n+1}}\frac{2^n \Gamma \left(\frac{n+1}{2}\right)}{\Gamma
   \left(\frac{1}{2}-\frac{n}{2}\right)}+O(
   b^{-7/2})\,.
\end{equation}
This corresponds to considering functionals with polynomials
\begin{equation}
q_n(p)  = (p/M)^n + \frac12 (n - 5) (p/M)^3- \frac12 (n -3) (p/M)^5\,.
\end{equation}
One can show that the oscillating behavior at large $b$ in \eqref{eq:asympPositivity} is produced by the upper extreme of integration: the faster $q_n(p)$ goes to zero as $p\to M$, the more suppressed are the oscillations. Our choice of functionals satisfies $q_n(p) = O(M-p)^2$.\footnote{In \cite{Caron-Huot:2021rmr} the same result is obtained by functions of the form $q_n(p)  = (M-p)^2 p^n$.}

Note that at this point the condition \eqref{eq:asympPositivity} is asymptotically achieved by taking the coefficient multiplying $q_{3/2}$ negative, since $\Gamma(-1/4)<0$ in \eqref{eq:noOscillations}.

\subsubsection{Reintroducing the IR cut-off}

Unfortunately, the above discussion does not hold for finite values of the dimensionless impact parameter $b$. As discussed in \cite{Caron-Huot:2021rmr}, there exists a tension between the conditions \eqref{eq:asympPositivity} and \eqref{eq:finiteAlow} -- the two conditions are mutually exclusive. This fact is made manifest when passing to the impact parameter space by taking the two dimensional Fourier transform of a function of the transverse momentum $\vec p$:

\begin{equation}
\widetilde \phi(b) = \int d^2 p  \frac{\phi(p)}{2\pi p}   e^{i \vec b \cdot \vec p/M}= \Gamma\left(\frac12\right) \int_0^\infty dp\, \phi(p) J_0(pb/M) \,,
\end{equation}
where again $J_0$ is the Bessel function which also appears in the large $\ell$ limit \eqref{eq:asymptoticsC}. Recalling the definition of our functional $\Lambda$, we can interpret the positivity conditions in the large $m,\ell$ limit as the condition 
\begin{equation}
\label{eq:condition1}
\widetilde \phi(b) >0 \quad \text{for any } b\geqslant 0\,.
\end{equation}
 On the other hand, the finiteness of the functional on the gravity pole \eqref{eq:finiteAlow} would require
\begin{equation}
\label{eq:condition2}
0= \lim_{p\to 0}\frac{\phi(p)}{p} \propto  \int_0^\infty d^2 b \widetilde\phi(b)   \,.
\end{equation}
Clearly the conditions \eqref{eq:condition1} and \eqref{eq:condition2} are incompatible.
As a consequence, in order to proceed further we must relax one of the two conditions. Given the presence of IR divergences in gravity,
it seems natural to introduce an IR regulator in the form of a maximal impact parameter $b_\text{max}$ that can be probed by the scattering process. If that were the case, we would only need to demand positivity of $C_{0,n}(b) $ for $b\leqslant b_\text{max}$.  In principle we could search for functionals subject to this reduced positivity condition and obtain bounds as a function of $b_\text{max}$.

However, is more convenient  to introduce an IR cutoff as a regulator $m_\mathrm{IR}>0$ at small momenta as in  \eqref{eq:functinalDef}. This modification gets rid of the restriction \eqref{eq:finiteAlow} on the polynomials $q_n$, since now all the integrals are finite, but at the same time makes the bounds on couplings explicitly dependent on $m_\mathrm{IR}$. More precisely, including a term $q_n$ with $n<1$ introduces factors of the form $(M/m_\mathrm{IR})^{1-n}$.\footnote{We checked that adding powers of $-1<n'<1$ does not lead to stronger bounds (for $n'\leqslant-1$ the integral \eqref{eq:intCnu} does not converge). This is due to the following: the new power would determine the leading behavior at large $b$, hence the sign of its coefficient $c_{n'}$ is fixed to be positive. But then the bound on, say, $g_2$ would look like $g_2 \geqslant - c_{n'} (\frac{M}{m_\mathrm{IR}})^{1-n'}$ which is optimized by taking $c_{n'}\sim0$.} 
A milder dependence on the cut-off can be obtained by only including $q_1$, which instead gives a logarithmic dependence in \eqref{eq:bounds}
\begin{equation}
\label{eq:log}
 \Lambda[\vec{\mathcal I} |_\mathrm{low}]  \supset c_{1,1} \int_{m_\mathrm{IR}}^M  \frac{q_1(p)}{p^2 M_\mathrm{P}^2}\frac{dp}{M}  \simeq \frac{c_{1,1}}{M^2 M_\mathrm{P}^2} \left(\log \left(\frac{M}{m_\mathrm{IR}}\right) -\frac14 +O\left(\frac{m_\mathrm{IR}}{M}\right)
 \right)\,.
\end{equation} 

In conclusion, our tentative choice for the polynomials in $\mathcal S_1$ is
\begin{equation}
\label{eq:S1}
\mathcal S_1 = \big\{\left(\frac{p}M\right)^n+\frac{1}{2} (n-5) \left(\frac{p}M\right)^3-\frac{1}{2} (n-3) \left(\frac{p}M\right)^5 ; \text{ with } n = 1, \frac32, \frac52, \frac72,\ldots \big\}\,.
\end{equation}

A second important effect of $m_\mathrm{IR}$ is that the cancelation of the oscillating terms in \eqref{eq:noOscillations} is not exact anymore: from the power $p^n\subset q_1(p)$ one gets a correction to \eqref{eq:Cnu} of the form ($\nu=n=1$)
\begin{equation}
\label{eq:0F1}
\frac{m_{\mathrm{IR}}}{b M}J_1\left(b\frac{m_{\mathrm{IR}}}M\right)\simeq \frac12 \left(\frac{m_{\mathrm{IR}}}{M}\right)^2 + O\left(\frac{ b m_{\mathrm{IR}}}{ M}\right)\,.
\end{equation}
In the large $b$ limit the leading decay is controlled by $b^{-5/2}$, coming from \eqref{eq:noOscillations} with $n=3/2$, which is comparable to the above correction for 
\begin{equation}
b\gtrsim b_{\text{max}}= \left(\frac{M}{m_{IR}}\right)^{4/5} \,.
\end{equation}
Hence, the price of regularizing the action of the functional on the graviton pole with an IR cutoff is to introduce a small negativity at large impact parameter. The smaller the cutoff, the farther away we push the negativity, but at the same time we make the bound on the low-energy coefficients less stringent, specifically the bounds will depend logarithmically on $m_\mathrm{IR}$.

\hspace{2em}

Finally, let us consider the other dispersion relations in \eqref{eq:vectorDispersion}: we see that their asymptotic behavior enters with both signs or in the out-of-diagonal components of a matrix. In order for them not to spoil the suppression of large-$b$ oscillations, we must not introduce new dominant contributions. To achieve this, it is enough to cancel the leading universal $b^{-3/2}$ power. Hence we choose:

\begin{align}
\label{eq:S234}
&\mathcal S_2= \mathcal S_4 = \{\left(\frac{p}M\right)^n - \left(\frac{p}M\right)^3   ; \text{ with } n = 1, \frac32, \frac52, \frac72,\ldots,  n_\text{max} \}\,,
\\
&\mathcal S_3 = \{\left(\frac{p}M\right)^n - \left(\frac{p}M\right)^3  ; \text{ with } n = \frac32, \frac52, \frac72,\ldots , n_\text{max}\}\,,
\\
&\mathcal S_5 = \{\left(\frac{p}M\right)^n - \left(\frac{p}M\right)^5 ; \text{ with } n =  \frac32, \frac52, \frac72,\ldots , n_\text{max}\}\,.
\end{align}
More details about our numerical setup can be found in Appendix~\ref{app:numerical}.

Putting all the pieces together, our approach will then be the following. We will numerically look for a functional of the form \eqref{eq:functinalDef} (with $m_\mathrm{IR}=0$) with the choice of functions $q_n$ as in \eqref{eq:S1} and \eqref{eq:S234} subject to the  conditions \eqref{eq:sdpb-conditions}. This is done by running the numerical semi-definite program solver \texttt{SDPB} \cite{Simmons-Duffin:2015qma,Landry:2019qug}. The output of the algorithm are the coefficients $c_{n,i}$ of  \eqref{eq:functinalDef}. Taken at face value, such a functional would give a divergent result when applied to $\vec{\mathcal I}|_\text{low}$. We then modify the functional by taking  $m_\mathrm{IR}>0$:
\begin{equation}
\Lambda |_{m_\mathrm{IR}=0} \longrightarrow \Lambda' = \Lambda |_{m_\mathrm{IR}>0} \,.
\end{equation}
The new functional satisfies $\Lambda'[\vec{\mathcal I}|_\text{low}] =\text{finite}$, and produces bounds with a logarithmic dependence on $m_\mathrm{IR}$. On the other hand, the positivity condition is violated at large impact parameter $b\gtrsim b_\text{max}$. This is acceptable since it does not make sense to probe infinitely large distances in a theory with an IR cut-off.

\subsection{Example bounds from simple functionals}
\label{sec:examplefunctionals}

In this section, we will derive two bounds on the four-derivative coefficients by considering two explicit functionals. This will give concrete examples of the considerations above, and produce bounds that share the qualitative features with those presented in figure~\ref{fig:g2VSh2sq}.

\subsubsection{Example 1: Global minimum of $g_2$}
\label{sec:WorkedExample1}

As a first example, we will derive a bound involving $g_2$ only, by finding a functional $\Lambda=\Lambda_1$ that is manifestly positive. For the sake of simplicity, we take a slightly different form of the functional and allow integer powers.\footnote{Restricting to half-integer powers is merely a trick to optimize the numerics.}
Here we shall only use the sum rules derived from $\mathcal I_g$.
Let us start by the following ansatz: 
\begin{equation}
\Lambda_1[f] = \int_{m_\mathrm{IR}}^M  \phi(p) f(p) dp\,, \qquad 
\phi(p)=(M-p)^3\sum_{n=1}^3 c_n\left(\frac pM\right)^n \,.
\end{equation}
We can fix two of the coefficients to assume that
\begin{align}
\lim_{m_{\mathrm{IR}}\to0}\Lambda_1[\mathcal I_g |_{\mathrm{low}, g_2}]=M^4, \qquad 
\lim_{m_{\mathrm{IR}}\to0}\Lambda_1[\mathcal I_g |_{\mathrm{low},\beta^2}]=0\,.
\end{align}
The first condition is just a normalization condition, while the second condition is chosen to produce a bound that is independent of $\beta^2$.
Solving for $c_2$ and $c_3$ gives
\begin{equation}
\label{eq:phiformtoy}
\phi(p)=(M-p)^3\frac{p}{M}\left(c_1+\frac pM\frac{2100-51c_1}8+\frac{p^2}{M^2}\frac{63(c_1-60)}8\right)\,.
\end{equation}
At this point we first look for a functional that satisfies all the positivity conditions \eqref{eq:sdpb-conditions} in the limit $m_\mathrm{IR}\to 0$; this is done in the next sub-section. Once we have found it, we can now use the same value of $c_1$ to define a functional $\Lambda'_1$ where now $m_\mathrm{IR}$ is kept small but finite. As explained in the previous section, the newly defined functional mildly violates the positivity conditions \eqref{eq:sdpb-conditions}  at large $m^2$, $\large \ell$ and large impact parameter. Neglecting this violation, we  get $ \Lambda'_1[\mathcal I_g|_{\mathrm{low}}]\gtrsim 0$, with
\begin{equation}
 \Lambda'_1[\mathcal I_g|_{\mathrm{low}}]= M^4 g_2 + \frac{M^2}{M_\mathrm{P}^2}\left(42-\frac{91}{30}c_1+c_1\log\left(\frac M{m_{\mathrm{IR}}}\right)\right) + O(m_{\mathrm{IR}}/M)\,,
\end{equation}
and in the parametric limit $m_{\mathrm{IR}}/M\to 0$ where the log dominates, we would get the bound
\begin{equation}
\label{eq:toyboundlimitpre}
 \Lambda_1[\mathcal I_g|_{\mathrm{low}}] \geqslant 0 \quad \Rightarrow \quad g_2 \geqslant -  \frac{c_1}{M^2M_{\mathrm{P}}^2}\log\left(\frac M{m_{\mathrm{IR}}}\right)\,.
\end{equation}


\begin{figure}
	\centering
	\begin{subfigure}[t]{0.45\textwidth}
		\centering
		\includegraphics[width=\textwidth]{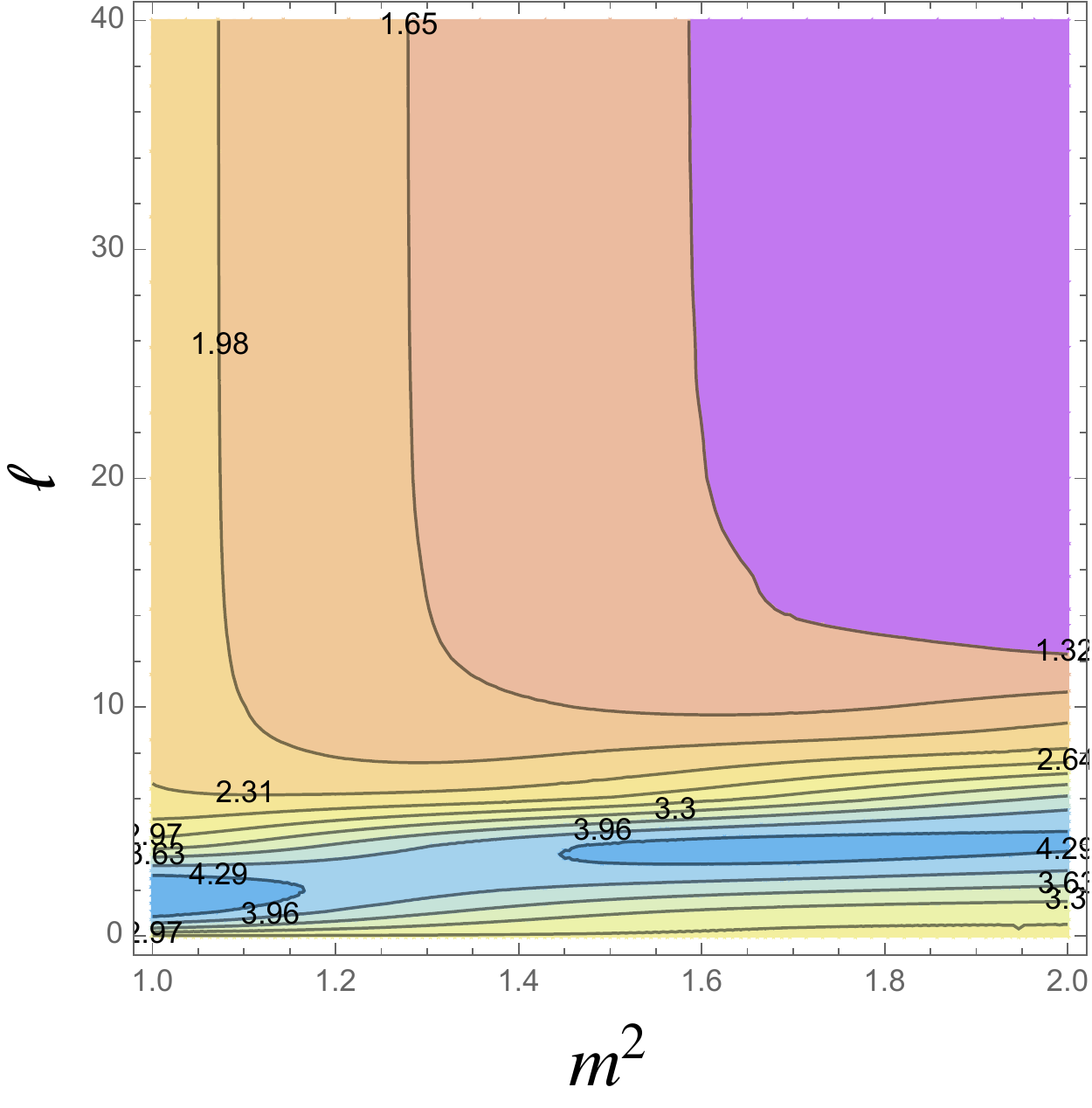}
	\end{subfigure}
	\hfill
	\begin{subfigure}[t]{0.45\textwidth}
		\centering
		\includegraphics[width=\textwidth]{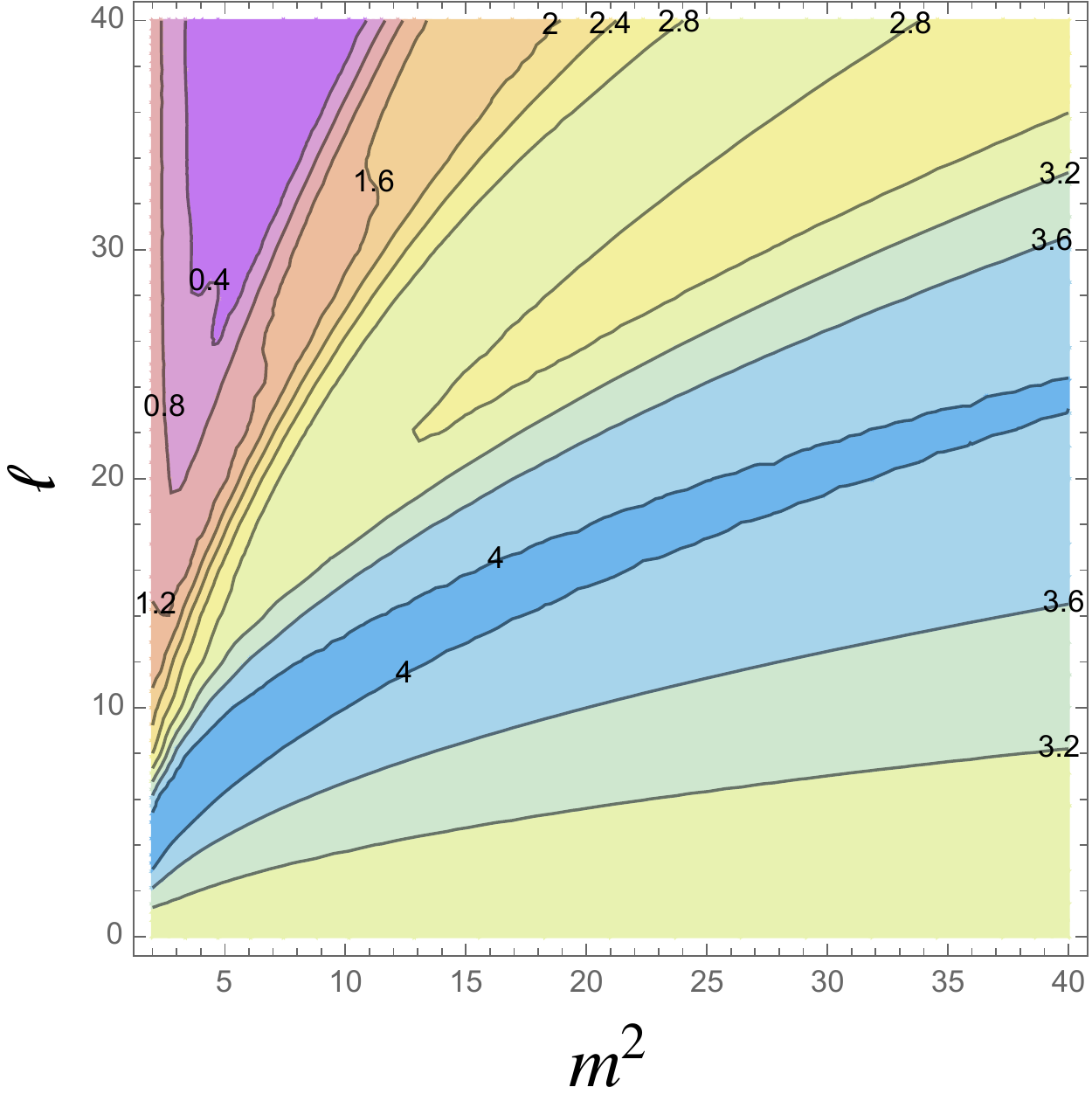}
	\end{subfigure}
	\hfill
	\caption{Action of the functional \eqref{eq:phiformtoy} (with $c_1 = 28.0033$) on $ I_{g,1}$, see \eqref{eq:vectorsV} and \eqref{eq:functionsI}, for finite values of $\ell$ and $m^2$. The contours are obtained by interpolating integer values of $\ell$. Similar plots can be obtained for $ I_{g,2}$.}
		\label{fig:FunctionalAction}
\end{figure}

\begin{figure}
	\centering
	\begin{subfigure}[t]{0.3\textwidth}
		\centering
		\includegraphics[width=\textwidth]{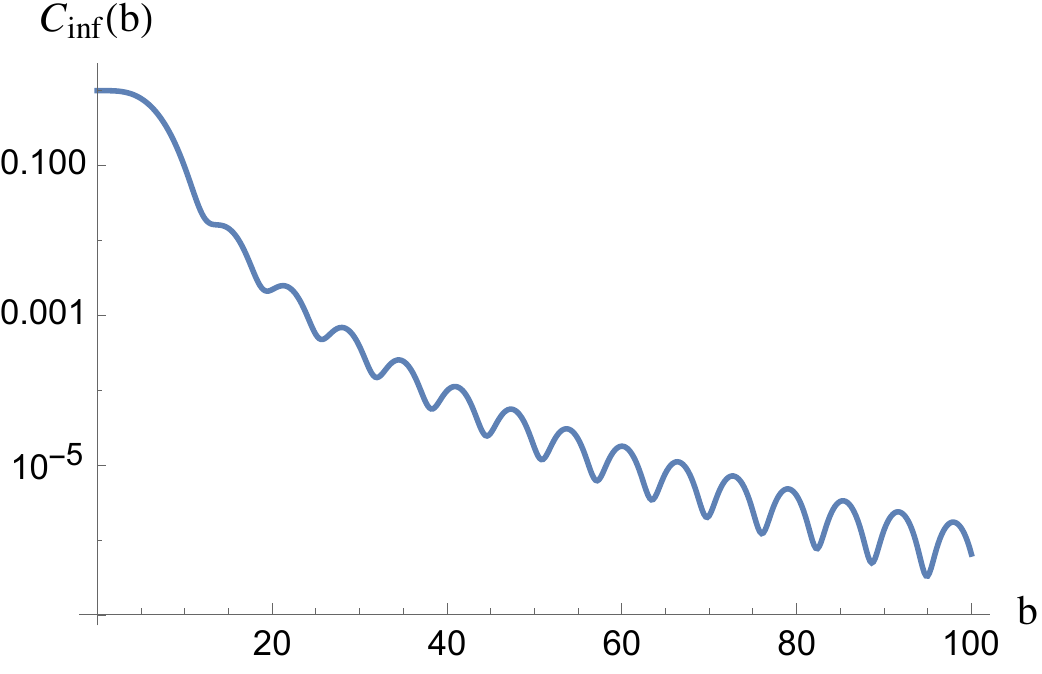}
	\end{subfigure}
	\hfill
	\begin{subfigure}[t]{0.3\textwidth}
		\centering
		\includegraphics[width=\textwidth]{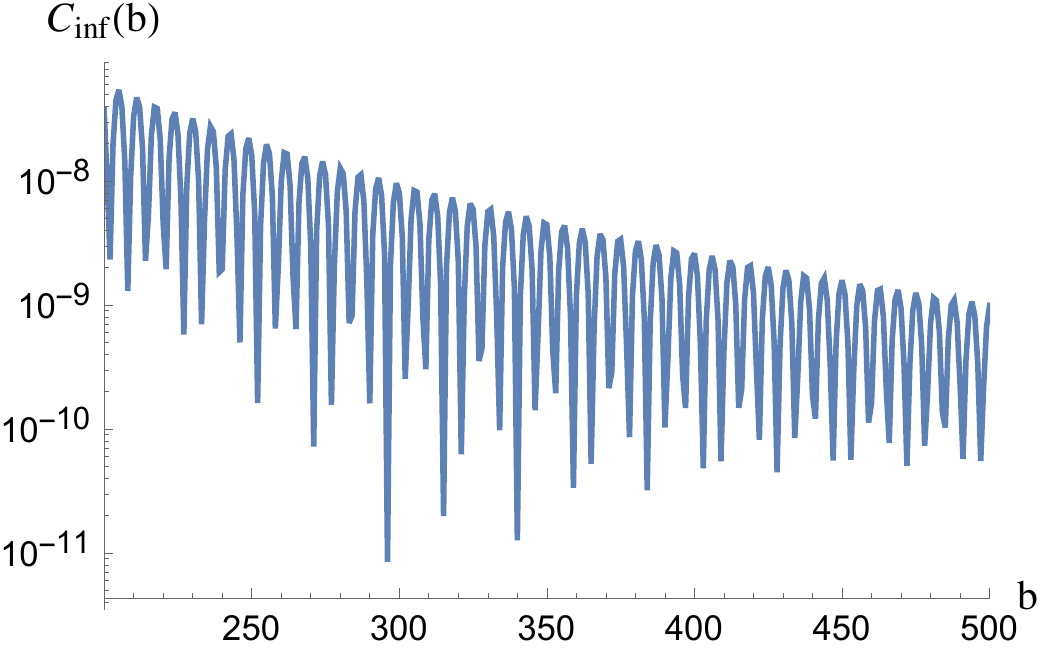}
	\end{subfigure}
		\hfill
	\begin{subfigure}[t]{0.3\textwidth}
		\centering
		\includegraphics[width=\textwidth]{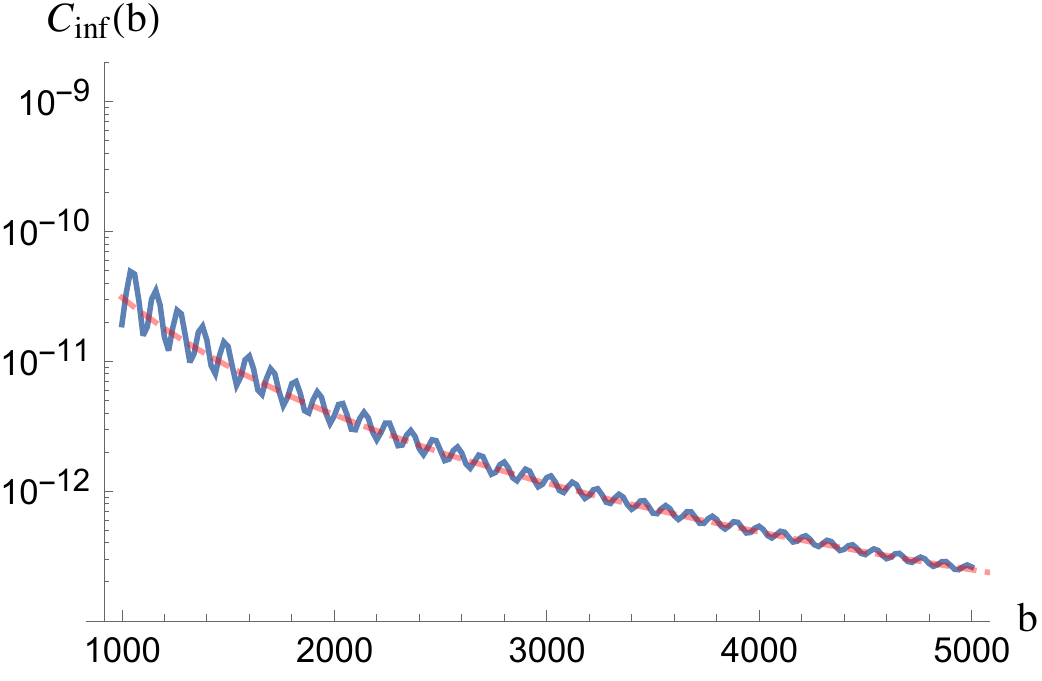}
	\end{subfigure}
	\hfill
	\caption{Action of the functional in the limit of large $m^2$, large $\ell$, as a function of $b$. Here $\delta=0.0033$. The oscillatory terms become important for $b\sim 10^2$ and then gets suppressed compared to the leading $b^{-3}$ behavior (dashed red line), as predicted by \eqref{eq:expansionCinf}.}
		\label{fig:Cinfplot}
\end{figure}

\subsubsection*{Optimizing over $c_1$}

We will now look for a functional of the form \eqref{eq:phiformtoy} that is positive on the high-energy part when $m_\mathrm{IR}=0$. 
The strongest bound is derived from the lowest positive value of $c_1$, which we denote $c_1^*$.
It turns out that it is the last condition in figure~\ref{fig:positivityregimes}, at large $m^2$ and $\ell$, that puts the strongest constraints on what $c_1$ can be used to find a positive functional. We will examine this limit to find 
\begin{equation}
c_1^*=28+\delta\,,
\end{equation}
for a small $\delta$, which has to be numerically determined.

In the high-energy expression \eqref{eq:I-low}, there are two different expressions that appear: $I_{g,1}$ and $I_{g,2}$ \eqref{eq:functionsI}. In the limit of large $\ell$ and $m^2$, for fixed impact parameter, these two expressions have the same asymptotic behavior \eqref{eq:largespin} 
\begin{equation}
\int_0^M dp\,\phi(p)I_{g,i}=  \frac{b^4}{16 \ell^4}C_{\mathrm{inf}}(b) +O(\ell^{-5}), \qquad i=1,2\,,
\end{equation}
with 
\begin{align}
  \label{eq:Cphiinf}
C_{\mathrm{inf}}(b) =&c_1 C_{0,1}(b)-\frac{75}{8} \left(c_1-28\right) C_{0,2}(b)+30 \left(c_1-42\right) C_{0,3}(b)+\left(2205-\frac{175
   c_1}{4}\right) C_{0,4}(b)\nonumber\\
   &+30 \left(c_1-56\right) C_{0,5}(b)-\frac{63}{8} \left(c_1-60\right) C_{0,6}(b)\, ,
   \end{align}
and $C_{0,n}$ defined in \eqref{eq:asymptoticsC}.
Given the expansion
\begin{equation}
\label{eq:expansionCinf}
C_{\mathrm{inf}}(0,b)=\frac{75(c_1-28)}{8}\frac{1}{(bM)^3}+O(b^{-9/2}), \qquad b\to\infty
\,,
\end{equation}
we see that for $c_1>28$, it is positive in the large $b$ limit. To find a functional that is positive also for finite $b$, we choose $c_1^*=28+\delta$, and find that $\delta$ can be taken as small as $\delta=0.0033$.\footnote{We obtained this number numerically. For smaller values of $\delta$ the subleading oscillating powers in \eqref{eq:expansionCinf} produces a negative region at some finite $b$.} Thus we get
\begin{equation}
\label{eq:g2globalsimpl}
g_2 \geqslant -  \frac{28+\delta}{M^2M_{\mathrm{P}}^2}\log\left(\frac M{m_{\mathrm{IR}}}\right)\,.
\end{equation}

By inspection, we can verify that the action of the functional is positive in the first entry  $\vec V^+_{0,m^2}, \vec V^+_{\ell,m^2} ,
\vec V^-_{\ell,m^2} ,
\vec V^o_{\ell,m^2}$ and also $C_{\mathrm{inf}}(b)\geqslant 0$ for any value of $b\geqslant 0$. This is shown in figures~\ref{fig:FunctionalAction} and \ref{fig:Cinfplot}.


\subsubsection{Example 2: Bounds with fixed relation between $g_2$ and $\beta^2$}
\label{sec:WorkedExample2}

\begin{figure}[ht]
  \centering
\includegraphics[width=0.6\textwidth]{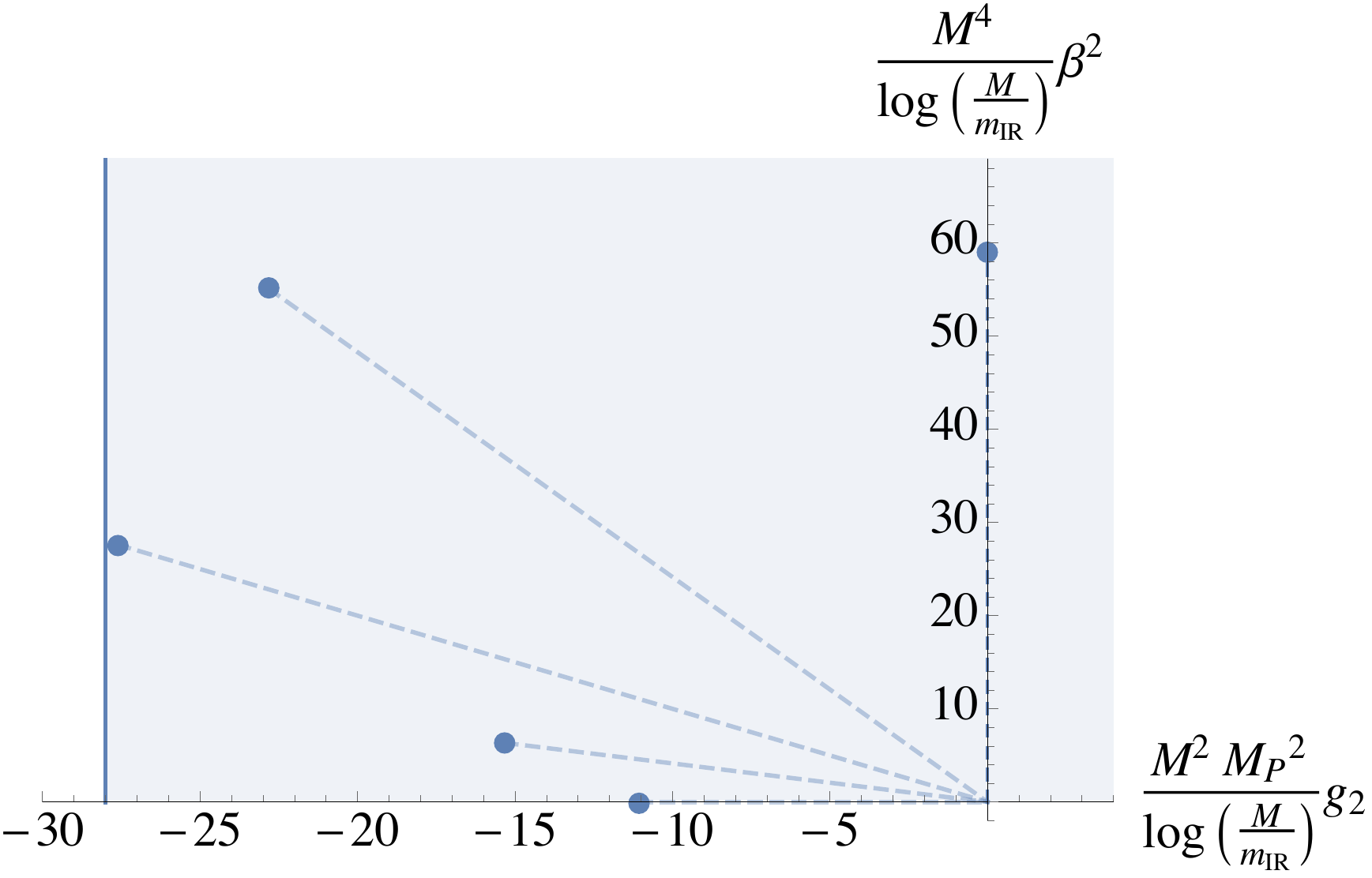}
\caption{Bounds in the plane $g_2M^2M^2_{\mathrm P}/\log(M/m_{\mathrm{IR}})$ vs $\beta^2M^4/\log(M/m_{\mathrm{IR}})$ in the limit $\log(M/m_{\mathrm{IR}})\gg1$. The shaded region corresponds to the bound obtained in \eqref{eq:g2globalsimpl} with $\delta=0.0033$. The circles and the corresponding dashed lines display upper bounds on $\alpha$ in \eqref{eq:alphaParam}.}\label{fig:combinedexplicit}
\end{figure}

Now we will instead look at a functional of the form
\begin{equation}
\label{eq:functionalfourconstants}
\int_{m_{\mathrm{IR}}}^M dp(M-p)^3\left(\left(c_1\tfrac pM+c_2\tfrac{p^2}{M^2}+c_3\tfrac{p^3}{M^3}\right)\mathcal I_g+\left(d_1\tfrac pM+d_2\tfrac{p^2}{M^2}+d_3\tfrac{p^3}{M^2}\right)\mathcal I_0\right)\,.
\end{equation}

We will consider bounds along rays with fixed ratio between $\beta^2$ and $g_2$. Specifically, we will define
\begin{equation}
	\label{eq:alphaParam}
g_2=\frac{\alpha}{M^4}\cos\theta,\qquad \beta^2=\frac{\alpha M_{\mathrm P}^2}{M^6}\sin\theta\,,
\end{equation}
and maximize the parameter $\alpha$ for a given $\theta$. We can normalize the functional so that the low-energy part gives
\begin{equation}
-\alpha +c_1\frac{M^2}{M_\mathrm{P}^2} \log(M/m_{\mathrm{IR}})+\frac{M^2}{M_\mathrm{P}^2} c_0+O(m_{\mathrm{IR}}/M)\,,
\end{equation}
where $c_0$ is a constant. This sets
\begin{equation}
\label{eq:c1c2rel}
c_3=-\frac{3 \left(3 c_1 (4 \sin\theta+7 \cos\theta )+c_2 (6 \sin\theta +7 \cos
   \theta)+420\right)}{10 \sin\theta+9 \cos\theta}\,,
\end{equation}
and for the constant we find
\begin{equation}
\label{eq:c0sol}
c_0=-\frac{c_1(1179\cos\theta+1208\sin\theta)-24c_2(3\cos\theta+4\sin\theta)+3780}{60(9\cos\theta+10\sin\theta)}
\,.
\end{equation}
The optimal upper bound on $\alpha$ is obtained for the smallest value of $c_1$. The algorithm will then be to minimize $c_1>0$ while varying $c_1,c_2$, $d_1$, $d_2$ and $d_3$.
The results are given in table~\ref{tab:explicitfunctionals}. One can see that they satisfy \eqref{eq:c1c2rel}.

\begin{table}
\centering
\caption{Functionals of the form \eqref{eq:functionalfourconstants}, and the corresponding bounds. $c_0$ is given by \eqref{eq:c0sol}.}\label{tab:explicitfunctionals}
\resizebox{\textwidth}{!}{
\renewcommand{\arraystretch}{2.25}
\begin{tabular}{ccccccc}
\hline
$\theta$&\multicolumn{6}{c}{Functional (bound: $\alpha\leqslant \frac{M^2}{M_{\mathrm P}^2}\left(c_1\log(M/m_{\mathrm{IR}})+c_0\right)$)} 
\\\hline
$\frac \pi2$ & $c_1=58.8508$ & $c_2=-95.2201$ & $c_3=-166.4664$ & $d_1=-855.8213$ & $d_2= 3129.6066$ & $ d_3=-2676.5321$
\\
$\frac 58\pi$ & $c_1=59.5901$ & $c_2=-96.4164$ & $c_3=-168.5580$ & $d_1=-866.5732$ & $d_2= 3168.9249$ & $ d_3=-2710.1583$
\\
$\frac 34\pi$ & $c_1=39.0239$ & $c_2=117.0110$ & $c_3=-377.2315$ & $d_1=-175.9453$ & $d_2= 643.4036$ & $ d_3=-550.2567$
\\
$\frac 78\pi$ & $c_1=16.5636$ & $c_2=49.6517$ & $c_3=-21.65410$ & $d_1=300.0060$ & $d_2= -760.9365$ & $ d_3=409.8483$
\\
$ \pi$ & $c_1=11.0348$ & $c_2=33.0782$ & $c_3=-14.4261$ & $d_1=199.92585$ & $d_2= -507.1523$ & $ d_3=273.2191$
\\\hline
\end{tabular}
}
\end{table}

With the results found so far we get, in the limit $\log(M/m_{\mathrm{IR}})\to\infty$, the results \eqref{eq:g2globalsimpl} and the results in table~\ref{tab:explicitfunctionals}. The combined effect of these found are shown in figure~\ref{fig:combinedexplicit}.

\subsection{More results}

In this section we present our best bounds on some of the couplings appearing in the low-energy part of the dispersion relations \eqref{eq:I-low}. We already showed in figure~\ref{fig:g2VSh2sq} the constraint on the parameters $g_2$ and $\beta^2$ obtained acting with a functional on the dispersion relations $\mathcal I_g, \mathcal I_0$ and $\mathcal I_{\beta^2}$. 

\begin{figure}[ht]	
  \centering
\includegraphics[width=0.7\textwidth]{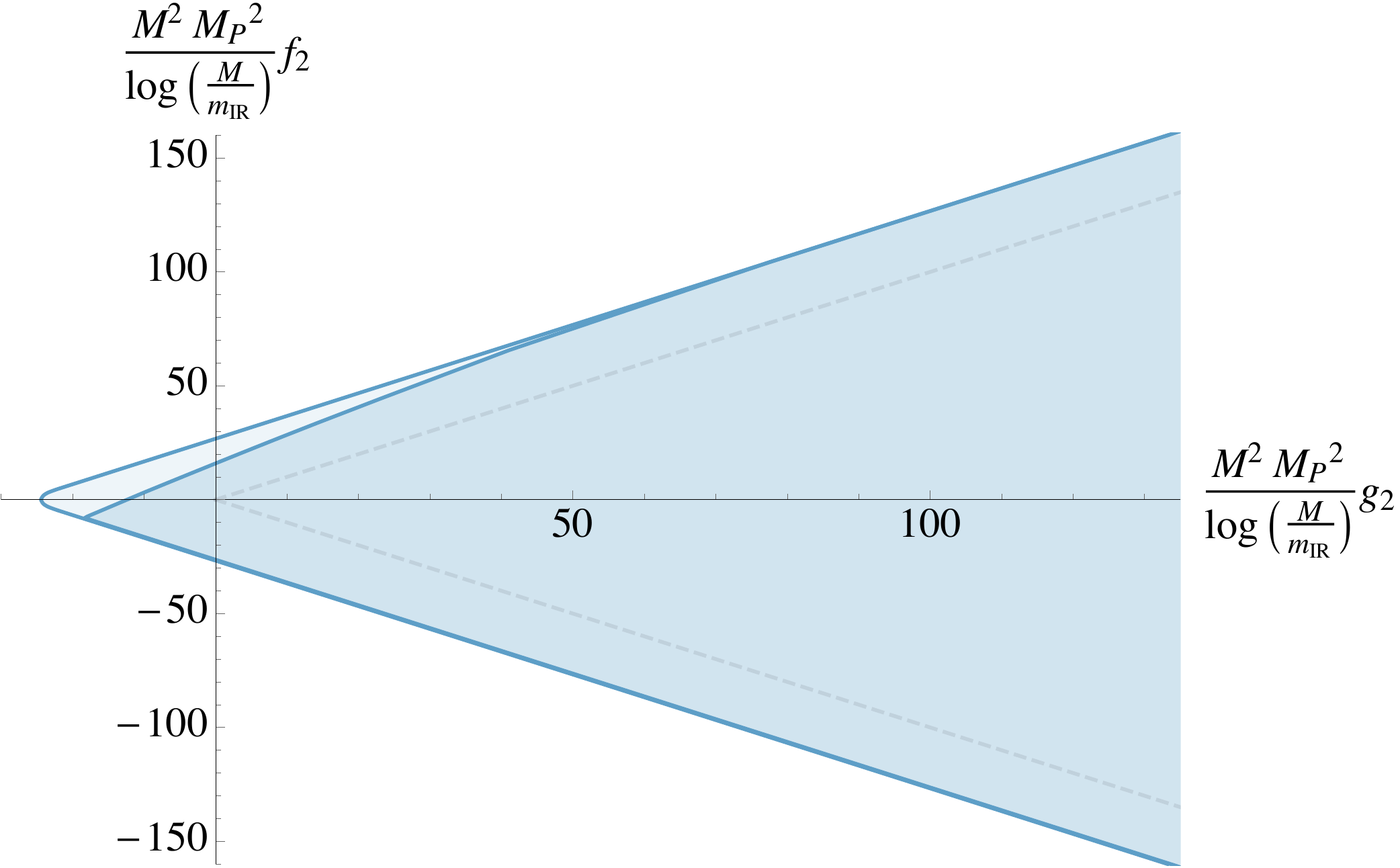}
\caption{Exclusion plot in the plane $(g_2, f_2)$ normalized to $M^2 M^2_\mathrm{P} \log(M/m_\textrm{IR})$. The shaded regions correspond to assuming $f_3\leq0$ (light blue) or $f_3\geq0$ (dark blue). The former contains the latter. The dashed gray lines indicate the bound without gravity $g_2=\pm f_2$. This plot used the dispersion relations $\mathcal I_g$, $\mathcal I_f$, $\mathcal I_0$, $\mathcal I_{\beta^2} $. The minimal value of $g_2$ is the same as in Fig.~\ref{fig:g2VSh2sq} and forces $f_2=0, f_3\leqslant 0$.} \label{fig:g2VSf2}
\end{figure}

Next, we include in our analysis the dispersion relation $\mathcal I_f$, which allows us to consider the coefficient $f_2$, also appearing in the black hole WGC. More precisely, the black hole WGC would require \mbox{$g_2 \pm f_2\geqslant 0$}, but, as shown in figure~\ref{fig:g2VSf2}, the presence of gravity in our setup allows again a violation of the inequality of order $(M/ M_\textrm{P})^2 \log (M/m_\textrm{IR})$.

In section~\ref{sec:posfromunit} we introduced a method to get bounds on the coefficients $h_2 = \beta/M_\mathrm{P}^2$ and $h_3$ appearing in the inelastic scattering amplitude such as $\mathcal M^{+++-}$, however until now we have not fully exploited this technology, except in absence of gravity in section~\ref{sec:resultswithoutgravity}. This because the dispersion relations considered so far only depend on positive spectral densities $|c_{\ell,X}^-|^2$, $|c_{\ell,X}^{+,1}|^2$, $|c_{\ell,X}^{+,2}|^2$, and  $|c_{\ell,X}^o|^2$.

\begin{figure}[ht]	
  \centering
\includegraphics[width=0.7\textwidth]{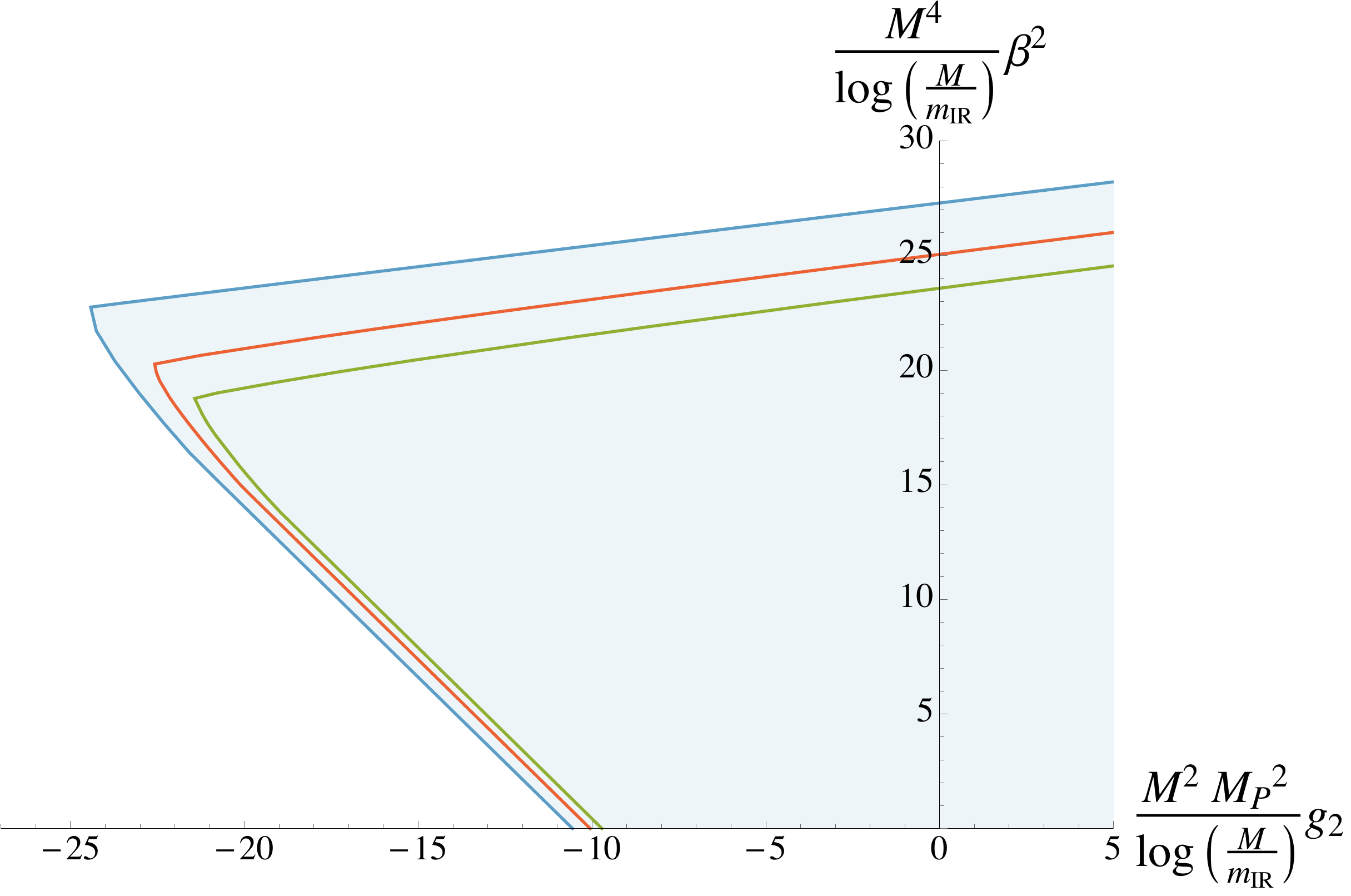}
\caption{Exclusion plot in the plane $(g_2, \beta^2)$ normalised appropriately and divided by $ \log(M/m_\textrm{IR})$. The shaded region reproduces the best bound from figure~\ref{fig:g2VSh2sq} while the green and red lines are obtained with finite values of $m_\textrm{IR} /M= 10^{-6},\, 10^{-10} $ and using the dispersion relations $\mathcal I_g$, $\mathcal I_h$, $\mathcal I_0$, $\mathcal I_{\beta^2}$. Decreasing the value of $m_\textrm{IR}$ the bounds shifts towards the blue bound, obtained in the limit $\log(M/m_\textrm{IR})\gg1$.  The green and red lines do not change when assuming $h_3\geqslant 0$ or $h_3\leqslant 0$.} \label{fig:g2VSbeta}
\end{figure}

Thus, as a final application we include the dispersion relation $\mathcal I_h$ (and drop $\mathcal I_f$) and consider again bounds on $g_2$ and $\beta$. Moreover, we fix $m_\textrm{IR}$ to a finite value and inspect the dependence of the bounds on such value. The results are shown in figure~\ref{fig:g2VSbeta}, for $m_\textrm{IR} = 10^{-6} M$, $10^{-10}M$. The inclusion of $\mathcal I_h$ does not substantially improve the bounds, while the finite value of $m_\textrm{IR}$ corresponds to a finite shift (which is less and less important since we are plotting the bounds divided by $\log(M/m_\textrm{IR}$).

The fact that the bound on $\beta$ does not change  in an appreciable way when including the new dispersion relation $\mathcal I_h$ is a bit surprising. This however is a consequence of the  fundamental input from the low-energy EFT which allowed us to relate $\beta \in \mathcal I_h$ and $\beta^2\in \mathcal I_g$. If we insisted on being agnostic about the interpretation of the low-energy couplings, the inclusion of $\mathcal I_h$ would still let us bound them separately.\footnote{As an example, by assuming $\beta^2\geq0$ but relaxing the relation between $\beta^2$ and the linear term contained in $\mathcal I_h$,  we obtained a bound on the linear term alone: $M^2 |\beta| \geqslant 267.1 \log{(M/m_\textrm{IR})}$, when $g_2=0$ and in the limit where the logarithmic term dominates. This bound is definitively weaker than the one showed in figure~\ref{fig:g2VSbeta} but nevertheless it exists.}

\subsection{Violations to the weak gravity conjecture}
\label{sec:interpretation}

The purpose of this section has been to derive bounds on the four-derivative corrections to Einstein--Maxwell theory. The most interesting bound, in our opinion, is the lower bound on $g_2$, which we find is given by
\begin{align}
    g_2 \geqslant -24.257 \frac{1}{M^2 M_{\mathrm{P}}^2} \log \frac{M}{m_\mathrm{IR}} \, .
\end{align}
This bound is determined with the optimization procedure described above and in the appendix, so our conclusion is the following:

\vspace{5mm} 
\emph{The assumptions of this paper, including unitarity, causality, and weak coupling, are not enough to prove the black hole WGC.}
\vspace{5mm}

\noindent In our opinion, this conclusion is not surprising. As we stated in the introduction, it was already anticipated by \cite{deRham:2019ctd, deRham:2020zyh} that gravity might weaken causality bounds by introducing time delays. Furthermore, it is consistent with the gravitational weakening of the dispersion-relation bounds for scalars in $d \geqslant 5$ reported in \cite{Caron-Huot:2021rmr}.

Using this bound requires that we make sense of the logarithmic divergence. Strictly speaking, if we demand that the cutoff may be taken to 0, this bound simply tells us that no constraint may be placed on $g_2$. However, we believe that it is possible to do better than this -- for instance, it was pointed out in \cite{Caron-Huot:2022ugt} that even with the conservative estimates $M \sim 1$ TeV and $m_{\mathrm{IR}}$ near the Hubble scale, the resulting $\log 10^{77}$ is not very large. Still, it would be nice to understand what the sharpest possible bounds are, but this will require further assumptions. We comment more on this direction in the conclusion.

Another important assumption is weak coupling, which ensures that EFT loops are suppressed in the amplitudes. This is a rather typical assumption and simply means that we are bounding classical, or tree-level amplitudes. However, in the presence of gravity, it becomes more subtle, because the coupling at low-energy might include factors of the high-energy coupling, or the mass of the high-energy particle. To make this more concrete, consider the EFT which arises from integrating out a charged particle such as an electron or charged scalar. The high-energy loop diagrams which contribute to this may include EM and gravitational couplings, and give rise to four-derivative coefficients which take the form
\begin{align}
\label{eq:g2QEDscenario}
g_2 &= \hat g_2\frac{\alpha^2}{M^4}+ \hat g_2'\frac{\alpha}{M^2M_{\mathrm P}^2}  +O\left(\frac1{M_{\mathrm P}^4}\right)
\,,
\\
f_2 &=\hat f_2\frac{\alpha^2}{M^4}+\hat f_2'\frac{\alpha}{M^2M_{\mathrm P}^2}  +O\left(\frac1{M_{\mathrm P}^4}\right)
\,,
\\
\label{eq:bQEDscenario}
h_2 &= \frac\beta{M_\mathrm{P}^2} = \hat \beta\frac{\alpha}{M^2M_{\mathrm P}^2}  +O\left(\frac1{M_{\mathrm P}^4}\right)
\,,
\end{align}
where the hatted variables are completely numerical constants, and the $O(1/M_{\mathrm P}^4)$ terms correspond to pure gravitational loops. Here $\alpha$ represents the strength of the electromagnetic coupling, for QED $\alpha=\frac{e^2}{4\pi}$.
For the cases of a spin-$\frac12$ or scalar particle, the values are referred to as QED or scalar QED (sQED), and for these cases the constants take known values \cite{Drummond:1979pp,Cheung:2014ega},
\begin{align}
&\text{QED}:  & \hat g_2 &=\frac{176}{45} & \hat g_2' &=-\frac{22}{45\pi} & \hat f_2 &=-\frac{16}{15}  & \hat f_2' &=0  & \hat \beta &= -\frac1{180\pi}
\,,
\\
&\text{sQED}:  & \hat g_2 &=\frac{32}{45} & \hat g_2' &=-\frac4{45\pi} & \hat f_2 &=\frac8{15}  & \hat f_2' &= 0 & \hat \beta &= \frac1{360\pi}
\,.
\end{align}

Our assumptions require that the entire amplitude is weakly coupled, meaning that $\alpha \ll 1$, but the meaning of the bounds \eqref{eq:globalg2min}--\eqref{eq:g2minAtbeta0} depends on the relative size of  $\alpha$ and $\mu = \frac{M^2}{M_{\mathrm P}^2}$. In this language, the bounds are
\begin{align}
    \frac{\mu}{M^4} \left( \frac{\alpha^2}{\mu} \hat g_2 + \alpha \hat g_2' \right) \geqslant \frac{\mu}{M^4}  \left( c_1 \log(M / m_\mathrm{IR}) + c_0 \right) \, ,
    \label{eq:hatted bound}
\end{align}
so let us comment on their meaning in the following regimes:
\begin{description}
\item[Regime $\boldsymbol{\mu\ll\alpha^2}$.] In this case, the $\hat g_2$ term is larger than everything else in~\eqref{eq:hatted bound}, so the bound reduces to the familiar
\begin{equation}
\hat g_2\geqslant 0 
\,.
\end{equation}
This is equivalent to the regime where gravity decouples, so we recover those bounds from section~\ref{sec:resultswithoutgravity}. If we integrate out a particle that is light relative to the coupling, \emph{i.e.}
\begin{align}
    M^2 \ll \alpha^2 M_{\mathrm{P}}^2 \, , \label{eq:particleWGC}
\,,
\end{align}
then our results mean that the WGC bounds will be satisfied. This is not so surprising, as such particles already (easily) satisfy the particle form of the WGC.\footnote{The connection between the particle form of the WGC and black hole WGC deep in the IR was also discussed in \cite{Hamada:2018dde}.}
\item[Regime $\boldsymbol{\mu\sim\alpha^2}$.] Let us set $\alpha^2=\mu$ and absorb any additional factor in the numerical constants. Then the bound \eqref{eq:hatted bound} that determines the minimum of $g_2$ takes the form
\begin{equation}
\label{eq:hatg2bound}
\hat g_2 + \alpha \hat g_2' > - c_1\log(M/m_{\mathrm{IR}}) + c_0
\,.
\end{equation}
In this case, the $\hat g_2'$ term is suppressed relative to the others, so our bound only effectively constrains $\hat g_2$. As a result, we find 
\begin{itemize}
\item For any fixed $M/m_{\mathrm{IR}}$, we cannot rule out the possibility that $\hat g_2 $ be negative by an amount given by the right-hand side of~\eqref{eq:hatg2bound}.
\item If $\hat g_2$ has a component that runs logarithmically with $\log(M/m_{\mathrm{IR}})$, we cannot rule out the possibility that this term can be negative, but it must be $>-c_1$.
\end{itemize}
\item[Regime $\boldsymbol{\mu\gg\alpha^2}$.] We are unable to probe this regime, which includes $\mu \sim \alpha$, simply because the left-hand side of ~\eqref{eq:hatg2bound} becomes suppressed compared to the right-hand side. Therefore the bounds are trivially satisfied.
\end{description}

\subsubsection{Constraints on $\beta^2$ and a species bound}

Let us also comment on the meaning of our bounds on $\beta^2$ in terms of $g_2$.\footnote{We thank Simon Caron-Huot for encouraging us to investigate this point.} The allowed region in the space of these two parameters, visible in figure~\ref{fig:g2VSh2sq}, is quite irregular, but at larger values of $g_2$, the slope appears to approach about $0.17$. However, in what follows we will ignore all numbers to focus on the scaling. We find
\begin{align}
    \frac{M^4 }{\log \frac{M}{m_{\mathrm{IR}}}}  \beta^2 \lesssim \frac{M^2 M_\mathrm{P}^2}{\log \frac{M}{m_{\mathrm{IR}}}} g_2 \, .
    \label{eq:specbound1}
\end{align}
In the spirit of the discussion above, let us assume for the moment that these coefficients are dominated by integrating out charged particles at 1-loop level. Then, again ignoring order one numbers, we have 
\begin{align}
    \beta \simeq \sum_{\mathrm{species } \ i} \frac{q^2_i}{m_i^2}
    \label{eq:specbound2} \, , \qquad \qquad g_2 \simeq \sum_{\mathrm{species } \ i} \frac{q^4_i}{m_i^4} + \frac{1}{M_\mathrm{P}^2} \sum_{\mathrm{species } \ i} \frac{q^2_i}{m_i^2}  \, .
\end{align}
In this case, $\beta$ comes from triangle-type diagrams with two photons and one graviton attached to the charged loop, and $g_2$ gets contributions from those diagrams as well as from simple boxes with all four photons attached to the charged loop.\footnote{Actually fermions and bosons contribute to $\beta$ in this sum with the opposite signs, which we will elaborate on below.
}

Let us make the simplifying assumption that there are $n$ different species of charged scalars and they all have an equal value of $z = \sqrt{2} q M_\mathrm{P} / m$. Then our schematic bound~\eqref{eq:specbound1} becomes
\begin{align}
    n^2 z^4 \lesssim \frac{M_\mathrm{P}^2}{M^2} (z^4 +  z^2 )n
\,.
\end{align}
Here it is possible that $z \gg 1$ (incidentally, this means that the species satisfy the particle WGC, though that is not relevant), in which case the bound on $\beta^2$ leads to a simple bound on the number of charged species, $n \lesssim M_\mathrm{P}^2 / M^2$. If $z \lesssim 1$, then a bound on the number of charged species still follows, but the actual value of $z$ begins to matter as well.

It is interesting that this bound is highly analogous to the ``species bound'' \cite{Dvali:2007hz, Dvali:2007wp}, which roughly states that the cutoff scale in a EFT with gravity and a large number of species, is given by 
\begin{align}
    M \simeq M_\mathrm{P} / \sqrt n \, .
\end{align}
Our bound may be interpreted as an analogous bound for charged particles. Adding an extra species with $z > 1$ contributes more to the $\beta^2$ term than to the $g_2$ term, so an upper bound on $\beta^2$ in terms of $g_2$ gives a limit on the number of such species. 

If we allow for both bosons and fermions, the bound is weaker because their contributions to $\beta$ have the opposite sign. However, some scenarios may still be ruled out this way. For instance, the Standard Model has charged bosons and fermions, but the fermions dominate due to the low mass of the electron. Thus if we imagine $n$ copies of the Standard Model coupled only through gravity and electromagnetism, then our upper bound on $\beta^2$ implies a bound on $n$. More generally, since the contribution to $\beta$ from a Dirac fermion is $(-2)$ times the contribution from a complex scalar, we see that bosonic and fermionic degrees of freedom have exactly equal and opposite contributions in this case. This suggests that an upper bound on $\beta^2$ might have an interpretation as a bound on fermion-boson asymmetry. It would be interesting to try to make this speculation more precise in the future.

\section{Conclusion}

In this paper, we have applied dispersion relations to $2 \to 2$ scattering amplitudes of photons in order to derive bounds on higher-derivative corrections to Einstein--Maxwell theory. In doing so, we overcame two main technical challenges. 
First, using an approach similar to \cite{Du:2021byy,Bern:2021ppb}, we arranged the helicity amplitudes in a $4 \times 4$ matrix indexed by their ingoing and outgoing states. This allowed us to derive bounds on inelastic amplitudes in terms of the elastic ones. 
This is important because the WGC inequalities given in~\eqref{eq:wgcbounds} depend linearly on $\beta$, but it is clear from~\eqref{eq:lowamplitudes} that the only amplitude which depends linearly on $\beta$ is $\mathcal{M}^{+++-}$, which is inelastic and cannot be bounded on its own.

The second, and more significant, technical issue addressed here is the so-called graviton pole: the appearance of terms in the low-energy amplitudes which diverge in the limit of small transverse momenta. These terms invalidate bounds derived by taking the forward limit of doubly-subtracted dispersion relations. 
One possible strategy is to include more subtractions to remove the pole from the sum rules, but this has the undesirable side-effect of also removing the four-derivative coefficients, which are relevant to the black hole WGC. In this paper, we used the doubly-subtracted dispersion relation, but we acted on it with more general functionals, rather than simply expanding in the forward limit. This method, developed in \cite{Caron-Huot:2021rmr} for scalars coupled to gravity, yields bounds on four-derivative coefficients. 

However, these bounds are, in general, weaker than the bounds which may be derived without gravity, \emph{i.e.} the forward limit bounds. This is exactly what we found here: as reviewed in section \ref{sec:resultswithoutgravity}, it is easy to prove the WGC bound in the limit where gravity decouples, but in the presence of the graviton pole, the strongest bounds we are able to derive allow for some violation of the WGC. 
This violation is proportional to the ratio $M^2 / M^2_{\mathrm P}$, so it vanishes in the $M_{\mathrm{P}} \to \infty$ limit where gravity decouples, as it should. 

This ``allowed violation'' also includes a logarithmic dependence on an IR cutoff, included to eliminate divergences associated with the well-known IR divergences plaguing massless amplitudes in four-dimensions. 
This does not seem like a fundamental issue, simply because gravitational scattering in four dimensions happens in the real world. Still, it would be nice to understand what if further assumptions might allow us to remove its dependence from our bounds. 
One promising possibility, used recently in \cite{Haring:2022cyf} to derive Froissart-like bounds in for gravitational amplitudes in $d \geqslant 5$, is to add assumptions about the behavior of the amplitude in particular semiclassical regimes. 
Specifically, it may be possible to derive rigorous bounds using functionals that are negative in a regime, if that regime is where the amplitude is controlled by semiclassical physics, such as the eikonal regime at large $b$ or the black hole regime at large $s$. 
Perhaps these, or other assumptions, will tame the divergences. Ultimately, a complete understanding may require reconsidering the meaning of the S-matrix for massless particles, perhaps along the lines of \cite{Kulish:1970ut}, which defines the physical asymptotic states by dressing the free states with a cloud of soft photons / gravitons. 

It is also interesting to consider situations where the cutoff is meaningful. The classic example of this is in AdS, where the role of the IR cutoff is played by the AdS radius $L$. Indeed, it was shown in \cite{Caron-Huot:2021enk} how flat space bounds may be uplifted to AdS, where the divergences are naturally regulated. 
This raises some interesting possibilities. The EFT inequalities for the black hole WGC in AdS were explored in \cite{Cremonini:2019wdk}, and also recently addressed in \cite{Cano:2022ord}. 
Relatedly, CEMZ-like bounds on the $W_{\mu \nu \rho \sigma} F^{\mu \nu} F^{\rho \sigma}$ were obtained in AdS using the analytic bootstrap \cite{Li:2017lmh} and boundary causality \cite{Afkhami-Jeddi:2018own}, the latter of which also considered AdS${}_4$ and found parametric bounds depending on the $\log \Delta_{\mathrm{gap}}$. Our results might be used to make these constraints precise. It would be very interesting to translate our bounds to AdS in order to do a more careful comparison with those works. 

A somewhat more speculative idea is that the IR cutoff may be bounded by basic properties of quantum gravity. This idea is based on the observation, due to Bekenstein \cite{Bekenstein:1973ur, Bekenstein:1974ax, Bekenstein:1980jp}, that the entropy contained in a volume is bounded by the region's surface area. The result is that any local EFT description must breakdown at very large length scales. In \cite{Cohen:1998zx} it is argued that, to satisfy this bound, EFTs should satisfy
\begin{align}
    M^2 \lesssim m_{\mathrm{IR}} M_{\mathrm{P}} \, .
\end{align}
In principle, this could be applied to the IR cutoff scale in this work, giving a natural way to bound it from below by the other two scales. It might be interesting to try to pursue this line of reasoning further.

Of course, these divergences may also be removed by working in more than four dimensions. This introduces new technical issues such as determining the higher-dimensional spinning partial waves, but it seems to us that this can be overcome. Another issue is that in higher dimensions, there are also curvature squared corrections such as $R_{\mu \nu \rho \sigma} R^{\mu \nu \rho \sigma}$, which are related to topological terms in 4d. In general, these terms will appear in the electric\footnote{Magnetic fields are no longer 1-forms in higher dimensions so they no longer couple to black holes. Bounds may be derived by instead considering black strings or branes: see \cite{Cremonini:2020smy} for the relevant caluclation in five dimensions. Interestingly, the Riemann squared contributed to the electric and magnetic bounds with the opposite sign, so it is possible that some positive linear combination of electric and magnetic bounds might be related to the photon amplitudes.} WGC bounds \cite{Kats:2006xp} but not in the the photon four-point function. Therefore we expect that one would need to consider graviton amplitudes as well to relate causality bounds to the WGC in $d > 4$. Bounding $R_{\mu \nu \rho \sigma} R^{\mu \nu \rho \sigma}$ could also have significant interest beyond the WGC, for instance for corrections to the ratio of shear viscosity over entropy \cite{Kovtun:2004de} (see \cite{Cremonini:2011iq} for a review).

More generally, it would be interesting to try to understand if quantum gravity requires more stringent assumptions about the S-matrix than does simple QFT. Indeed, this is related to the basic idea of the Swampland, which is that there are some consistent EFTs which nonetheless cannot arise as a low-energy limit of a theory of quantum gravity. 
In this paper, we show how including quantum gravity \emph{weakens} the possible bounds on scattering amplitudes, so one might wonder if or how quantum gravity can introduce stronger constraints than those of the traditional S-matrix program. 
One promising hint discussed in \cite{Haring:2022cyf} is that certain smeared amplitudes admit \emph{singly} subtracted dispersion relations if one adds assumptions about the behavior of the amplitudes in certain semi-classical limits. Exploring whether these or other assumptions can lead to stronger bounds is an important question that we leave to the future.

\acknowledgments

We thank Callum Jones, Shruti Paranjape, Simon Caron-Huot, Brando Bellazzini and Sasha Zhiboedov for useful discussions and comments on the manuscript. We also thanks David Simmons-Duffin for discussions on the numerical implementation.
This project has received funding from the European Research Council (ERC) under the European Union's Horizon 2020 research and innovation programme (grant agreement no.~758903). 

\appendix

\section{More details on the derivation of sum rules}
\label{app:moredetailssumrules}

The goal of this appendix is to explain in detail the crucial steps that lead to equation
\eqref{eq:hIJintocs}, which we for completeness repeat here:
\begin{equation}
h^{IJ}(\ell,m^2)=\sum_Q\sum_{\ell}\sum_X \vc c_{\ell,X}^\dagger V^{IJ}_Q(\ell,m^2) \vc c_{\ell,X}
\,,
\end{equation}
where the sum over $X$ is a sum over any additional labels that index the states with spin and parity indicated by $Q$.
In this equation, $h^{IJ}$ are the components of a matrix of (the imaginary part of) partial wave densities, and the high-energy contribution is given by summing over spin and integrating along the positive cut, see \eqref{eq:Hhigh}. Explicitly
\begin{align}
&\mt h=\frac1{(m^2-s)(m^2+u)}\begin{pmatrix}
\rho_\ell^{++--}d^\ell_{0,0}(\theta) & \rho_\ell^{++-+}d^\ell_{0,2} (\theta)& \rho_\ell^{+++-}d^\ell_{0,-2}(\theta)  &   \rho_\ell^{++++}d^\ell_{0,0} (\theta)
\\
\rho_\ell^{+---}d^\ell_{2,0} (\theta)& \rho_\ell^{+--+} d^\ell_{2,2}(\theta) & \rho_\ell^{+-+-}d^\ell_{2,-2}(\theta)  & \rho_\ell^{+-++}d^\ell_{2,0}(\theta)
\\ 
\rho_\ell^{-+--}d^\ell_{-2,0}(\theta)   & \rho_\ell^{-+-+}d^\ell_{-2,2}(\theta)   & \rho_\ell^{-++-}d^\ell_{-2,-2}(\theta)   & \rho_\ell^{-+++}d^\ell_{-2,0}(\theta)   
\\
\rho_\ell^{----}d^\ell_{0,0}(\theta)   & \rho_\ell^{---+}d^\ell_{0,2}(\theta)   & \rho_\ell^{--+-}d^\ell_{0,-2}(\theta)   & \rho_\ell^{--++}d^\ell_{0,0}(\theta)  
\end{pmatrix}
\\\nonumber&
+
\frac1{(m^2+u)(m^2+s+u)}\!
\begin{pmatrix}
\rho_\ell^{+--+}d^\ell_{2,2}(\theta)   &\rho_\ell^{++-+}d^\ell_{0,2}(\theta)   & \rho_\ell^{+-++}d^\ell_{2,0}(\theta)  & \rho_\ell^{++++}d^\ell_{0,0}(\theta)  
\\
\rho_\ell^{+---}d^\ell_{2,0}(\theta)  & \rho_\ell^{++--}d^\ell_{0,0}(\theta)  & \rho_\ell^{+-+-}d^\ell_{2,-2}(\theta)  & \rho_\ell^{+++-}d^\ell_{0,-2}(\theta)  
\\ 
\rho_\ell^{---+}d^\ell_{0,2}(\theta)  & \rho_\ell^{-+-+}d^\ell_{-2,2}(\theta)  & \rho_\ell^{--++}d^\ell_{0,0}(\theta)  & \rho_\ell^{-+++}d^\ell_{-2,0}(\theta)  
\\
\rho_\ell^{----}d^\ell_{0,0}(\theta)  & \rho_\ell^{-+--}d^\ell_{-2,0}(\theta)  & \rho_\ell^{--+-}d^\ell_{0,-2}(\theta)  & \rho_\ell^{-++-}d^\ell_{-2,-2}(\theta)  
\end{pmatrix}
\!,
\end{align}
where $\theta=\arccos(1+\frac{2u}{m^2})$. The first of these terms comes from the direct-channel cut, and the second term comes from the crossed-channel cut.

We will use the following simple expression for the Wigner $d$ functions, given in \cite{Caron-Huot:2022ugt}:
\begin{align}
\nonumber
d_{h,h'}^{\ell}(\arccos(x))&=\frac1{\Gamma(h-h'+1)}\sqrt{\frac{\Gamma(\ell+h+1)\Gamma(\ell-h'+1)}{\Gamma(\ell-h+1)\Gamma(\ell+h'+1)}}\left(\tfrac{1+x}2\right)^{\frac{h+h'}2}\left(\tfrac{1-x}2\right)^{\frac{h-h'}2}\\&\quad\times{_2F_1}\left(h-\ell,\ell+h+1;h-h'+1;\tfrac{1-x}2\right)
\,.
\label{eq:WignerDsimple}
\end{align}
We will also use the identities $d^\ell_{h_1,h_2}(\theta)=d^\ell_{-h_2,-h_1}(\theta)$ and $d^\ell_{0,h}(\theta)=d^{\ell}_{0,-h}(\theta)$, valid for even $h$, to simplify the expressions further.

Following the logic in the main text, we will make use of the generalized optical theorem to write the imaginary part $\rho^{\lambda_1\lambda_2\lambda_3\lambda_4}_\ell$ of the partial wave densities as a sum over three-point functions of exchanges states
\begin{equation}
16(2\ell+1) \rho^{\lambda_1\lambda_2\lambda_3\lambda_4}_\ell(m^2)=\sum_{X}  c^{\lambda_1\lambda_2}_{\ell,X}\left(c^{-\lambda_3-\lambda_4}_{\ell,X}
\right)^*
\,.
\end{equation}
In our notation, $ c^{\lambda_i \lambda_j}_{\ell,X}$ is defined by
\begin{align}
    c^{\lambda_i \lambda_j}_{\ell,X}= \sqrt{8(2\ell+1) } \langle X\ell | T | s \ell \lambda_i \lambda_j \rangle
\,.
\end{align}
Boson exchange symmetry and parity symmetry imposes the following constraints on the $ c_{\ell,X}^{\lambda_i \lambda_j}$:
\begin{align}
\label{eq:parityonalpha}
c_{\ell,X}^{\lambda_1\lambda_2}&=P_Xc_{\ell,X}^{-\lambda_2-\lambda_1}
\,,
\\
\label{eq:exchangeonalpha}
c_{\ell,X}^{\lambda_1\lambda_2}&=(-1)^{\ell}c_{\ell,X}^{\lambda_2\lambda_1}
\,.
\end{align}
Here we demanded that the theory respects parity invariance, and hence the exchanged states can be assigned a definite parity $P_X$ in addition to spin $\ell=\ell_X$.
The constraints from \eqref{eq:parityonalpha} and \eqref{eq:exchangeonalpha} take different solutions depending on the assumptions on $P_a$ and $\ell_a$. 1) $P_X=1$, $\ell$ even, 2) $P_X=-1$, $\ell$ even, 3) $P_X=1$, $\ell$ odd. Note that the fourth possibility, odd parity and odd spin, admits no solution.

\subsubsection*{Even parity and even spin}

Here the solutions take the form
\begin{equation}
c_{\ell,X}^{-+}=c_{\ell,X}^{+-}, \qquad c_{\ell,X}^{--}=c_{\ell,X}^{++}
\,,
\end{equation}
which gives
\begin{equation}
h^{IJ}|_{P_X=1,\ell\text{ even}}=\begin{pmatrix}
c^{++}_{\ell,X} &c^{+-}_{\ell,X}
\end{pmatrix}
\begin{pmatrix}
[V_+^{IJ}]^{11} &
[V_+^{IJ}]^{12}
\\
[V_+^{IJ}]^{12} &
[V_+^{IJ}]^{22}
\end{pmatrix}
\begin{pmatrix}
(c_{\ell,X}^{++})^* \\ (c^{+-}_{\ell,X})^*
\end{pmatrix}
\,.
\end{equation}
In this equation, 
\begin{align}
[\mt V_+]_{11}&=\begin{pmatrix}
\frac{d^\ell_{0,0}(\theta)}{\left(m^2-s\right) \left(m^2+u\right)} & 0 & 0 & \frac{\left(2 m^2+u\right)d^\ell_{0,0}(\theta) }{\left(m^2-s\right) \left(m^2+u\right) \left(m^2+s+u\right)} \\
 0 & \frac{d^\ell_{0,0}(\theta)}{\left(m^2+u\right) \left(m^2+s+u\right)} & 0 & 0 \\
 0 & 0 & \frac{d^\ell_{0,0}(\theta)}{\left(m^2+u\right) \left(m^2+s+u\right)} & 0 \\
 \frac{ \left(2 m^2+u\right)d^\ell_{0,0}(\theta)}{\left(m^2-s\right) \left(m^2+u\right) \left(m^2+s+u\right)} & 0 & 0 & \frac{d^\ell_{0,0}(\theta)}{\left(m^2-s\right) \left(m^2+u\right)} 
\end{pmatrix}\!
,
\\
[\mt V_+]_{12}&=
\begin{pmatrix}
0 & 1 & 1 & 0 \\
 1 & 0 & 0 & 1 \\
 1 & 0 & 0 & 1 \\
 0 & 1 & 1 & 0
\end{pmatrix}
\frac{(2 m^2+u)d_{2,0}^\ell(\theta)}{2 \left(m^2-s\right) \left(m^2+u\right) \left(m^2+s+u\right)}
\,,
\\
[\mt V_+]_{22}&=
\begin{pmatrix}
\frac{d^\ell_{2,2}(\theta)}{\left(m^2+u\right) \left(m^2+s+u\right)} & 0 & 0 & 0 \\
 0 & \frac{d^\ell_{2,2}(\theta)}{\left(m^2-s\right) \left(m^2+u\right)} & \frac{\left(2 m^2+u\right)d^\ell_{2,-2}(\theta) }{\left(m^2-s\right) \left(m^2+u\right) \left(m^2+s+u\right)} & 0 \\
 0 & \frac{ \left(2 m^2+u\right)d^\ell_{2,-2}(\theta)}{\left(m^2-s\right) \left(m^2+u\right) \left(m^2+s+u\right)} & \frac{d^\ell_{2,2}(\theta)}{\left(m^2-s\right) \left(m^2+u\right)} & 0 \\
 0 & 0 & 0 & \frac{d^\ell_{2,2}(\theta)}{\left(m^2+u\right) \left(m^2+s+u\right)} 
\end{pmatrix}\!
,
\end{align}

For the Wigner $d$ functions, \eqref{eq:WignerDsimple} reduces to the explicit expressions
\begin{align}
d_{0,0}^\ell(\arccos x)&=\mathcal P_\ell(x)={_2F_1}\left(-\ell,\ell+1,1,\frac{1-x}2\right)
\,,
\\
d_{2,2}^\ell(\arccos x)&=\frac{(1+x)^2}4{_2F_1}\left(2-\ell,3+\ell,1,\frac{1-x}2\right)
\,,
\\
d_{2,0}^\ell(\arccos x)&=\frac{1-x^2}8\sqrt{(\ell-1)\ell(\ell+1)(\ell+2)} {_2F_1}\left(2-\ell,\ell+3;3;\frac{1-x}{2}\right)
\,,
\\
d_{2,-2}^\ell(\arccos x)&=\frac{(1-x)^2}{96}(\ell-1)\ell(\ell+1)(\ell+2) {_2F_1}\left(2-\ell,\ell+3;5;\frac{1-x}{2}\right)
\,.
\end{align}
In the first line the Wigner $d$ function is given by the usual Legendre polynomial.

The case $\ell=0$ is special, in this case only the upper left corner $[\mt V_+]_{11}$ survives, and we denote it by 
\begin{align}
\mt V_0&=[\mt V_+]_{11}|_{\ell=0}
\nonumber
\\&=\begin{pmatrix}
 \frac{1}{\left(m^2-s\right) \left(m^2+u\right)} & 0 & 0 & \frac{2 m^2+u}{\left(m^2-s\right) \left(m^2+u\right) \left(m^2+s+u\right)} \\
 0 & \frac{1}{\left(m^2+u\right) \left(m^2+s+u\right)} & 0 & 0 \\
 0 & 0 & \frac{1}{\left(m^2+u\right) \left(m^2+s+u\right)} & 0 \\
 \frac{2 m^2+u}{\left(m^2-s\right) \left(m^2+u\right) \left(m^2+s+u\right)} & 0 & 0 & \frac{1}{\left(m^2-s\right) \left(m^2+u\right)}
\end{pmatrix}
\,.
\end{align}

\subsubsection*{Odd parity, even spin}

For odd parity, only even spins contribute and $c_{\ell,X}^{+-}=c_{\ell,X}^{-+}=0$. Moreover, 
\begin{equation}
c_{\ell,X}^{--}=-c_{\ell,X}^{++}\,,
\end{equation}
We get
\begin{equation}
h^{IJ}|_{P_X=-1,\ell\text{ even}} = c_{\ell,X}^{++} V_-^{IJ}(c_{\ell,X}^{++})^* \,,
\end{equation}
where
\begin{equation}
\mt V_-=
\begin{pmatrix}
\frac{d^\ell_{0,0}(\theta)}{\left(m^2-s\right) \left(m^2+u\right)} & 0 & 0 & -\frac{\left(2 m^2+u\right)d^\ell_{0,0}(\theta) }{\left(m^2-s\right) \left(m^2+u\right) \left(m^2+s+u\right)} \\
 0 & \frac{d^\ell_{0,0}(\theta)}{\left(m^2+u\right) \left(m^2+s+u\right)} & 0 & 0 \\
 0 & 0 & \frac{d^\ell_{0,0}(\theta)}{\left(m^2+u\right) \left(m^2+s+u\right)} & 0 \\
 -\frac{\left(2 m^2+u\right)d^\ell_{0,0}(\theta) }{\left(m^2-s\right) \left(m^2+u\right) \left(m^2+s+u\right)} & 0 & 0 & \frac{d^\ell_{0,0}(\theta)}{\left(m^2-s\right) \left(m^2+u\right)} 
\end{pmatrix}
\!.
\end{equation}

\subsubsection*{Even parity, odd spin}

For even parity and odd spin, we have $c_{\ell,X}^{--}=c_{\ell,X}^{++}=0$ and
\begin{equation}
c_{\ell,X}^{-+}=-c_{\ell,X}^{+-}\,,
\end{equation}
We get
\begin{equation}
h^{IJ}|_{P_X=1,\ell\text{ odd}} = c_{\ell,X}^{+-} V_o^{IJ}(c_{\ell,X}^{+-})^*\, ,
\end{equation}
where
\begin{equation}
\mt V_o=
\begin{pmatrix}
 \frac{d^\ell_{2,2}(\theta)}{\left(m^2+u\right) \left(m^2+s+u\right)} & 0 & 0 & 0 \\
 0 & \frac{d^\ell_{2,2}(\theta)}{\left(m^2-s\right) \left(m^2+u\right)} & -\frac{ \left(2 m^2+u\right)d^\ell_{2,-2}(\theta)}{\left(m^2-s\right) \left(m^2+u\right) \left(m^2+s+u\right)} & 0 \\
 0 & -\frac{ \left(2 m^2+u\right)d^\ell_{2,-2}(\theta)}{\left(m^2-s\right) \left(m^2+u\right) \left(m^2+s+u\right)} & \frac{d^\ell_{2,2}(\theta)}{\left(m^2-s\right) \left(m^2+u\right)} & 0 \\
 0 & 0 & 0 & \frac{d^\ell_{2,2}(\theta)}{\left(m^2+u\right) \left(m^2+s+u\right)}
\end{pmatrix}.
\end{equation}

\subsection*{Sum rules}

In general, we will write down sum rules for EFT coefficients on the form
\begin{align}
\alpha =
 & \int_{M^2}^\infty \frac{dm^2}{m^2} \bigg( \sum_X|c_{0,X}^{0}|^2   V^0_{\alpha}  +\sum_{\ell=2,4,\ldots}\sum_X(\vc c_{\ell,X}^+)^\dagger  \mt V^+_{\alpha}\vc c^+_{\ell,X} \nonumber  \\
&+  \sum_{\ell=0,2,\ldots}\sum_X |c^-_{\ell,X} |^2    V^-_\alpha +   \sum_{\ell=3,5,\ldots}\sum_X  |c^o_{\ell,X} |^2    V^o_{\alpha} \bigg)
\,,
\label{eq:genSumRule}
\end{align}
where $\alpha$ represents a generic EFT coefficient. Specifying a sum rule amounts to giving the functions $V_\alpha^0$, $\mt V_\alpha^+$, $V^-_\alpha$ and $V^o_\alpha$. 

As explained in the main text, sum rules are derived by choosing a real vector $\vc v$ and powers $s^pu^q$, \eqref{eq:low-highpq}, or alternatively $s^0$ and integrating against a function of $p=\sqrt{-u}$, \eqref{eq:low-highint}.
As an example, choose $\vc v=(1,0,0,0)^T$, and $s^pu^0$ for $p\geqslant 0$. This gives a sum rule for $\tilde g_{p,0}$, where we defined $\tilde g_{p,q}$ to be the term in $g(s|t,u)$ that is proportional to $s^{2+p-q}u^q$:\footnote{This parametrization of the $g$ amplitude agrees with the one used in \cite{Arkani-Hamed:2020blm}, where $\tilde g_{p,q}$ was denoted $a_{p,q}$. Relating to the parametrization of \eqref{eq:lowamplitudes}, $\tilde g_{0,0}=g_2$, $\tilde g_{1,0}=g_3-\frac23\frac{\beta^2}{M_{\mathrm{P}}^2}$, $\tilde g_{2,0}=g_{4,1}+2g_{4,2}$ and $\tilde g_{2,1}=2g_{4,2}$.}
\begin{align}
    V^0_{\tilde g_{p,0}} \ &= \ \frac1{m^{2p+4}}    \, , \quad 
    &
    \mt V^+_{\tilde g_{p,0}}\ &= \  \frac1{m^{2p+4}} 
    \begin{pmatrix}
     1 & 0 \\
     0 & (-1)^p \\
    \end{pmatrix}  \, 
\,,  \nonumber\\
  V^-_{\tilde g_{p,0}} \ &=\  \frac1{m^{2p+4}}   \, , \quad 
    &
   V^o_{\tilde g_{p,0}}  \ &= \ \frac{(-1)^p}{m^{2p+4}} 
\,.
   \label{eq:tildegp0rule}
\end{align}
The rule for $p=0$, with $\tilde g_{0,0}=g_2$, is reproduced in \eqref{eq:g2sumrule} in the main text, and is not valid in the presence of gravity.
Note that the sum rule \eqref{eq:tildegp0rule} immediately implies that the bound
\begin{equation}
0\leqslant \frac{\tilde g_{p,0}}{g_2}\leqslant \frac{1}{M^{2p}}, \quad \text{$p$ even}\,, \qquad  -\frac1{M^{2p}}\leqslant \frac{\tilde g_{p,0}}{g_2}\leqslant \frac{1}{M^{2p}}, \quad \text{$p$ odd}\,,
\label{eq:gtilderule}
\end{equation}
valid without gravity.

In a similar way, by picking the same $\vc v$ and looking at a suitable linear combination of the powers $s^{p-1}u^1$ and $s^pu^0$, one finds
\begin{align}
    V^0_{\tilde g_{p,1}} \ &= 0  \, , \quad 
    &
    \mt V^+_{\tilde g_{p,1}}\ &= \  \frac1{m^{2p+4}} 
\begin{pmatrix}
\ell(\ell+1) &0 
\\
0  & (-1)^p(p+6-\ell(\ell+1))
\end{pmatrix} \, 
\,,  \\
  V^-_{\tilde g_{p,1}} \ &=\  \frac{\ell(\ell+1)}{m^{2p+4}}  \, , \quad 
    &
   V^o_{\tilde g_{p,1}}  \ &= \ \frac{(-1)^p}{m^{2p+4}} (p+6-\ell(\ell+1))
\,.
\end{align}

To systematically derive a basis of sum rules, we note that for a matrix $\mt M$, we have $\vc v^T\mt M\vc v=\mathrm{Tr}(\mt w\mt  M)$ for $\mt w=\vc v\vc v^T$. Using this fact, a basis of sum rules can be found by considering all linearly independent symmetric matrices $\mt w$. For any given $s^pu^q$, this would give ten different sum rules, however typically not all of the sum rules are linearly independent and one has to find a basis among the sum rules. Any sum rule in such a basis with the low-energy side being zero constitutes a null constraint.

Before proceeding to integral sum rules, let us explain how to make contact with the formalism used in \cite{Henriksson:2021ymi}. In that paper, no dispersion relation for the $h$-type amplitude was used, which means that $V_+^{IJ}$ is diagonal, \emph{i.e.} $[V_+^{IJ}]^{12}=0$ in all sum rules. Then \eqref{eq:genSumRule} takes the form
\begin{align}
 \alpha &= \int_{M^2}^\infty \frac{dm^2}{m^2} \bigg( \sum_{\ell=0,2,\ldots}\sum_X \left|(\vc c_{\ell,X}^+)^{11}\right|^2 [V^+_{\alpha}]^{11} + \sum_{\ell=2,4,\ldots}\sum_X \left |(\vc c_{\ell,X}^+)^{22}\right|^2 [V^+_{\alpha}]^{22} \nonumber  \\
&+  \sum_{\ell=0,2,\ldots}\sum_X |c^-_{\ell,X} |^2    V^-_\alpha +   \sum_{\ell=3,5,\ldots}\sum_X  |c^o_{\ell,X} |^2    V^o_{\alpha} \bigg)
\,.
\end{align}
For a given sum rule for $g$-type and $f$-type Wilson coefficients $\alpha$, the entries $[V_\alpha^+]^{11}$, $[V_\alpha^+]^{22}$, $V_\alpha^-$ and $V_\alpha^o$ agree exactly with the expressions for $2V_\alpha^+$, $V_\alpha^e$, $2V_\alpha^-$ and $V_\alpha^o$ in \cite{Henriksson:2021ymi}.

\subsection*{Improved integral sum rules}

Consider taking the contraction with $\vc v=(1,0,0,0)^T$, and picking the power $s^0$, keeping $u$ general. This gives for the low-energy side
\begin{equation}
\left.\vc v^T\mt L\vc v\right|_{s^0} = -\frac1{M_{\mathrm{P}}^2}\frac1u+g_2-u g_3-\frac{10\beta^2}{3M_{\mathrm{P}}^2}u+u^2\tilde g_{2,0}-u^3\tilde g_{3,0}+u^4\tilde g_{4,0}-u^5\tilde g_{5,0}+\ldots
\,,
\end{equation}
with a corresponding high-energy expressions
\begin{align}
\left. v^IV_0^{IJ}  v^J\right|_{s^0} & = \frac1{m^2(u+m^2)}
\,,
&
\left. v^I\mt V_+^{IJ}  v^J\right|_{s^0} &= \begin{pmatrix}
\frac{\mathcal P_\ell(1+\frac{2u}{m^2})}{m^2(u+m^2)} & 0
\nonumber\\
0 &
\frac{\tilde{\mathcal P}_\ell(1+\frac{2u}{m^2})}{m^4}
\end{pmatrix}
\,,
\\
\left. v^IV_-^{IJ}  v^J\right|_{s^0} & =  \frac{\mathcal P_\ell(1+\frac{2u}{m^2})}{m^2(u+m^2)}
\,,
&
\left. v^IV_o^{IJ}  v^J\right|_{s^0} & = \frac{\tilde{\mathcal P}_\ell(1+\frac{2u}{m^2})}{m^4} 
\,,
\label{eq:highenergyimprovedpre}
\end{align}
where $\mathcal P_\ell(x)={_2F_1}(-\ell,\ell+1,1,\frac{1-x}2)$ and $\tilde{\mathcal P}_\ell(x)={_2F_1}(2-\ell,3+\ell,1,\frac{1-x}2)$.
In principle, one could construct integral sum rules by integrating this expression at $u=-p^2$ against a function $\phi(p)$. A much more practical method is to first subtract an infinite tower of sum rules for $(-1)^ku^k\tilde g_{k,0}$ with $k\geqslant 1$, using \eqref{eq:tildegp0rule}. This idea was advocated in \cite{Caron-Huot:2021rmr}.
In this manner, we find the formal equality
\begin{equation}
\mathcal I_g|_{\mathrm{low}}=\mathcal I_g|_{\mathrm{high}}
\,,
\end{equation}
where 
\begin{align}
\mathcal I_g |_{\mathrm{low}} &=- \frac1{M_{\mathrm{P}}^2}\frac 1u+ g_2- 4\frac{\beta^2}{M_{\mathrm P}^2} u
\,,
\end{align}
and

\begin{align}
\mathcal I_g |_{\mathrm{high}} &= \int_{M^2}^\infty \frac{dm^2}{m^2}\bigg( \sum_X |c_{0,X}^{0}|^2   V^0_{g}  +\sum_{\ell=2,4,\ldots}\sum_X 
(\vc c_{\ell,X}^+)^\dagger   \mt V^+_{g}\vc c_{\ell,X}^+ \nonumber  \\
&
\quad \qquad \qquad
+  \sum_{\ell=0,2,\ldots}\sum_X
 |c^-_{\ell,X} |^2   { V^-_{g}} +   \sum_{\ell=3,5,\ldots}\sum_X 
  |c^o_{\ell,X} |^2   { V^o_{g}} 
\label{eq:IgDispersion}
\,,
\end{align}

with
\begin{align}
V_g^0&= \frac1{m^2(u+m^2)}+\frac{u}{m^4(u+m^2)}
\\
\mt V_g^+&=\begin{pmatrix}
\frac{\mathcal P_\ell(1+\frac{2u}{m^2})}{m^2(u+m^2)} & 0
\,,
\\
0 &
\frac{\tilde{\mathcal P}_\ell(1+\frac{2u}{m^2})}{m^4}
\end{pmatrix}+\begin{pmatrix}
\frac{u}{m^4(u+m^2)} & 0
\\
0 &\frac{u}{m^4(u-m^2)}
\end{pmatrix}
\,,
\\
V_g^-&=
\frac{\mathcal P_\ell(1+\frac{2u}{m^2})}{m^2(u+m^2)}
+
\frac{u}{m^4(u+m^2)}
\,,
\\
V_g^o&=
\frac{\tilde{\mathcal P}_\ell(1+\frac{2u}{m^2})}{m^4}
+
\frac{u}{m^4(u-m^2)}
\,.
\end{align}
Here in each expression, the first term is the original sum rule in \eqref{eq:highenergyimprovedpre}, and the second term is the result of the (negative) sum of forward sum rules using \eqref{eq:tildegp0rule}.
In section~\ref{sec:resultswithgravity} we present four more improved sum rules of this type. They are constructed in a completely analogous way, deriving from other choices of $\vc v$. Specifically, $\mathcal I_f$ is found by looking at a linear combination of sum rules from $(1,0,0,1)^T$ and $(1,0,0,-1)^T$, $\mathcal I_h$ by a linear combination of rules from $(1,1,0,0)^T$ and $(1,-1,0,0)^T$, and $\mathcal I_{\beta^2}$ by a linear combination of rules from $(0,1,1,0)^T$ and $(0,1,-1,0)^T$. Finally $\mathcal I_0$ is an integral null constraint, found by adding to $\mathcal I_g$ a linear combination of sum rules from $(0,1,0,0)^T$.

\section{Plots without gravity}
\label{app: Plots no grav}
\begin{figure}
	\centering
	\begin{subfigure}[t]{0.45\textwidth}
		\centering
		\includegraphics[width=\textwidth]{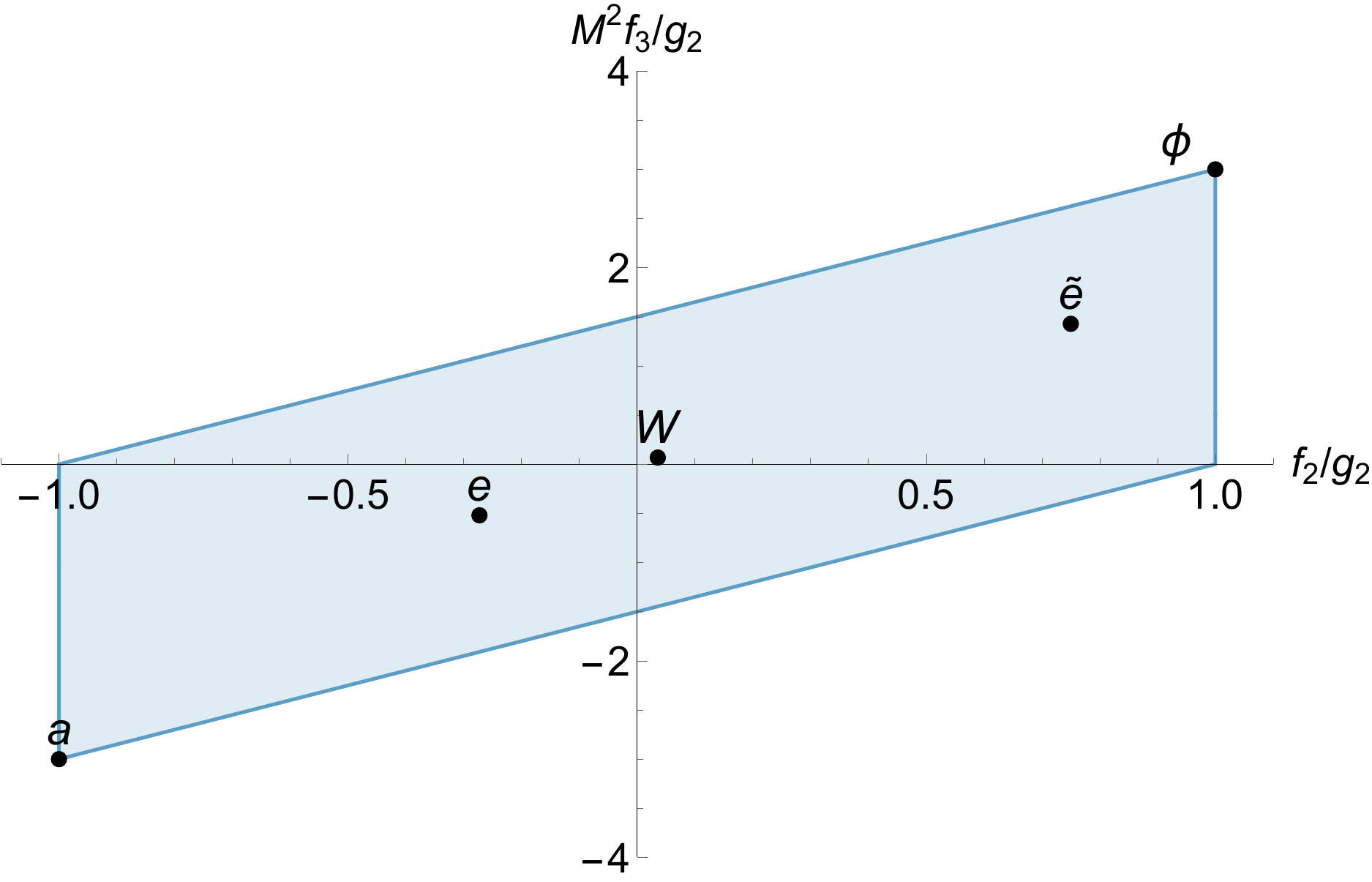}
	\end{subfigure}
	\hfill
	\begin{subfigure}[t]{0.45\textwidth}
		\centering
		\includegraphics[width=\textwidth]{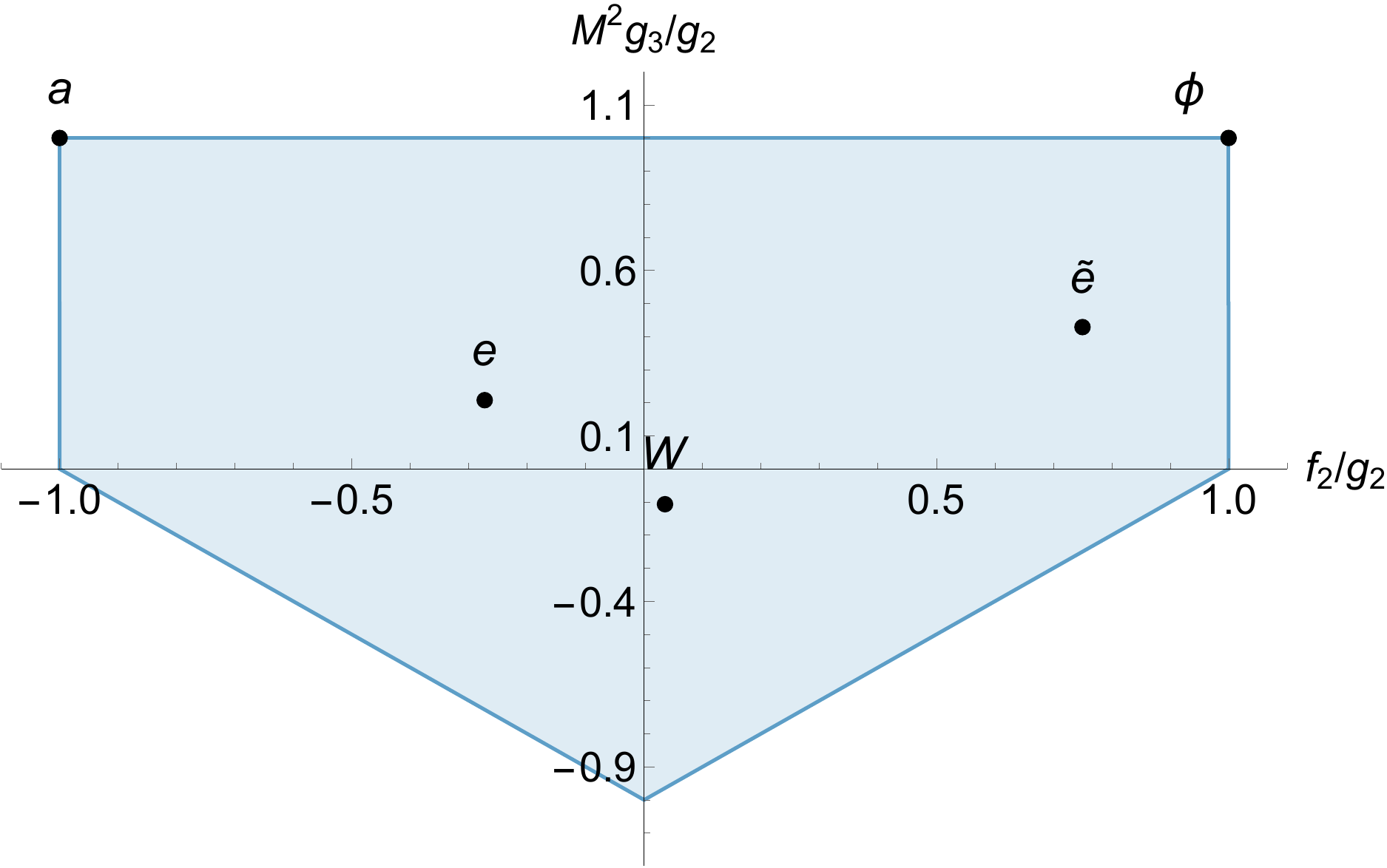}
	\end{subfigure}
	\hfill
		\caption{Allowed regions in the space $(M^2 f_3/g_2,f_2/g_2)$ on the left, $(M^2 g_3/g_2,f_2/g_2)$ on the right. They both give stronger bounds on the vertical direction w.r.t. \cite{Henriksson:2021ymi}.}
	\label{fig:app1}
\end{figure}

\begin{figure}
	\centering
	\begin{subfigure}[t]{0.45\textwidth}
		\centering
		\includegraphics[width=\textwidth]{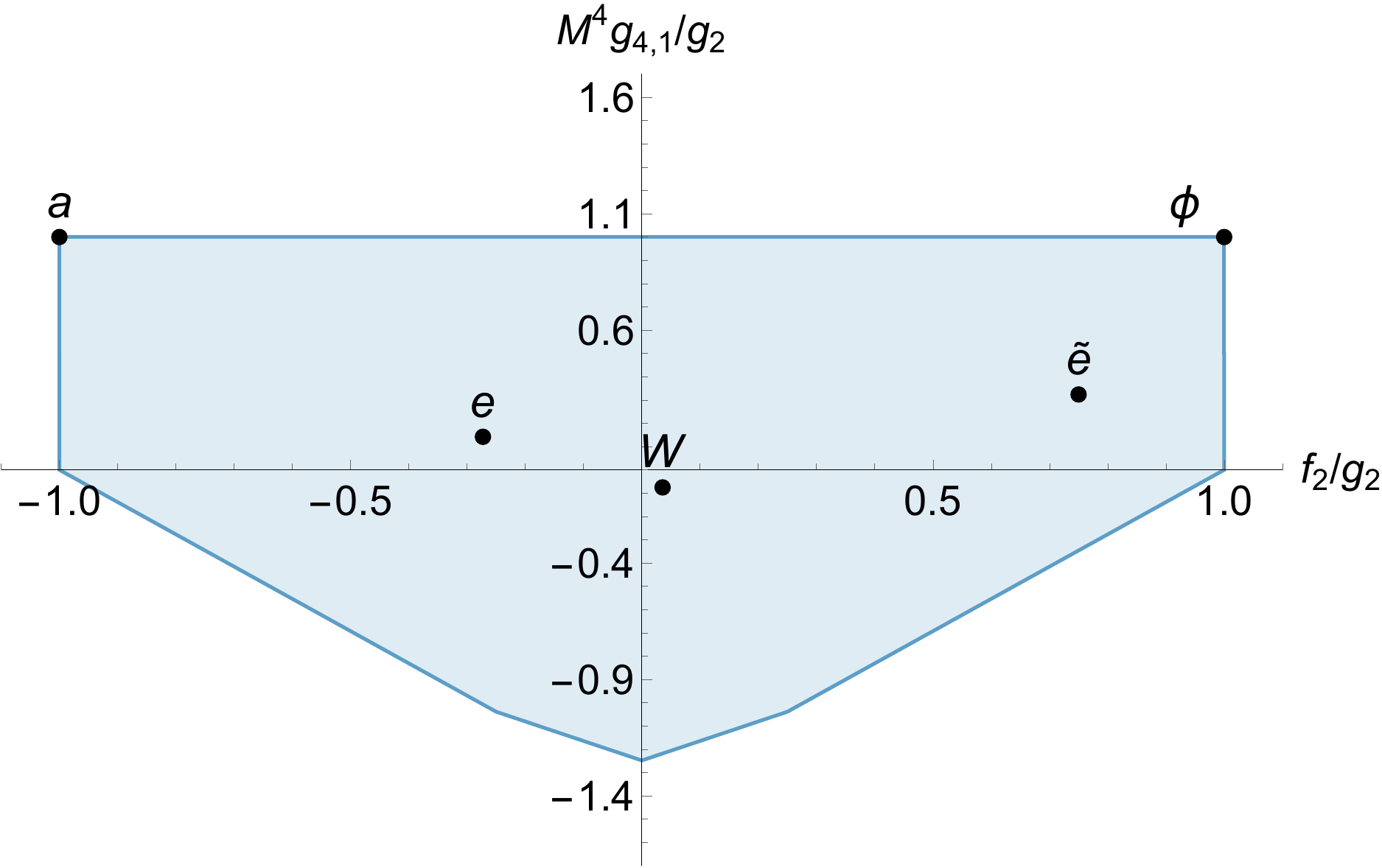}
	\end{subfigure}
	\hfill
	\begin{subfigure}[t]{0.45\textwidth}
		\centering
		\includegraphics[width=\textwidth]{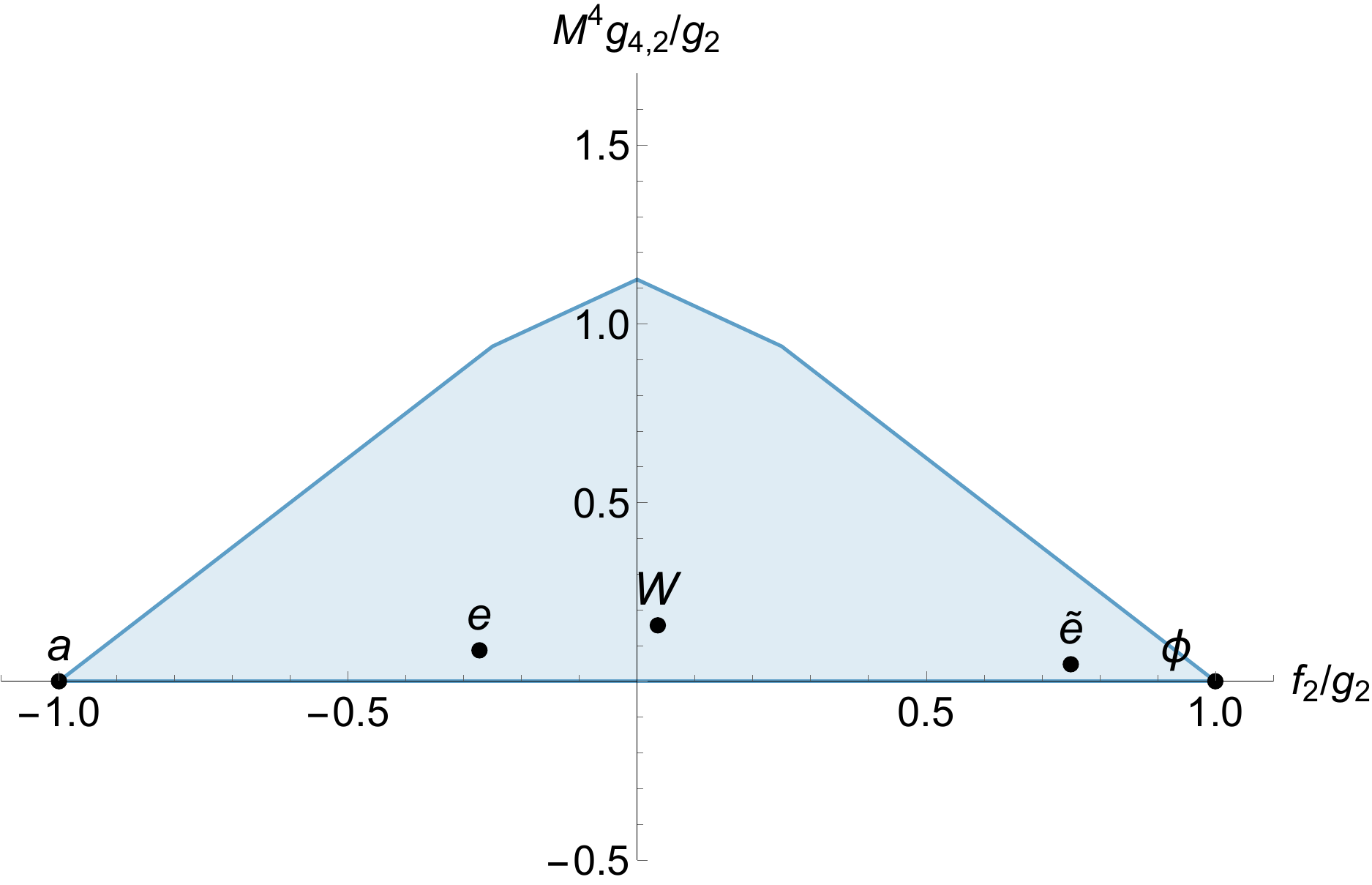}
	\end{subfigure}
	\hfill
		\caption{Allowed regions in the space $(M^4 g_{4,1}/g_2,f_2/g_2)$ on the left, $(M^4 g_{4,2}/g_2,f_2/g_2)$ on the right.}
	\label{fig:app2}
\end{figure}

\begin{figure}
	\centering
	\begin{subfigure}[t]{0.45\textwidth}
		\centering
		\includegraphics[width=\textwidth]{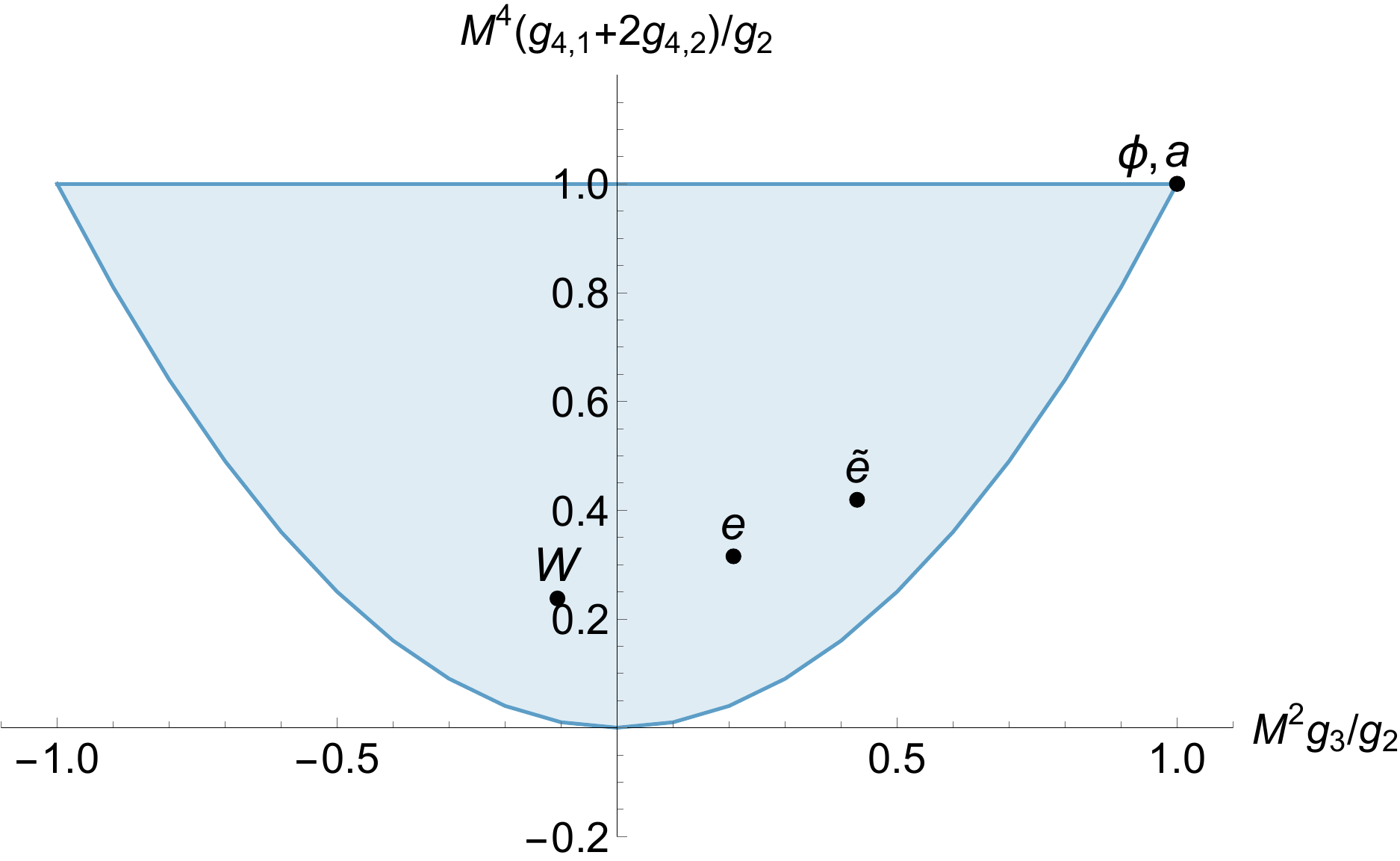}
	\end{subfigure}
	\hfill
	\begin{subfigure}[t]{0.45\textwidth}
		\centering
		\includegraphics[width=\textwidth]{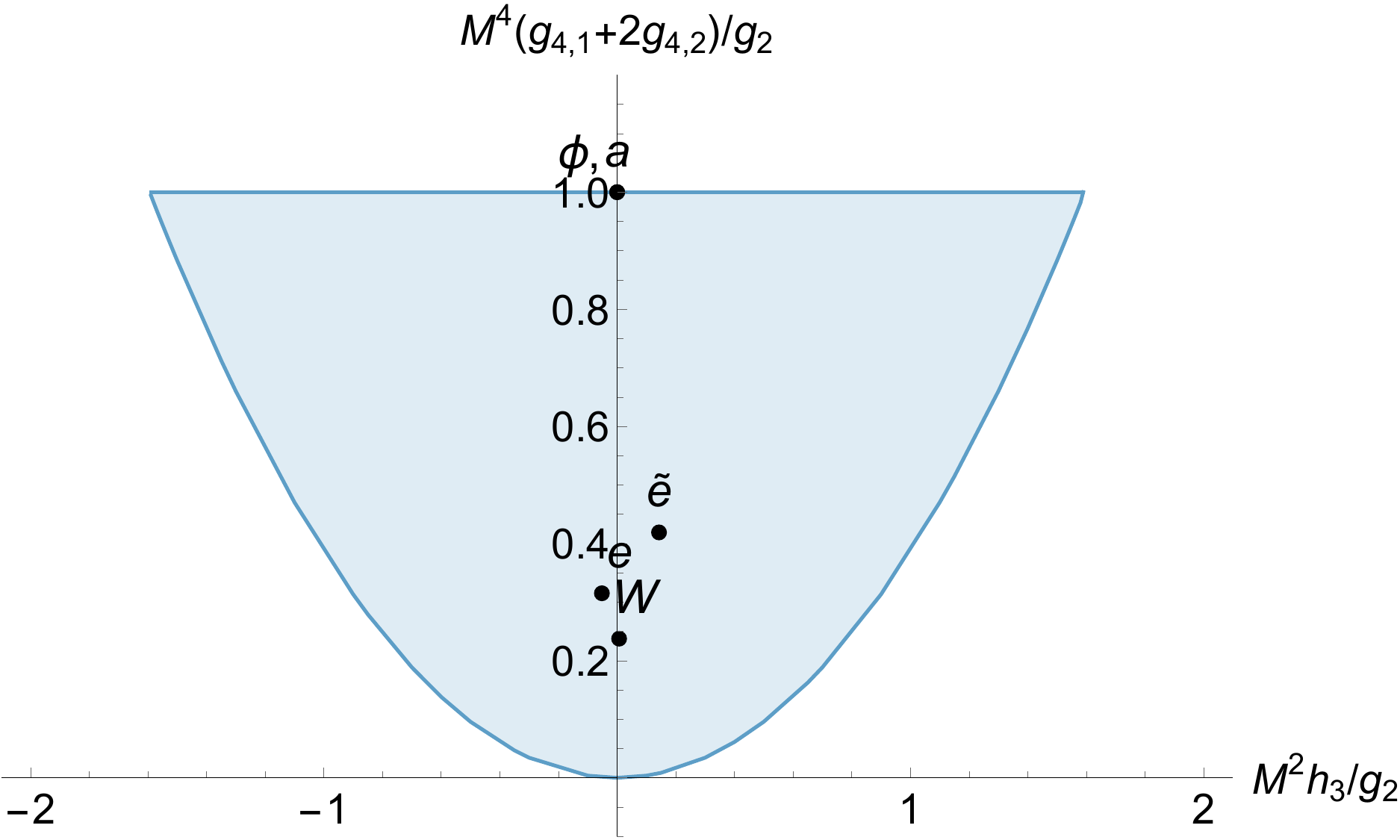}
	\end{subfigure}
	\hfill
		\caption{On the left, allowed region in the space $(M^4 (g_{4,1}+2g_{4,2})/g_2,M^2 g_3/g_2)$. We notice that it rules out a kink we were speculating about in \cite{Henriksson:2021ymi}. On the right, a bound in the $(M^4 (g_{4,1}+2g_{4,2})/g_2,M^2 h_3/g_2)$ space.}
	\label{fig:app3}
\end{figure}

\begin{figure}
	\centering
	\begin{subfigure}[t]{0.45\textwidth}
		\centering
		\includegraphics[width=\textwidth]{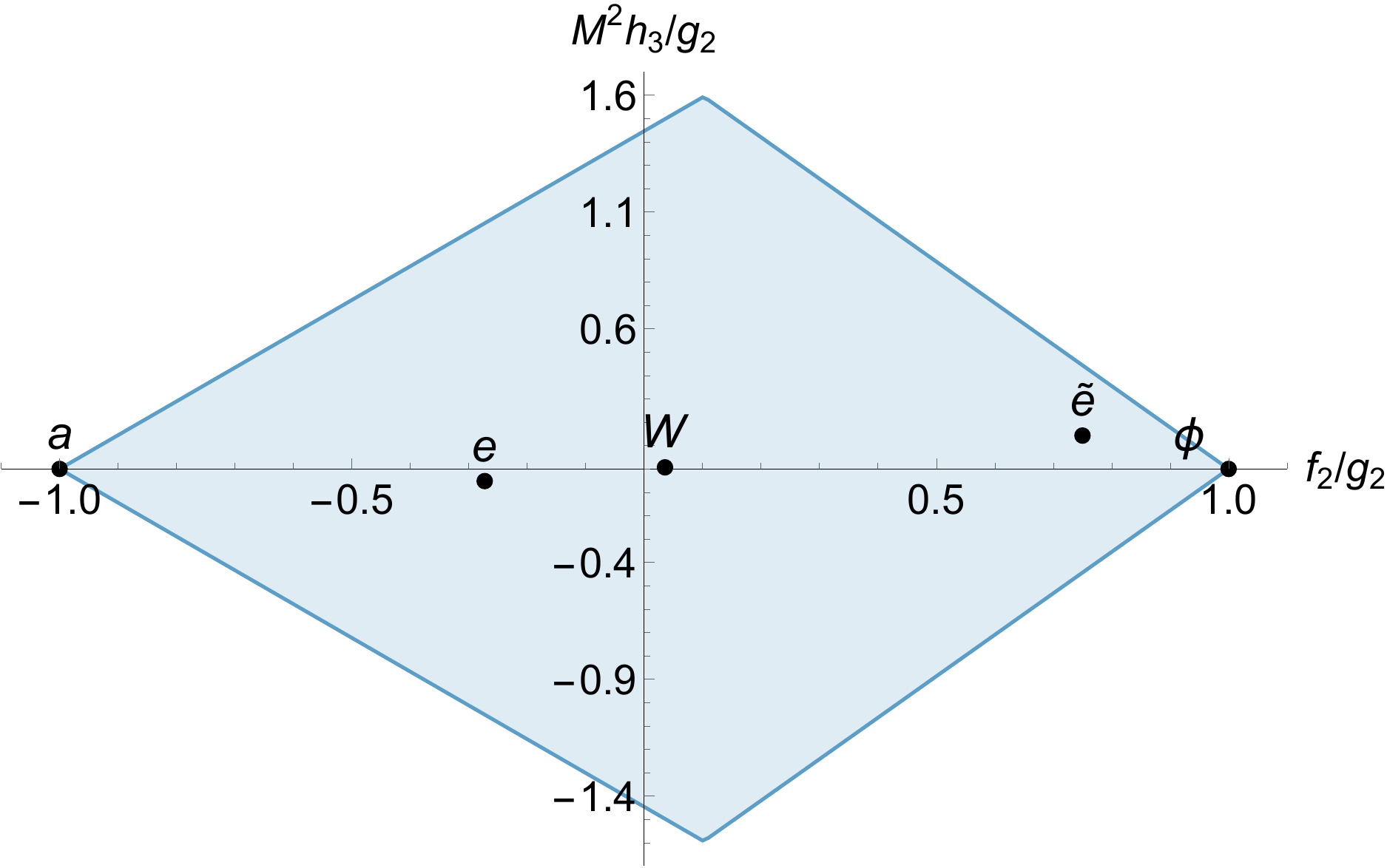}
	\end{subfigure}
	\hfill
	\begin{subfigure}[t]{0.45\textwidth}
		\centering
		\includegraphics[width=\textwidth]{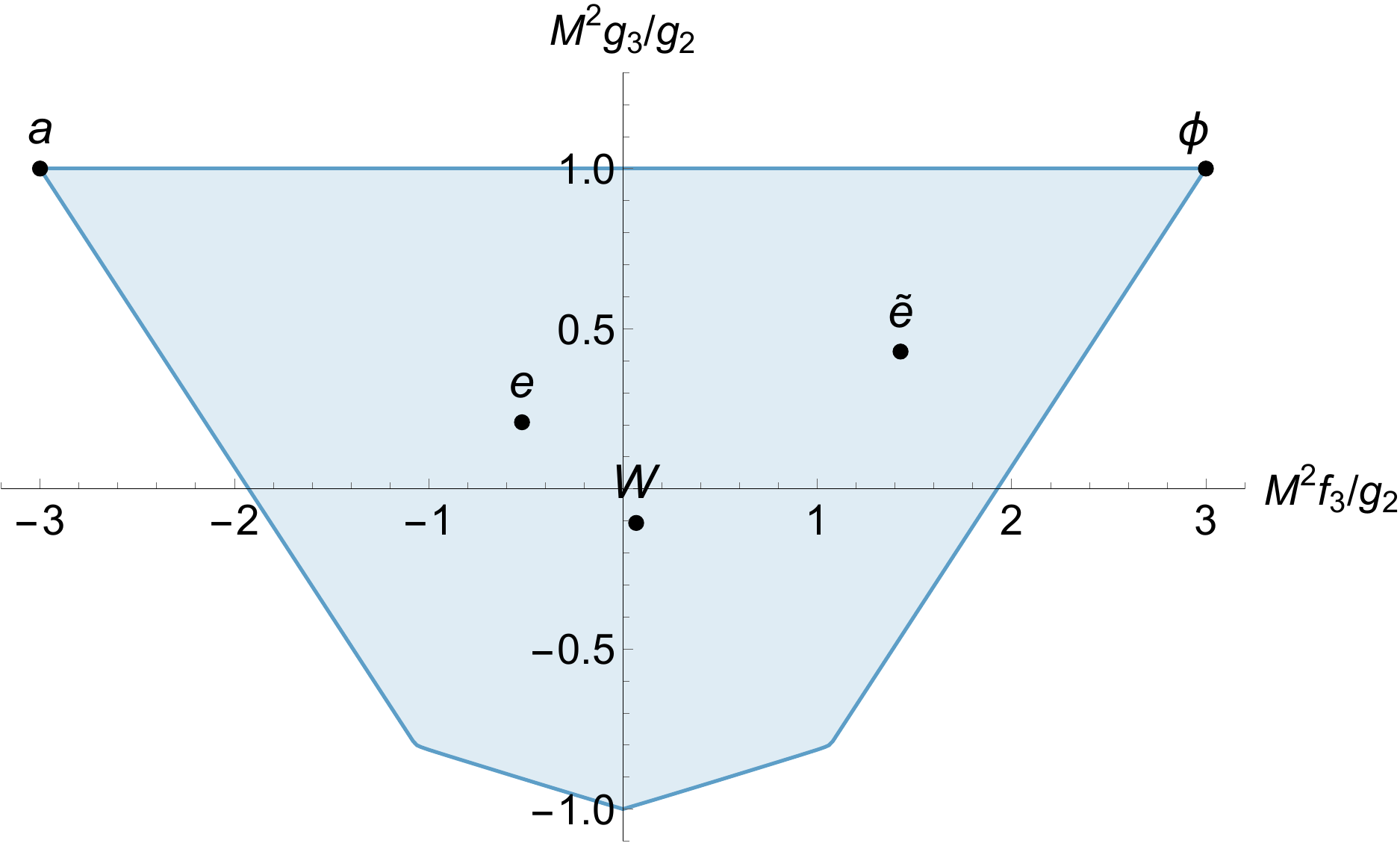}
	\end{subfigure}
	\hfill
	\caption{Allowed regions in the space $(M^2 h_3/g_2,f_2/g_2)$ on the left, $(M^2 g_3/g_2,M^2 f_3/g_2)$ on the right.}
	\label{fig:app4}
\end{figure}

Here we put results for additional bounds without gravity, involving different combinations of the low-energy ciefficients, see figures~\ref{fig:app1}, \ref{fig:app2}, \ref{fig:app3} and \ref{fig:app4}. Comparing with the results in \cite{Henriksson:2021ymi}, we see that in general there is an improvement of one side of the bounds due to the presence of the new null constraint of order $m^{-6}$. Indeed, we notice that the tree-level completions from table \ref{tab:EFTcoefValues_trees} (scalar and axion) now always saturate our bounds, while the loop-level ones lie inside them.

\begin{table}[ht]
	\centering
	\caption{EFT coefficients of partial UV completions both tree- (scalar, axion) and loop-level (QED, sQED, $W^{\pm}$) from~\cite{Henriksson:2021ymi}.} \label{tab:EFTcoefValues_trees}
	{
		\renewcommand{\arraystretch}{1.5}
		\begin{tabular}{|c|cc|ccc|ccc|}
			\hline
			&  \multicolumn{2}{c|}{ $\Delta=8$} & \multicolumn{3}{c|}{ $\Delta=10$}& \multicolumn{3}{c|}{ $\Delta=12$} 
			\\
			Completion& $f_2$ & $g_2$ & $f_3$ & $g_3$ & $h_3$ & $f_4$ & $g_{4,1}$ & $g_{4,2}$ 
			
			\\\hline\hline
			Scalar
			&  $\frac{4g^2}{M^4}$&$\frac{4g^2}{M^4}$
			& $\frac{12g^2}{M^6}$ & $\frac{4g^2}{M^6}$ & $0$
			& $\frac{2g^2}{M^8}$ & $\frac{4g^2}{M^8}$ & $0$
			
			\\\hline
			Axion
			&  $-\frac{4g^2}{M^4}$&$\frac{4g^2}{M^4}$
			& $\!-\frac{12g^2}{M^6}$ & $\frac{4g^2}{M^6}$ & $0$
			& $\!-\frac{2g^2}{M^8}$ & $\frac{4g^2}{M^8}$ & $0$

			\\\hline\hline
			QED 
			& $-\frac{\alpha^2}{15m_e^4}$  & $\frac{11\alpha^2}{45m_e^2}$
			& $\!-\frac{2\alpha^2}{63m_e^6}$ & $\frac{4\alpha^2}{315m_e^6}$ & $-\frac{\alpha^2}{315m_e^6}$
			& $\!-\frac{\alpha^2}{945m_e^8}$ & $\frac{41\alpha^2}{18900m_e^8}$ & $\frac{\alpha^2}{756m_e^8}$ 
			
			\\\hline
			Scalar QED & $\frac{\tilde\alpha^2}{30m_{\tilde e}^4}$&$\frac{2\tilde\alpha^2}{45m_{\tilde e}^4}$
			& $\frac{\tilde\alpha^2}{63m_{\tilde e}^6}$& $\frac{\tilde\alpha^2}{210m_{\tilde e}^6}$& $\frac{\tilde\alpha^2}{630m_{\tilde e}^6}$
			& $\frac{\tilde\alpha^2}{1890m_{\tilde e}^8}$& $\!\frac{17\tilde\alpha^2}{18900m_{\tilde e}^8}\!$& $\frac{\tilde\alpha^2}{7560m_{\tilde e}^8}$ 
			
			\\\hline
			${W^\pm}$ sector & $\frac{\alpha^2}{10m_W^4}$&$\frac{14\alpha^2}{5m_W^4}$ & $\frac{\alpha^2}{21m_W^6}$ & $\!\!-\frac{47\alpha^2}{630m_W^6}$ & $\frac{\alpha^2}{210m_W^6}$ & $\frac{\alpha^2}{630m_W^8}$ & $\!\!-\frac{83\alpha^2}{6300m_W^8}$ & $\frac{23\alpha^2}{840m_W^8}$ 
			
			\\\hline
		\end{tabular}
	}
\end{table}


\section{Details about numerical implementation}
\label{app:numerical}

In this section we provide technical details about the numerical implementation of the algorithm discussed in section \ref{sec:algorithm} in the main text. As discussed there, we need to impose positivity conditions \eqref{eq:sdpb-conditions} for all $m^2\geqslant M^2$ and on an infinite sequence of values of $\ell$. Let us discuss the various regimes, as outlined in figure~\ref{fig:positivityregimes} in section~\ref{sec:positivity}. All the parameters introduced in the following discussion are reported in table~\ref{tab:params}.
\begin{table}[ht]
	\centering
	\caption{Parameters used in the numerical implementation.} \label{tab:params}
	{
		\renewcommand{\arraystretch}{1.5}
		\begin{tabular}{cccccccccc}
			\hline
				$ \ell_\text{max}$ &  $m^2_\text{max}$ & $B_\text{max}$ &  $\delta_b$   & $\delta_m$ &  $\delta_y$ & $n_M$ & $n_L$  & $n_b$ & $ n_\text{max}$
			\\\hline
			 40 &  5 & 40  & $10^{-2}$   & $10^{-3}$ & 1/200 &10 & 10  & 10 & 15/2  \\
			 \hline
		\end{tabular}

		\begin{tabular}{c|c}
			\hline
				& $\phantom{-.}$ non-standard  \texttt{SDPB}  parameter $\phantom{--}$   \\
			\hline
				 \texttt{precision} & 1024\\
				 \texttt{dualityGapThreshold} &  $10^{-10}$ \\
				 \texttt{initialMatrixScaleDual} & $10^{80}$\\
				 \texttt{maxIteration} & 1000\\
			\hline
		\end{tabular} 
	}
\end{table}
\begin{enumerate}
\item Finite $m^2$, finite $\ell$: we impose positivity on the following set of spins:
\begin{equation}
\label{eq:ells}
\ell \in \{ 0,2,3,4,5,\ldots, 40, 50, 60, 70,80,90,100,200\}\,.
\end{equation}
For each value of $\ell$, we include a discrete set of values for $m^2$:
\begin{equation}
m^2 \in \{ M^2, M^2+ \delta_m , M^2 + 2\delta_m ,\ldots , m^2_\text{max}\,\}\,.
\end{equation}
We checked that decreasing the interval of the discretization does not change the bounds substantially.
\item Large $m^2$, finite $\ell$: we computed a polynomial approximation of the high-energy part of the dispersion relation \eqref{eq:functionsI} integrated against $p^n$ for any $\ell$ as in \eqref{eq:ells}. This is done by Taylor expanding these expressions around $m^2=\infty$ up to a certain order $n_M$, multiplying by the appropriate power of $m^2$ and substituting $m^2=m^2_\text{max}(1+x)$. The positivity condition is equivalent to the positivity of the resulting polynomial for any $x\geqslant 0$. This condition can be easily implemented in \texttt{SDPB}. We also checked that our approximation reproduces the original function to high precision for all $m\geqslant m_\text{max}$.
\item Finite $m^2$, large $\ell$: we computed a polynomial approximation of the high-energy part of the dispersion relation \eqref{eq:functionsI} integrated against $p^n$ for fixed values of $m^2$. This is done by Taylor expanding these expressions around $\ell=\infty$ up to a certain order $n_L$, multiplying by the appropriate power of $\ell$ and substituting $\ell=(500+x)$. Since we are interested in the large $\ell$ behavior, in all expressions \eqref{eq:functionsI}, we can neglect hypergeometric functions, which are rapidly oscillating in this limit.
We then require positivity of the resulting polynomial for any $x\geqslant 0$. The chosen values of $m^2$ are
\begin{equation}
m^2 = \frac{M^2}{1-y}\,, \quad \text{with} \quad y\in\{\delta_y, 2\delta_y, \ldots  ,1-\delta_y\}\,.
\end{equation}

\item Large $m^2$, large $\ell$, fixed $b=2\ell/m$: this limit is controlled by \eqref{eq:largespin}. We impose positivity on a discrete set of values of $b$:
\begin{equation}
b \in \{ \delta_b,2\delta_b,\ldots, B_\text{max}\}\,.
\end{equation}
To take into account the large $b$ regime, we Taylor expanded \eqref{eq:largespin}  around $b=\infty$ up to a certain order $n_b$ as in \eqref{eq:leadingAsymp}. Following \cite{Caron-Huot:2021rmr,Caron-Huot:2022ugt}, we rewrite an expression containing oscillating terms in terms of a $2\times2$ matrix. For instance
\begin{align}
C_{0,n}(b)\simeq& \frac1{b^{\frac32+n_b}}\left(A_n(b) + B_n(b) \cos(b) + D_n(b)\sin(b)\right)\\
\simeq &\frac1{b^{\frac32+n_b}} \begin{pmatrix} \cos(\frac{b}2) & \sin(\frac{b}2)  \end{pmatrix}\cdot
 \begin{pmatrix} A_n(b) + B_n(b) & D_n(b) \\ D_n(b) &A_n(b) - B_n(b)  \end{pmatrix}
 \cdot  \begin{pmatrix} \cos(\frac{b}2) \\ \sin(\frac{b}2)  \end{pmatrix} 
\,,\nonumber
\end{align}
where $A_n,B_n,D_n$ are polynomials in $b$. Given our choices of functional, they only contain integer powers of $b$.\footnote{If one were to use different functionals, then the polynomials $A_n$, $B_n$, $D_n$ would contain also fractional powers of $b$, One could deal with this issue by redefining $b\rightarrow (\tilde b)^k$, with appropriate $k$. The only complication is an increase of the degree of the polynomials.} We then replace the positivity condition with the stronger requirement
\begin{align}
\sum a_n C_{0,n}(b) \geqslant 0\ \longrightarrow \ \sum a_n  \begin{pmatrix} P^{(n)}_1(b) + P^{(n)}_2(b) & P^{(n)}_3(b) \\ P^{(n)}_3(b) &P^{(n)}_1(b) - P^{(n)}_2(b)  \end{pmatrix} \succcurlyeq 0
\,.
\end{align}
A similar replacement can be done for any combination of $C_{\nu,n}(b)$. On the other hand the positivity conditions  \eqref{eq:sdpb-conditions} also involve semi-definite conditions on linear combination of matrices. In the large $m^2,\ell$ limit these matrices also contain oscillating terms, as shown in  \eqref{eq:largespin}. Again we can replace each element of the matrices with a two by two matrix. For instance:
\begin{align}
& \begin{pmatrix}C_{0,n}\\0&C_{0,n}\end{pmatrix} \rightarrow \begin{pmatrix}
  {\footnotesize  \begin{pmatrix} A_n(b) + B_n(b) & D_n(b) \\ D_n(b) &A_n(b) - B_n(b)  \end{pmatrix}} & {\footnotesize  \begin{pmatrix} 0&0\\0&0 \end{pmatrix}}
 \\ {\footnotesize  \begin{pmatrix} 0&0\\0&0 \end{pmatrix}} 
 &  {\footnotesize  \begin{pmatrix} A_n(b) + B_n(b) & D_n(b) \\ D_n(b) &A_n(b) - B_n(b)  \end{pmatrix}}  \end{pmatrix} , \nonumber \\
& \begin{pmatrix}0&C_{2,n}\\C_{2,n}&0\end{pmatrix} \rightarrow \begin{pmatrix} 
   {\footnotesize  \begin{pmatrix} 0&0\\0&0 \end{pmatrix}} &
 {\footnotesize  \begin{pmatrix} A'_n(b) + B'_n(b) & D'_n(b) \\ D'_n(b) &A'_n(b) - B'_n(b)  \end{pmatrix}} \\
    {\footnotesize  \begin{pmatrix} A'_n(b) + B'_n(b) & D'_n(b) \\ D'_n(b) &A'_n(b) - B'_n(b)   \end{pmatrix}} &
     {\footnotesize  \begin{pmatrix} 0&0\\0&0 \end{pmatrix}}
    \end{pmatrix}.
\end{align}
We then demand positivity of linear combinations of the resulting $4\times 4$ matrices, which is again a stronger condition but has the advantage of being in the form of a polynomial matrix, which can be fed to \texttt{SDPB}.
\end{enumerate}

Finally, let us list the procedures followed to obtain each figure:
\begin{itemize}
\item Figure~\ref{fig:g2VSh2sq}: we used the dispersion relations $\mathcal I_g, \mathcal I_0$ and $\mathcal I_{\beta^2}$. This is equivalent to considering $\Lambda[\vec v] = \sum_i \Lambda_i[v_i]$, with $i=1,4,5$.  The low-energy part only depends on $g_2$ and $\beta$. Just as we did in section~\ref{sec:WorkedExample2}, obtained bounds along rays in the $(g_2,\beta^2)$ plane:
we looked for a functional satisfying (see \eqref{eq:I-low})
\begin{align}
& \frac{1}{M^4 }\cos\theta \Lambda_1[1] + \frac{M^2_\textrm{P}}{M^6}\sin\theta \left(\Lambda_1[4p^2]-\Lambda_5[p^2]\right)=1 \,,\\
&\text{maximize} \quad - \Lambda_1[p^{-2}]
\,.
\end{align}

\item Figure~\ref{fig:g2VSf2}: we used the dispersion relations $\mathcal I_g, \mathcal I_f, \mathcal I_0$ and $I_{\beta^2}$. This is equivalent to considering $\Lambda[\vec v] = \sum_i \Lambda_i[v_i]$, with $i=1,2,4,5$.  The low-energy part only depends on $g_2$, $\beta$, $f_2$ and $f_3$. This time we obtained bounds along rays in the $(g_2,f_2)$ plane while leaving $\beta^2$ and $f_3$ undetermined. Hence we looked for a functional satisfying (see \eqref{eq:I-low})
\begin{align}
& \frac{1}{M^4}\cos\theta \Lambda_1[1] + \frac{1}{M^4}\sin\theta\Lambda_2[1] =1\,,\\
& -\left(\Lambda_1[4p^2]+5\Lambda_2[p^2]-\Lambda_5[p^2]\right) \geqslant 0 \,,\\
&  \pm \Lambda_2[p^2] \geqslant0 \,,\\
&\text{maximize} \quad - \Lambda_1[p^{-2}]\,.
\end{align}
The second and third condition ensures that we can discard the contribution of $\beta^2$ and $f_3$, provided that the latter has fixed sign. Hence the two distinct allowed regions. The overall allowed region is the union of the two regions.

\item Figure~\ref{fig:g2VSbeta}: we used the dispersion relations $\mathcal I_g, \mathcal I_h, \mathcal I_0$ and $\mathcal I_{\beta^2}$. This is equivalent to considering $\Lambda[\vec v] = \sum_i \Lambda_i[v_i]$, with $i=1,3,4,5$.  The low-energy part only depends on $g_2$, $\beta$, $\beta$ and $h_3$. This time we cannot get bounds along rays since both $\beta$ and $\beta^2$ appear. Hence we scanned over $\beta$, and got bounds on $g_2$. We looked for a functional satisfying (see \eqref{eq:I-low})
\begin{align}
&  \Lambda_1[1]  = 1\,,\\
&  \pm \Lambda_3[p^2] \geqslant 0 \,,\\
&\text{maximize} -\left( \Lambda_1[p^{-2}]  + \beta^2 \left(\Lambda_1[4p^2]-\Lambda_5[p^2]\right) +\beta\Lambda_3[1] \right)\,.
\end{align}
The second condition ensures that we can discard the contribution of $h_3$, provided that it has fixed sign. It turns out that the allowed region is independent of the choice of the sign.

\end{itemize}

\bibliography{cite.bib}

\bibliographystyle{JHEP.bst}

\end{document}